# Characterization and decomposition of the natural van der Waals heterostructure SnSb$_2$Te$_4$ under compression

*Juan A. Sans,*[a,*] *Rosario Vilaplana,*[b] *E. Lora Da Silva,*[a] *Catalin Popescu,*[c] *Vanesa P. Cuenca-Gotor,*[a] *Adrián Andrada-Chacón,*[d] *Javier Sánchez-Benitez,*[d] *Oscar Gomis,*[b] *André L. J. Pereira,*[a,e] *Plácida Rodríguez-Hernández,*[f] *Alfonso Muñoz,*[f] *Dominik Daisenberger,*[g] *Braulio García-Domene,*[h] *Alfredo Segura,*[h] *Daniel Errandonea,*[h] *Ravhi S. Kumar,*[i] *Oliver Oeckler,*[j] *Julia Contreras-García,*[k] *and Francisco J. Manjón*[a]

[a] Instituto de Diseño para la Fabricación y Producción Automatizada, MALTA-Consolider Team, Universitat Politècnica de València, Valencia, Spain
[b] Centro de Tecnologías Físicas, MALTA-Consolider Team, Universitat Politècnica de València, Valencia, Spain
[c] ALBA-CELLS, Barcelona, Spain
[d] Departamento de Química-Física, MALTA-Consolider Team, Universidad Complutense de Madrid, Madrid, Spain
[e] Grupo de Pesquisa de Materiais Fotonicos e Energia Renovavel - MaFER, Universidade Federal da Grande Dourados, Dourados, MS 79825-070, Brazil
[f] Departamento de Física, MALTA-Consolider Team, Instituto de Materiales y Nanotecnología, Universidad de La Laguna, Tenerife, Spain
[g] Diamond Light Source Ltd, Oxon, England
[h] Departamento de Física Aplicada-ICMUV, MALTA-Consolider Team, Universidad de Valencia, Valencia, Spain
[i] Department of Physics, University of Illinois at Chicago, Chicago IL 60607-7059, USA
[j] Institut für Mineralogie, Kristallographie und Materialwissenschaft, Universität Leipzig, Germany
[k] CNRS, UMR 7616, Laboratoire de Chimie Théorique, F-75005, Paris, France

**Abstract**

This joint experimental and theoretical study of the structural, vibrational and electrical properties of rhombohedral SnSb$_2$Te$_4$ at high pressure unveils the internal mechanisms of its compression. The equation of state and the internal polyhedral compressibility, the symmetry and behavior of the Raman-active modes and the electrical behavior of this topological insulator under compression have been discussed and compared with the parent binary α-Sb$_2$Te$_3$ and SnTe compounds and with related ternary compounds. Our X-ray diffraction and Raman measurements together with theoretical calculations, which include topological electron density and electronic localization function analysis, evidence the presence of an isostructural phase transition around 2 GPa and a Fermi resonance around 3.5 GPa. The Raman spectrum of SnSb$_2$Te$_4$ shows vibrational modes that are forbidden in rocksalt SnTe; thus showing a novel way to experimentally observe the forbidden vibrational modes of some compounds. Additionally, since SnSb$_2$Te$_4$ is an incipient metal, like its parent binary compounds, we establish a new criterion to identify the recently proposed metavalent bonding in complex materials when different bond characters coexist in the system. Finally, SnSb$_2$Te$_4$ exhibits a pressure-induced decomposition into the high-pressure phases of its parent binary compounds above 7 GPa, which is supported by an analysis of their formation enthalpies. We have framed the behavior of SnSb$_2$Te$_4$ within the extended orbital radii map of $BA_2$Te$_4$ compounds, which paves the way to understand the pressure behavior and stability ranges of other layered van der Waals-type compounds with similar stoichiometry.

**Popular Summary**

The search for new materials with exciting electronic properties has triggered the recent discovery of topological insulating materials, such as SnSb$_2$Te$_4$, which is also an incipient metal showing metavalent bonding. Pressure is a valuable tool and can be employed to explore the structural stability, since it involves variation of the interatomic distances, and therefore the tuning of the properties of these materials, similar to the capability to host dopants, which may the properties of materials. This work provides new insight on the behaviour of $SnSb_2Te_4$ under extreme conditions of pressure. It sheds light on the lattice dynamics under compression, revealing the nature of the interatomic interactions given in this compound, even the recently discovered metavalent bonds. Moreover, this work provides revealing proofs of how this compound decomposes into their binary constituents once reached a critical pressure. Our study also identifies the vibrational modes of the parent binary materials from the Raman spectrum of the ternary compound, which is pointed out as a new method to overstep limitations given by the selection rules in the observation of the vibrational properties in materials without Raman-active modes. In a general context, we have provided an updated version of the orbital radii map, describing the whole family of $BA_2X_4$ compounds, which allows us to extend our results in $SnSb_2Te_4$ to at least another compound ($SnBi_2Te_4$). Thus, it can be expected that chemical pressure might induce their decomposition into binary constituents above certain dopants threshold in both compounds. Finally, we provide a new criterion to identify the recently proposed metavalent bonding in complex materials, like $SnSb_2Te_4$, where different types of bonds can be found.

I. INTRODUCTION

The search for topological features in materials including topological insulators (TIs) and topological superconductors (TSs) is currently one of the hot topics in the Material Science field because of its interest in fundamental physics and applications in spintronics and quantum computation **[1-3]**. Recently, several tools have been designed to identify theoretically-predicted TI candidates and create an extensive database of compounds that exhibit these properties **[4-6]**, defining respective limitations and filters required by theoretical calculations to avoid false-positive predictions **[7]**. The discovery of 3D-TI properties in layered tetradymite $A_2X_3$ compounds, such as $Bi_2Te_3$, $Bi_2Se_3$ and $Sb_2Te_3$ compounds **[3,8,9]** and the trivial insulating behavior of $Sb_2Se_3$ **[10]** have opened the question about which is the electronic and structural origin and the limiting factors of this conduct, compared to other similar materials, such as the layered tetradymite-like $BA_2X_4$ compounds.

Binary $A_2X_3$ layered compounds are usually *p*-type semiconductors with a narrow gap, leading to a high electrical conductivity **[11]**. In particular, $Sb_2Te_3$ and $Bi_2Te_3$ are the best thermoelectric materials near room temperature found to date **[12,13]**, since they are layered materials formed by heavy atoms that feature a very small thermal conductivity. Additionally, the hybridization between the valence and the conduction band states favored by a large spin-orbit coupling (SOC) and a small bandgap, leads to the formation of Dirac cones in the electronic band structure that is responsible for their 3D-TI properties **[8]**. The TI properties observed in $A_2X_3$ binary compounds with layered tetradymite structure, **[8, 14-16]** have triggered the exploration of ternary $BA_2X_4$ compounds based on those materials because they are expected to show richer physics than respective binary counterparts, plus the possibility to tune their properties by selecting appropriate *B* atoms. In fact, 3D-TI behavior has been predicted in many $BA_2X_4$ compounds **[11,14-19]** and rhombohedral SnSb$_2$Te$_4$ with layered tetradymite-like structure has been found to be a *p*-type 3D-TI **[11]**.

The tetradymite $R\overline{3}m$ structure [space group (s.g.) 166, Z=3] of binary $A_2X_3$ compounds is formed by a block composed by five layers (*X*I–*A*–*X*II–*A*–*X*I) called *quintuple* layer (*QL*), being *A* and *X* the cation and the anion, respectively, and *X*I and *X*II being the two non-equivalent anions in the unit cell. These *QL*s are piled up along the hexagonal *c* axis and are linked by van der Waals (vdW) forces leading to the 3D material **[8]**. In ternary $BA_2X_4$ compounds the tetradymite $R\overline{3}m$ structure is formed by replacing the central anion (*X*II) of $A_2X_3$ compounds with a three-atoms block (*X*II-*B*-*X*II). In this way, an atomic block of seven layers (*X*I–*A*–*X*II–*B*–*X*II–*A*–*X*I) called *septuple* layer (*SL*), with a central block formed by a central *B* cation surrounded by two additional anions, is formed **[20]**. In these ternary compounds, the tetradymite structure can be described by two octahedra formed by *A* and *B* cations surrounded by six *X*I/*X*II and *X*II anions, respectively. These ternary compounds with tetradymite-like layered structure defines a new family of materials, named "natural van der Waals heterostructures", which can be prepared by nanosheet form and promise novel and interesting properties **[21]**.

Due to their technological interest, the thermoelectric properties of some layered tetradymite-like $BA_2X_4$ compounds have been studied **[21-25]** and experimental results have shown that there is a certain disorder in the crystal structure **[17, 26-31]**. In particular, SnBi$_2$Te$_4$ and SnSb$_2$Te$_4$ are the $BA_2X_4$ materials showing the smaller cation exchange in their atomic sites **[23,31,32]**. Noteworthy, some $BA_2X_4$ compounds, like SnSb$_2$Te$_4$ and GeSb$_2$Te$_4$, have revealed strong properties as phase change materials since they are able to rapidly change between an amorphous and a crystalline state by light irradiation or current application. In particular, SnSb$_2$Te$_4$ crystallizes predominantly in the rhombohedral structure $R\overline{3}m$ (s.g. 166, Z=3) while a smaller fraction crystallizes in the metastable rocksalt-type structure $Fm\bar{3}m$ (s.g. 225, Z=4). Additionally, it can be stabilized in an amorphous phase with average octahedral coordination in the short-range order [33]. In this context, it has been shown that SnSb$_2$Te$_4$ and GeSb$_2$Te$_4$ have strong similarities being both p-type semiconductors, whose electrical conductivity is governed by defects [32]. Moreover, SnSb$_2$Te$_4$ is, like GeSb$_2$Te$_4$ and its binary parent compounds cubic rocksalt SnTe (c-SnTe) and Sb$_2$Te$_3$, an incipient metal showing the recently proposed metavalent bonding [34,35]. Therefore, SnSb$_2$Te$_4$ is a ternary chalcogenide whose study could help to understand metavalent bonding in complex ternary materials.

High pressure (HP) studies have been conducted in rocksalt-type SnSb$_2$Te$_4$ and GeSb$_2$Te$_4$ [36] and in rhombohedral SnBi$_2$Te$_4$ [37]. A pressure-induced amorphization (PIA) was reported in rocksalt-type SnSb$_2$Te$_4$ (GeSb$_2$Te$_4$) upon compression above 11 (15) GPa [36], whereas a pressure-induced electronic topological transition (ETT) has been suggested to occur in rhombohedral SnBi$_2$Te$_4$ [37]. However, many questions have yet to be addressed for the rhombohedral $BA_2X_4$ compounds and in particular for SnSb$_2$Te$_4$. For instance, "how does the presence of the new SnTe$_6$ octahedron in rhombohedral SnSb$_2$Te$_4$ affect the properties of the host Sb$_2$Te$_3$ structure?" and "how does rhombohedral SnSb$_2$Te$_4$ behave under compression?". In this scenario, we wonder if pressure on rhombohedral SnSb$_2$Te$_4$ leads to: i) a simple compression of the material remaining in the original structure; ii) an isostructural phase transition (IPT) followed by a structural phase transition (PT) towards a different structure, like in parent compound Sb$_2$Te$_3$; iii) an ETT, similar to its counterpart SnBi$_2$Te$_4$; iii) a PIA, reported in rocksalt-type SnSb$_2$Te$_4$ or iv) a pressure-induced decomposition (PID).

In this work, we report the room-temperature structural, vibrational and electrical properties of SnSb$_2$Te$_4$ under compression from an experimental and theoretical point of view by means of angle-dispersive X-ray diffraction (ADXRD), Raman scattering (RS) and transport measurements complemented with DFT *ab-initio* calculations. A good agreement has been found between both experimental and theoretical descriptions. The behavior of this material under compression has been compared to that of its parent binary compounds ($Sb_2Te_3$ and c-SnTe), where we pay special attention to the evolution of the interlayer vdW interaction under compression, considering this feature the key element to understand the behavior of the *c/a* ratio under compression and the stability pressure range of its low-pressure (LP) phase. This material can be considered a good candidate with incipient metal properties along the whole of stability pressure range, because both parent binary compounds are incipient metals **[35]**, forming metavalent interactions between Sb-Te and Sn-Te bonds. We will further demonstrate that $SnSb_2Te_4$ undergoes a pressure-induced IPT near 2 GPa followed by a PID above 7 GPa. The contextualization of this result on the framework of the ternary $BA_2X_4$ compounds can shed light on their behavior under compression. Finally, we will show that the Raman spectrum of $SnSb_2Te_4$ and its comparison with the theoretical vibrational properties of $SnSb_2Te_4$ and those of its parent compounds has revealed that: i) there is a Fermi resonance around 3.5 GPa, similar to what occurs in c-SnTe, and ii) the Raman spectrum of $SnSb_2Te_4$ shows some vibrational modes similar to those of forbidden c-SnTe. This result evidences a novel procedure to experimentally observe the forbidden vibrational modes of some materials. Finally, we will show that $SnSb_2Te_4$ is an incipient metal along its whole stability pressure range, because of the presence of metavalent interactions between Sb-Te and Sn-Te bonds. In this context, metavalent bonding will be analyzed with electron density analysis thus providing a new method to distinguish it from ionic, covalent and metallic interactions in complex compounds where very different atomic interactions can be found.

## II. RESULTS AND DISCUSSION

### A. X-ray diffraction under pressure

#### 1. LP phase

As already commented, SnSb$_2$Te$_4$ crystallizes in rhombohedral $R\bar{3}m$ phase following a GeSb$_2$Te$_4$ structure-type with four atoms at the independent Wyckoff sites (Sn at 3a and Sb, Te1 and Te2 at 6c). A clear scheme of the polyhedral arrangement for the GeSb$_2$Te$_4$-type structure is shown in **Figure 1**, together with a description of the different layers composing this compound. The experimental ADXRD pattern at room conditions (see **Figure S1** in Supplemental Material (SM)) has been fitted with a von Dreele-type Le Bail refinement, which yields a hexagonal unit-cell volume of 662.7(7) Å$^3$ with lattice parameters $a$= 4.2977(1) Å and $c$= 41.43(4) Å. These values are in good agreement with those reported in the literature and obtained from theoretical simulations (see **Table I**). The $c$-lattice parameter is found to be 10 times longer than the $a$-lattice parameter, which is a clear feature of the layered nature of this phase.

Two ADXRD experiments were carried out by employing helium (up to 37 GPa) and silicone oil (up to 12 GPa) as a pressure-transmitting medium (PTM). The latter experiment up to 12 GPa (experiment 2) was performed in order to conduct a more detailed study of the LP phase. ADXRD patterns of SnSb$_2$Te$_4$ at different pressures from experiment 2 are shown in **Figure S1** in SM.

In **Figure S1** it is possible to observe a clear shift of the Bragg reflections of the LP phase of SnSb$_2$Te$_4$ towards higher *2θ* angles with increasing pressure. This feature occurs due to the monotonous decrease of the interatomic distances and a clear change in the diffraction patterns above 7 GPa suggesting that the LP phase is no longer stable. We want to stress here that the lack of Rietveld refinement of our measurements prevents us from obtaining the experimental atomic positions in SnSb$_2$Te$_4$. Consequently, we have used the atomic positions of our calculations to discuss the pressure dependence of many parameters in order to understand the behavior of SnSb$_2$Te$_4$ under compression. The good agreement between the experimental and theoretical pressure dependence of the unit cell volume, lattice parameters and the *c/a* ratio discussed in the following paragraphs, assures us of the accuracy of our theoretical data for further considerations.

The pressure dependence of the experimental and theoretical unit cell volume and lattice parameters of SnSb$_2$Te$_4$ of the two experiments is displayed in **Figures 2a and 2b**, respectively. The slight discrepancy observed in both figures between the values obtained using helium and silicone oil as PTM can be explained by the typical pressure uncertainty in these experiments, considered to be around 0.5 GPa. The experimental unit cell volume (of both experiments) fitted to a third-order Birch-Murnaghan equation of state (BM-EoS), given by the positive trend of the F-f plot calculated from EoSFIT [38]. Experimental BM-EoS are displayed in **Figure 2a**, yields a zero-pressure unit cell volume, $V_0$, and bulk modulus, $B_0$, of 663.1(6) Å$^3$ and 31.6(4) GPa, respectively. These values are in good agreement with our calculations (see **Table I**).

The similar features of the layered structure of SnSb$_2$Te$_4$ and its parent binary compound α-Sb$_2$Te$_3$ suggests that the compressibility of the former must be related to that of the latter. More precisely, the compression of the former should be a combination of the compression of the SbTe$_6$ polyhedral units, also present in α-Sb$_2$Te$_3$, and of the SnTe$_6$ polyhedral units (a quasi-regular SnTe$_6$ octahedron) in the center of the *SL*, which are also present in c-SnTe. This hypothesis is supported by the similar volume compressibility of both SnTe$_6$ and SbTe$_6$ octahedral units in SnSb$_2$Te$_4$ and those occurring in its two parent binary compounds as shown in **Figure S2** in SM.

The evolution of the theoretical interatomic distances of SnSb$_2$Te$_4$ at HP (**Figure S3** in SM) shows that the Sn-Te bond distance inside the quasi-regular SnTe$_6$ octahedron compresses at a similar rate than the Sn-Te distances in c-SnTe with $Fm\bar{3}m$ structure ($B_0 \approx$ 50 GPa) [39]. Since the Sn-Te distance is the second less compressible interatomic distance of the SnSb$_2$Te$_4$ structure, it can be assumed that the compressibility of SnSb$_2$Te$_4$ is mostly determined by the compressibility of α-Sb$_2$Te$_3$. This hypothesis is confirmed by both experimental and theoretical data, shown in **Table I**. Note that α-Sb$_2$Te$_3$ shows an experimental average $B_0$ of 36.1(9) GPa **[40]** that is in very good agreement with the experimental bulk modulus of SnSb$_2$Te$_4$. In conclusion, the compressibility of SnSb$_2$Te$_4$ is dominated by the α-Sb$_2$Te$_3$ component due to the small compressibility of the intercalated SnTe$_6$ octahedron.

On the other hand, the experimental $B_0$ of SnSb$_2$Te$_4$ is similar to that of the isostructural SnBi$_2$Te$_4$ ($B_0$= 35(2) GPa) **[37]**. This result implies that the compressibility of these two layered $BA_2X_4$ compounds is almost independent of the $A$ cation and fully dependent on geometrical factors of the layered structure. A close look at the pressure dependence of the theoretical interatomic distances in **Figure S3** shows a much larger compression of the Te1-Te1 interlayer distance than that found for the Sn-Te and Sb-Te intralayer distances. Therefore, the bulk compressibility of SnSb$_2$Te$_4$ (also for SnBi$_2$Te$_4$) is dominated by the compressibility of the vdW gap between the *SL*s. The compressibility of the Te1-Te1 interlayer distances is similar for both SnSb$_2$Te$_4$ and α-Sb$_2$Te$_3$, in good agreement with the similar bulk compressibilities of both compounds, shown in **Table I**. In conclusion, the compressibility of SnSb$_2$Te$_4$ is dominated by the strong compression of the Te-Te interlayer distance governed by vdW interactions and consequently is similar to that of α-Sb$_2$Te$_3$ and SnBi$_2$Te$_4$.

As regards the evolution of the experimental and theoretical lattice parameters at HP (see **Figure 2b**), these can be fitted to a modified BM-EoS in order to reproduce their sub-linear behavior. **Table II** summarizes the axial bulk modulus, $B_{0i}$, and the axial compressibility as $\kappa_i = \frac{1}{3B_{0i}}$ obtained for each lattice parameter. It can be observed that $\kappa_c \sim 2\kappa_a$ at room conditions. This behavior is consistent with the much larger compression of the Te1-Te1 interlayer distance when compared to the Sn-Te and Sb-Te intralayer distances (**Figure S3**). Such a hypothesis is further substantiated by the evolution of the different interplanar distances along the *c*-axis (see **Figure S4** in SM), which allows us to identify the different contributions to the compressibility of the *c*-axis. The strong reduction of the interplanar compressed Te1-Te1 distance when compared with the behavior of the overall interplanar distances corroborates that the Te1-Te1 interplanar distance dominates the compression of the *c*-axis. Moreover, a fit of the interplanar Te1-Te1 distance with a modified BM-EoS yields a bulk modulus of 21(1) GPa which is similar to that of the *c*-axis (23.8(4) GPa). Therefore, we can conclude that the interplanar Te1-Te1 distance determined by the vdW interaction is the main source for the compressibility of the unit cell along the *c*-axis and that the stronger reduction of the Te1-Te1 distance than the Sb-Te1 or Sb-Te2 distance is also the responsible for the much larger compressibility of the *c*-axis than of the *a*-axis.

Additional support to the interpretation provided in the previous paragraph is obtained by comparing the experimental and theoretical $B_{0i}$ and $\kappa_i$ of SnSb$_2$Te$_4$ with those of SnBi$_2$Te$_4$ **[37]** and α-Sb$_2$Te$_3$ **[40,41]** presented in **Table II**. The close similarity of the values of the $B_{0i}$ and $\kappa_i$ in the three compounds supports the presence of similar mechanisms of compression and a similar strength of the interatomic interactions in all of them.

It is noteworthy of mentioning that the compression of the *c*-axis in SnSb$_2$Te$_4$ does not imply a compression of all the interplanar distances along the *c*-axis. Despite the decrease of the Sb-Te1 bond distance for SnSb$_2$Te$_4$ (see **Figure S3**), the interplanar Sb-Te1 distance expands under compression and the same occurs for Sb$_2$Te$_3$ (see **Figure S4**). This feature clearly suggests an increase of the angle between the *ab*-plane and the Sb-Te1 bond at HP that distorts the SbTe$_6$ octahedron as clearly observed in **Figure S5a** in SM. We also intend to highlight that although the Sb-Te1 bond distance shows the same pressure dependence of the SnSb$_2$Te$_4$ and $\alpha$-Sb$_2$Te$_3$, the presence of the short Sn-Te bond in SnSb$_2$Te$_4$ triggers a slight increase of the Sb-Te2 bond distance, and consequently a small decrease of the Sn-Te bond, in SnSb$_2$Te$_4$ with respect to c-SnTe. Consequently, there is a slightly larger and weaker Sb-Te2 bond in SnSb$_2$Te$_4$ than in $\alpha$-Sb$_2$Te$_3$. This variation caused by the chemical pressure induced by the combination of the binary parent compounds (c-SnTe and $\alpha$-Sb$_2$Te$_3$) in the formation of SnSb$_2$Te$_4$, can also be understood through the variation of the orientation of the lone electron pair (LEP) of Sb, which can induce changes in the regularity of the SbTe$_6$ polyhedral unit. In any case, this distortion is small because the compressibility of the Sb-Te2 bond distance for both SnSb$_2$Te$_4$ and $\alpha$-Sb$_2$Te$_3$ compounds follows a similar pattern.

A good agreement between experimental and theoretical data is also found for the evolution of the $c/a$ ratio at HP (see the inset of **Figure 2b)**. This ratio shows a clear change of tendency above 2 GPa similar to that observed in $\alpha$-Sb$_2$Te$_3$ **[40,41]** and SnBi$_2$Te$_4$ **[37]**. The minimum of the $c/a$ ratio in SnSb$_2$Te$_4$ can also be explained by analyzing the pressure dependence of respective interplanar distances along the $c$-axis (see **Figure S4**). In particular, the evolution of the Te1-Te1 interplanar distance in SnSb$_2$Te$_4$ shows a strong (normal) compression below (above) 2 GPa. This behavior is typically associated with the weak character of the vdW bonds at LP and its hardening due to the increase of the covalent character of the Te-Te bonds at HP **[42]**. A similar conduct of the $c/a$ ratio is observed for $\alpha$-Sb$_2$Te$_3$ since the Sb-Te interplanar distances evolve in this compound in a similar manner as in SnSb$_2$Te$_4$ (see **Figure S4**). Therefore, we can conclude that the change of the compression rate of the interlayer space causes the Te1-Te1 interplanar distance causes the $c/a$ ratio minimum observed in SnSb$_2$Te$_4$ and $\alpha$-Sb$_2$Te$_3$. The evolution of the vdW character of the interlayer bonds at HP will be addressed in depth later on, when we discuss the analysis regarding the dependence of the electronic topology at HP.

The big change of the slope of the *c/a* ratio in SnSb$_2$Te$_4$ and related compounds at HP is likely related to a pressure-induced IPT. We must recall that these chalcogenides are mainly composed by group-15 cations acting with their lowest valence state and featuring a strong cationic LEP stereoactivity that has a deep effect in the formation of the layered structure. The presence of the cationic LEP modifies the electronic distribution of the charges in the crystal and distorts the geometry of the bonds. Moreover, the cationic LEP is mainly oriented along the *c*-axis in these compounds and contributes to the vdW interaction between the layers and to the strong compressibility of the *c*-axis at LP since the LEP is extremely compressible at LP. This scenario has been already observed in a number of group-15 sesquichalcogenides **[43,44]**. Additionally, it has been widely reported that both the cationic LEP and the vdW interaction become more incompressible at HP due to the progressive decrease of the LEP activity and the increase of the strength of the interlayer forces. Therefore, it can be considered that these changes in compressibility of electronic 'density-clouds' at relatively LP lead to a new isostructural phase with a different *c/a* ratio in SnSb$_2$Te$_4$ and related chalcogenides, and these can be understood as an IPT of electronic origin, as we will further show.

A different way of understanding the structural behavior of SnSb$_2$Te$_4$ at HP is by studying the compression of the two octahedral units forming the SnSb$_2$Te$_4$ heterostructure: the regular octahedron around Sn and the slightly deformed octahedron around Sb (see **Figure 1**). **Figure S6** in SM shows the evolution of the theoretical quadratic elongation of both octahedra at HP. This parameter was defined by Robinson *et al* **[45]** to analyze the distance of the inner atom of a polyhedron with respect to the central position, which is an indirect measurement of the irregularity of the polyhedral unit. In our case, both SnTe$_6$ (SbTe$_6$) octahedron shows practically no changes in the quadratic elongation up to 2 GPa and an increase above 2 (4) GPa. This result suggests that both octahedra are almost insensitive to pressure while there is a strong compression of the vdW interlayer gap below 2 GPa. Nevertheless, SnTe$_6$ octahedra show a considerable increase of the polyhedral distortion above 2 GPa; once the vdW gap becomes as incompressible as the octahedral units. Therefore, results of **Figures S5** and **S6** support the occurrence of a pressure-induced IPT near 2 GPa in SnSb$_2$Te$_4$.

Finally, in order to measure the differences of the bond distances in both octahedra with respect to those in a regular coordination, we have plotted the evolution of the theoretical distortion index of both octahedra at HP **[46]** (**Figure S7** in SM). It can be observed that the regular SnTe$_6$ octahedron remains regular during the whole compression process while the distorted SbTe$_6$ octahedron becomes slightly more regular under compression. This is a consequence of the evolution of the Sb-Te1 and Sb-Te2 distances under compression, which tend to follow a similar pattern, but at a smaller rate. In summary, we have shown that the own definition of the SnSb$_2$Te$_4$ heterostructure as a layered distribution of *SL*s makes it essential to analyze the polyhedral units and the inter- and intra-layer compressibility in order to understand respective behavior at HP.

To close this section, we have plotted the experimental quadratic elongations of the *AX*$_6$ and *BX*$_6$ octahedra of all *BA*$_2$*X*$_4$ compounds known to have the $R\bar{3}m$ structure at room pressure according to the ICSD database (see **Figure 3**). In this way, we can try to shed light on the HP behavior of the compounds of this ternary family. It can be observed that SnSb$_2$Te$_4$ is one of the compounds with the smallest quadratic elongation of the *AX*$_6$ octahedron (SbTe$_6$) and conversely with the largest quadratic elongation of the *BX*$_6$ octahedron (SnTe$_6$). Therefore, according to the trend observed in the theoretical quadratic elongation of both octahedra at HP (see **Figure S6**), it can be deduced that SnSb$_2$Te$_4$ will behave at HP similar to PbSb$_2$Te$_4$ at LP, while SnBi$_2$Te$_4$ tends to behave at HP as GeSb$_2$Te$_4$ at LP.

As regards to **Figure 3**, more conclusions can be drawn. *A priori*, it could be thought that the $AX_6$ quadratic elongation should decrease in the sequence As-Sb-Bi due to the larger stereoactivity of the LEP in As than in Sb and Bi. This fact is due to the stronger hybridization of the cationic s-p levels in As than in Sb and Bi. On the other hand, it could be also thought that the $BX_6$ quadratic elongation should decrease in the sequence Pb-Sn-Ge due to the larger difference between the s-p levels in Pb than in Sn and Ge **[47]**. Unexpectedly, the above arguments seem not to be true since SnBi$_2$Te$_4$ (with a common *B* cation with SnSb$_2$Te$_4$) has a larger $AX_6$ quadratic elongation, whereas GeSb$_2$Te$_4$ (with a common *A* cation with SnSb$_2$Te$_4$) has a smaller $BX_6$ quadratic elongation and a much larger $AX_6$ quadratic elongation. Therefore, results of **Figure 3** suggest that there is a close relationship between the quadratic elongation of the $AX_6$ and $BX_6$ octahedra in these compounds and that not all combinations of $AX_6$ and $BX_6$ quadratic elongations are possible within the $R\bar{3}m$ structure. This hypothesis is supported by the lack of compounds PbAs$_2$Te$_4$ and SnAs$_2$Te$_4$ with $R\bar{3}m$ structure, which should be located in the upper right region in the diagram; i.e. far away from the red line marking the average of the possible $AX_6$ and $BX_6$ quadratic elongations. A similar reasoning can be applied to explain why PbSb$_2$Se$_4$, SnSb$_2$Se$_4$, PbBi$_2$S$_4$ and SnSb$_2$S$_4$ cannot crystallize in the $R\bar{3}m$ structure **[48-50]**. In this context, it must be recalled that the LEP activity of group 15 cations increases when the anion follows the sequence Te-Se-S-O **[47]**. Therefore, the strong LEP stereoactivity of As in tellurides should be similar to that of Sb cation in selenides and sulfides and to that of Bi in sulphides and oxides, taking into account the energetic difference between the p-levels of the anion and the s-levels of the cation. Consequently, this strong LEP stereoactivity allows us to explain why the above mentioned compounds do not crystallize in the $R\bar{3}m$ phase.

2. *Pressure-induced decomposition (PID)*

Above 7 GPa, a clear change of the experimental XRD patterns occurs. The disappearance of Bragg reflections around 11 and 14.6 degrees and the appearance of new peaks at 8.6 and 10.1 degrees (at 10.8 GPa) without showing a relevant peak broadening suggests the absence of a PIA. Thus, the only two possibilities are a PT or a PID.

In order to probe possible crystalline HP phases after a PT, we have resorted to the structure field map of $BA_2Te_4$ compounds reported by Zhang *et al.* **[51]**, with ternary compounds systematically ordered according to the cationic orbital radii, $R_B$ and $R_A$ **[52]**. This structure field map (see **Figure 4**) allows us to understand and predict the structures of ternary $BA_2Te_4$ compounds at LP by knowing $R_B$ and $R_A$. Moreover, it can help in predicting the HP phases if we know how orbital radii behave at HP. In the field map of Zhang *et al.*, materials with $R_B \in$ [1.6, 1.9] a.u. and $R_A \in$ [1.0, 2.0] a.u. crystallize in the b37 structure (s.g. $R\bar{3}m$); however, new tellurium-based ternary chalcogenides formed by different group-14 *B* cations (*B* = Ge, Sn and Pb) group-15 *A* cations (*A* = As, Sb and Bi) also crystallize in s.g. $R\bar{3}m$ according to the ICSD database. This are the cases of $GeAs_2Te_4$ (with $R_B$=1.415 a.u. and $R_A$=1.560 a.u.) and $PbBi_2Te_4$ (with $R_B$=1.997 a.u. and $R_A$=2.090 a.u.). We have included these two compounds in the structure field map because they limit the new borders of materials with s.g. $R\bar{3}m$. In this way, we can extend the stability ranges of the b37 structure to $R_B \in$ [1.4, 2.0] a.u. and $R_A \in$ [1.0, 2.1] a.u. with respect to the structure field map reported by Zhang *et al.* **[52]**.

The new structure field map of $BA_2Te_4$ compounds allows us to consider several possible HP phases for a compound with an initial $R\bar{3}m$ structure, such as spinel b4 (s.g. *P-42m* or *I-4*) or d3 (s.g. *C2/m*). It is noteworthy to highlight that one of the parent binary compounds of $SnSb_2Te_4$ (α-$Sb_2Te_3$) has the $R\bar{3}m$ structure and undergoes a PT towards a *C2/m* structure (β-$Sb_2Te_3$) **[53]**. Therefore, the latter phase could be a good candidate for the ternary compound $SnSb_2Te_4$ to crystallize in. The same reasoning applies to $SnBi_2Te_4$ since α-$Bi_2Te_3$ also has the $R\bar{3}m$ structure and undergoes a PT towards a *C2/m* structure (β-$Bi_2Te_3$) **[53]**. Nevertheless, all our attempts to identify the possible HP phase through theoretical simulations failed.

Subsequently, we have considered the possibility of a PID. For this purpose, we have calculated the formation enthalpy of $SnSb_2Te_4$ in s.g. $R\bar{3}m$ and compared it with those of the HP phases of its binary compounds (**Figure 5**). We have found that above 6.3 GPa the existence of the HP phases of β-SnTe (s.g. *Pnma*) and β-$Sb_2Te_3$ (s.g. *C2/m*) are energetically more favorable than the LP phase of $SnSb_2Te_4$. This result indicates that the ternary compound $SnSb_2Te_4$ should decompose into their parent binary materials above 6.3 GPa. In fact, we have found that the ADXRD patterns observed above 7 GPa in $SnSb_2Te_4$ (**Figure S1**) exhibit a perfect fit with the HP phases of the binary parent compounds. A similar theoretical result has been obtained for $SnBi_2Te_4$ (see **Figure 5**), thus suggesting that also PID should be observed in $SnBi_2Te_4$ above 7 GPa.

PIA is a common phenomenon occurring at room temperature, but not PID. In particular, $Ge_2Sb_2Te_5$ has been found to undergo PIA at room temperature **[54]**. The reason is that PIA is mainly due to either a frustrated PT to a HP phase or to a frustrated PID into daughter compounds **[55]**. This frustration mainly occurs due to kinetic features governed by temperature. In fact, PID is usually observed in complex compounds mostly at high temperature **[56-59]** or in some molecular materials, such as $H_2S$, where PID occurs at room temperature and at very HP **[60]**.

SnSb$_2$Te$_4$ is one of the few compounds exhibiting PID at room temperature and relatively LP (7 GPa). This result might be related to the large stability of their polyhedral units that prevail after the PID.

### B. Raman scattering measurements under pressure

In order to corroborate the results found in the structural study under pressure, we performed unpolarized HP-RS measurements. Since there is one formula unit (7 atoms) in the primitive unit cell SnSb$_2$Te$_4$, this compound has twenty-one normal vibrational modes at Γ with the following mechanical decomposition **[61]**:

$$\Gamma = 3\ A_{1g}(R) + 3\ A_{2u}(IR) + 3\ E_u(IR) + 3\ E_g(R) + A_{2u} + E_u$$

where E modes are doubly degenerated, A$_{1g}$ and E$_g$ modes are Raman-active (R) and A$_{2u}$ and E$_u$ are infrared-active (IR) except for one A$_{2u}$ and one E$_u$ mode that correspond to the three acoustic phonons considering that E-type modes are doubly degenerated. Therefore, there are six Raman-active modes ($\Gamma_{Raman}$= 3A$_{1g}$ + 3E$_g$) and six IR-active modes ($\Gamma_{IR}$= 3 A$_{2u}$ + 3 E$_u$). The assignment of the vibrational modes to atomic movements was done through the interface of Jmol Interface for Crystallographic and Electronic Properties (J-ICE) **[62]** and is discussed in the SM (see **Figures S8** to **S17**).

**Figure S18** in SM shows the unpolarized RS spectrum of SnSb$_2$Te$_4$ at ambient conditions together with the position of the theoretically predicted frequencies at ambient conditions (see vertical bottom tick marks). The RS spectrum at ambient conditions shows five of the six theoretically predicted Raman-active modes. A rather good agreement between the experimental and theoretical Raman-active mode frequencies at room pressure are observed (see also **Table S1**).

Consequently, we have made a tentative assignment of experimental Raman modes taking into account the predicted frequencies and pressure coefficients as discussed below.

As regards the vibrational modes of SnSb$_2$Te$_4$, it is well known that in layered materials vibrational modes at the Γ point can be classified into interlayer modes (in the low-frequency region) and intralayer modes (in medium- and high-frequency regions). Moreover, layered compounds crystallizing either in rhombohedral, hexagonal or tetragonal space groups, show A (or B) and E modes. In SnSb$_2$Te$_4$, there are two pure interlayer modes (E$_g^1$ and A$_{1g}^1$), which have the lowest frequencies, and the other ten Raman- and IR-active modes are intralayer modes and have both medium and high frequency values (see **Table S1** and **Table S2**). The two interlayer modes, also known as rigid layer modes, correspond to out-of-phase movements of the neighbor layers both along the *a-b* plane (E$_g^1$ mode) and along *c*-axis (A$_{1g}^1$ mode).

**Figure S18** also shows the unpolarized HP-RS measurements up to 9 GPa. The RS spectra exhibit a monotonous shift of the Raman modes towards higher frequencies at HP, except for two modes with negative slope near ambient pressure (see **Figure 6**). These two modes can be unambiguously assigned to metallic Te **[63]** and their appearance can be explained by the extreme sensitivity of some selenides and tellurides to visible laser radiation. In fact, Raman modes of trigonal Se and Te with negative slope have been recently identified in HP-RS studies of As$_2$Te$_3$ **[44]** and HgGa$_2$Se$_4$ **[64]**.

The evolution of the experimental and theoretical Raman-active modes in SnSb$_2$Te$_4$ at HP (see **Figure 6** and **Table S1**) shows a good agreement between the experimental and theoretically simulated frequencies at HP within a 5% accuracy interval. This feature is a clear sign of the good description reached by the *ab-initio* simulations for this compound. For the sake of completeness, the evolution of the theoretical IR-active modes in SnSb$_2$Te$_4$ is given in **Figure S19** and **S20** in SM.

Vibrational modes similar to those of SnSb$_2$Te$_4$ in the low- and high-frequency range are also observed in the parent binary compound α-Sb$_2$Te$_3$ (see **Figure S21** and **Tables S1** and **S2**). However, there are four vibrational modes of SnSb$_2$Te$_4$ in the medium-frequency region (two Raman-active and two IR-active) that do not have correspondence on the α-Sb$_2$Te$_3$ compound, because they are characteristic of the *SL* in ternary layered *BA*$_2$*X*$_4$ compounds. These extra modes of SnSb$_2$Te$_4$ arises from the vibrations involving the central Sn-Te2 bonds; in particular, out-of-phase vibrations of the Te2 atoms of the SnTe$_6$ polyhedral unit (see **Figures S10 and S11**).

The rocksalt structure of c-SnTe contains no Raman-active modes, but one IR-active modes with T$_{1u}$ symmetry that splits into a doubly degenerate TO mode and a single LO mode. It is interesting to note that the two extra modes observed in SnSb$_2$Te$_4$ can be related to the IR-active modes of c-SnTe (see **Figure S22**). In this context, it can be observed that phonons of similar nature in SnSb$_2$Te$_4$ and their parent compound α-Sb$_2$Te$_3$ are located around similar frequencies (**Figure S21**), so the position of the two medium-frequency modes in SnSb$_2$Te$_4$ is expected to be located at similar frequencies than in c-SnTe. In particular, there is a clear correlation between the A$_{1g}^2$ mode of SnSb$_2$Te$_4$ and the LO-type IR-active mode of c-SnTe at room pressure [65] (see **Table S2**) that is also extended at HP (see **Figure S22**). Therefore, our Raman results for SnSb$_2$Te$_4$ open a new way to characterize the lattice dynamics of binary compounds with forbidden Raman modes, like those crystallizing in the rocksalt structure as c-SnTe, by means of RS measurements in more complex compounds containing similar octahedral arrangements as those in the binary compounds. A similar example could to be the case of rocksalt-type MgO, whose longitudinal optical (LO)-type IR-active mode (738 cm$^{-1}$) **[66]** is consistent with the frequency of the highest Raman-active mode in MgTiO$_3$ (715 cm$^{-1}$), which crystallizes in s.g. R$\bar{3}$ **[67]**.

An anomalous decrease of the experimental and theoretical frequency of Raman-active modes A$_{1g}^2$ and E$_g^2$ (see **Figure 6**) and of the theoretical frequency of IR-active modes E$_u^2$ and A$_u^2$ (see **Figure S19**) of SnSb$_2$Te$_4$ is observed above 3.0 GPa. The softening of these vibrational modes, mainly related to Sn-Te vibrations could be *a priori* ascribed to the pressure-induced IPT around 2 GPa, similar to that occurring in α-Sb$_2$Te$_3$ between 2.5 and 3.5 GPa **[68,69]**. A close look at our theoretical simulations (see **Figures S10 and S11**) reveal that the E$_g^2$ and E$_u^2$ modes in SnSb$_2$Te$_4$ are mainly Sn-Te bending modes with a slight stretching contribution, while the A$_{1g}^2$ and A$_{2u}^2$ are mainly Sn-Te stretching modes. Since stretching modes mainly depend on the bonding force constant and the bonding distance, the softening of these vibrational modes involving Sn-Te2 bonds could likely be related either to a strong decrease of the Sn-Te2 bonding force constant (depending on the charge density) or to a strong increase of the bond distance. However, neither an increase of the Sn-Te2 bond distance (**Figure S3**) nor a decrease of the Sn-Te2 bond charge density (**Figure S26**) have been observed in the whole studied pressure range. Another possibility to explain the softening is that there is a change of the character of the stretching modes so that they become more bending-like than stretching-like. To prove that, we have looked at the angle between the *ab*-plane and the Sn-Te2 bond (see **Figure S5b**). It can be observed that a subtle change in this angle occurs around 4 GPa but the increase of the angle value suggests that these vibrational modes acquire an increasing stretching character, which is incoherent with the loss of bonding strength.

After all these considerations, we have concluded that the softening of the medium-frequency phonons in SnSb$_2$Te$_4$ could only be explained by a Fermi resonance effect **[70]**. A Fermi resonance occurs when there is a strong anharmonic interaction of a first-order phonon with a two-phonon combination containing a high density of states. In such a case, a shift of the frequency of the first-order vibrational mode as well as a change of its intensity and width occurs. The Fermi resonance is a rare phenomenon in solids that has been observed in copper halides, molecules and defect modes **[71-74]**. In SnSb$_2$Te$_4$, the combination of A$_g^1$ and E$_g^1$ at Γ almost coincides in frequency with the E$_g^2$ mode at 2 GPa; i.e. the pressure range close to the onset of the softening of this mode in SnSb$_2$Te$_4$ (**Figure 6**). The anomalous softening of the Sn-Te related modes in SnSb$_2$Te$_4$ is also reproduced, even more clearly, by the theoretically predicted evolution of the IR-active LO-mode of c-SnTe at HP (see **Figure S22**). Note that the frequency of this mode coincides in frequency at Γ with twice the value of the transverse optical (TO)-mode when a sudden change of trend occurs (**Figure S22**). Therefore, we conclude that c-SnTe exhibits a Fermi resonance in the IR-active LO mode that is also reproduced in SnSb$_2$Te$_4$. This result gives further support to our previous interpretation of the relationship between the vibrational modes of SnSb$_2$Te$_4$ and c-SnTe. In summary, we attribute the softening of A$_g^2$ and E$_g^2$ of SnSb$_2$Te$_4$ at HP to a pressure-induced Fermi resonance caused by the coincidence of the frequencies of these firs-order Raman modes and the combination of A$_g^1$ and E$_g^1$ modes along the whole BZ.

To close this section, we have calculated the phonon dispersion curves in SnSb$_2$Te$_4$ at 0, 2 and 4 GPa (see **Figure S23**) in order to understand the nature of the IPT close to 2 GPa previously discussed. As observed, our calculations do not show softening of any of the phonon branches. This result suggests that the pressure-induced IPT found above 2 GPa, cannot be assigned to a 2$^{nd}$-order IPT and must be of higher-order.

In summary, our analysis of the lattice dynamics of SnSb$_2$Te$_4$ at HP shows the layered character of this compound and a good agreement between our experimental and theoretical data. The Raman-active modes of SnSb$_2$Te$_4$ have been explained in relation to its binary parents α-Sb$_2$Te$_3$ and c-SnTe and it has been proved that modes that cannot be observed by Raman scattering in c-SnTe can be observed with this technique in SnSb$_2$Te$_4$ containing similar SnTe$_6$ polyhedra than c-SnTe. Additionally, the anomalous behavior of two Raman-active and two IR-active modes above 3.5 GPa in the medium-frequency region, which are characteristic of the *SLs* and related to Sn-Te vibrations, has been attributed to a Fermi resonance that also occurs in c-SnTe near 2 GPa. Finally, we have proved, with the help of the phonon dispersion curves for SnSb$_2$Te$_4$ at different pressures, that the IPT occurring in SnSb$_2$Te$_4$ close to 2 GPa is not of 2$^{nd}$-order but of higher order, as expected for an IPT of electronic origin.

### C. Electrical resistance measurements under pressure

In the parent binary compound α-Sb$_2$Te$_3$, a pressure-induced IPT was reported and it was argued to be associated to a pressure-induced ETT around 3.5 GPa [40]. This result has motivated us to carry out resistance measurements and theoretical calculations of the electronic band structure at different pressures in order to verify if a pressure-induced ETT could be also observed in SnSb$_2$Te$_4$. This kind of measurements have been previously validated by other works [75] up to 50 GPa.

Two abrupt variations in the pressure dependence of the electrical resistance of SnSb$_2$Te$_4$ above 2 and 8 GPa have been observed (see **Figure 7**). The variation above 8 GPa can be ascribed to the PID of the sample, already commented in section A, which is similar to that observed also above 8 GPa in SnBi$_2$Te$_4$ [37]. This result confirms our formation enthalpy analysis (**Figure 5**) that suggest that SnBi$_2$Te$_4$ likely undergoes also a PID. Furthermore, the low resistance measured above 8 GPa can be ascribed to the metallic nature of the HP phases of Sb$_2$Te$_3$ and SnTe [39,76].

Below 8 GPa, we can distinguish two different ranges in the evolution of the electrical resistance of SnSb$_2$Te$_4$ below and above 2 GPa. At the LPs, SnSb$_2$Te$_4$ evidences a very low electrical resistance (constant from room pressure to 2 GPa), whose behavior and values are typical of a degenerate semiconductor. Previous results on the literature **[32]** have shown that these results correspond to a lack of stoichiometric vacancies, which leads to the formation of a p-type degenerate semiconductor whose carriers are created by cation vacancies. Above 2 GPa, there is a drastic increase of the electrical resistance, which traditionally has been associated in the literature to the creation of structural defects along an ongoing phase transition **[77]**. These defects might create a donor levels that modify the carrier concentration of the material. In SnSb$_2$Te$_4$, the IPT occurring around 2 GPa might be the responsible for the creation of donor levels associated to defects. These could be able to trap p-carriers, thus helping to pass from a degenerate to a non-degenerate semiconductor or acting as scattering centers that decrease carrier mobility. In any case, the decrease in carrier mobility would be an indication of a decrease of the p-type character of the electrical conduction in SnSb$_2$Te$_4$, similar to that occurs in ZnTe between 7-11 GPa **[77]**. Thus, the decrease of the conductivity evidenced by SnSb$_2$Te$_4$ above 2 GPa could be exploited to overcome one of the main problems of TIs; i.e. the non-observation of surface carrier conductivity, which would be masked by bulk carrier conductivity **[78]**.

As regards electronic band structure calculations, we obtain that SnSb$_2$Te$_4$ is an indirect bandgap semiconductor with a bandgap energy $E_g$= 0.12 eV at room pressure, similar to that previously obtained **[16]**, and that the bandgap energy decreases with increasing pressure leading to a closening of the bandgap above 4.5 GPa (see **Figure S24**). Since our calculations based on Density-Functional Theory (DFT) are known to yield underestimated bandgaps when (semi-)local functionals are employed, the value of the real bandgap is expected to be above 0.2 eV at room pressure **[11]** and the real metallization must occur at higher pressures. In fact, the predicted lack of metallization at LP and the negative slope of the bandgap are compatible with the above-mentioned explanation of the evolution of the electrical resistance at HP. However, we cannot neglect the fact that the lack of hydrostatic conditions above 2.5 GPa due to the use of a solid PTM may blur the interpretations of the changes observed in the electrical properties of the material.

In summary, our electrical measurements in $SnSb_2Te_4$ show a small resistance at LP, typical of a degenerate semiconductor, and an increase of the electrical resistance above 2.0 GPa, attributed to the generation of defects due to the pressure-induced IPT. Finally, the decrease of the electrical resistance above 8 GPa is attributed to the metallic nature of the HP phases of $Sb_2Te_3$ and SnTe due to the decomposition of $SnSb_2Te_4$ above 7 GPa. The behavior of the electrical resistance in $SnSb_2Te_4$ is consistent with our calculations of the electronic band structure that do not show metallization up to 8 GPa when the value of the theoretical bandgap is corrected.

### D. Electronic topology under pressure

#### 1. *Bader Charge analysis.*

In recent HP works of $\alpha$-$Sb_2Te_3$, the occurrence of a pressure-induced IPT was studied on the light of the electron density evolution under pressure **[69]**. Additionally, the pressure-induced ETT was interpreted on the light of the evolution of the Bader charges at HP **[68]**. Since the occurrence of a pressure-induced ETT depends on the location of the Fermi level, it is rather doubtful that the pressure at which an ETT is observed can be determined with the study of the theoretical Bader charge analysis, which is independent of the location of the Fermi level. However, a change in the evolution of the Bader charges at HP can be indicative of a pressure-induced IPT, not related with a variation of the Fermi level.

Figure 8a shows the pressure dependence of the Bader charges for each of the constituent atoms of $SnSb_2Te_4$. The Bader charges of the internal atoms of the *SLs* (Sn and Te2 atoms), related to the $SnTe_6$ polyhedral units, are larger than those of the external atoms of the SL (Sb and Te1) related to the $SbTe_6$ polyhedral units, and describe a monotonous trend with respect to pressure following an almost linear behavior with a small kink near 2 GPa. On the other hand, the pressure evolution of the Bader charges of the most external atoms of the *SLs* cannot be fitted to a single linear trend and two linear fits below and above 2 GPa are required.

The comparison between the polyhedral net charges contained in the two octahedral units that compose this ternary material (**Figure 8b**) reveals a clear change of trend above 2 GPa (**Figure 8c**). An abrupt charge transfer from $SbTe_6$ towards $SnTe_6$ occurs up to 2 GPa and ceases above this pressure. This evolution can be explained as follows: At LP, the closening of the inter-layer gap leads to a strong charge redistribution of the external $SbTe_6$ unit that lead to a charge transfer to the $SnTe_6$ unit. Above 2 GPa, the vdW space is already closed and no further charge is transferred between both $SnTe_6$ and $SbTe_6$ units so their Bader charges evolve similarly under pressure. In summary, the charge redistribution between the polyhedral units is consistent with Te1 atoms being involved in the $SbTe_6$ octahedral units and being the ones responsible for the vdW interactions in the inter-layer gap that define the IPT around 2 GPa in $SnSb_2Te_4$.

2. *Non-Covalent Interaction analysis.*

We have performed Non-Covalent Interaction (NCI) simulations to probe the evolution of the low electron density regions at the inter-layer space between two neighbouring *SLs*. **Figure S25** in SM highlight the vdW interactions at the inter-layer (Te1-Te1) space at LP and their evolution towards more localized interactions at HP. A 2D data profile of the NCI of the inter-layer space between two *SLs* is represented in **Figure 9**. This figure can be interpreted as follows: At LP, the electron density cloud is very flat and delocalized. However, above 2 GPa bonds become more localized; thus clear bonds among polyhedra of neighbor *SL*s. Overall, all results point at charge localization with increasing pressure so that the Te1-Te1 vdW nature of the inter-layer space is no longer governing the response of the solid above 2 GPa.

3. *Electronic density analysis.*

A more quantitative analysis of the electronic topology leads us to use the electron charge density and its Laplacian, $\nabla^2\rho(\vec{r})$, at the bond critical point (BCP), which can be defined by the point of minimum electron density between two bonded atoms (following the electron density gradient). This method allows us to determine the bonding character of the four bounds found in SnSb$_2$Te$_4$ (Te1-Te1, Sb-Te1, Sb-Te2 and Sn-Te2).

With regards to the Te1-Te1 interaction, we can distinguish its vdW character at LPs, determined by a low $\rho(\vec{r})$ and a positive value of the $\nabla^2\rho(\vec{r})$ at the BCP (see **Figure S26**). An increase of both parameters at HP is coherent with the decrease of vdW character **[79]** we have already observed and commented in the previous section. At LP, Te1 atoms are bonded to three Sb atoms and at close vicinity with other three Te1 atoms, belonging to the neighboring layer, at much larger distance. At HP, the strong decrease of the interlayer Te1-Te1 distance gives rise to a stronger interaction between the Te atoms belonging to different layers, thus leading to a sixfold coordination of the Te1 atoms; such an interaction supports the evidence of the formation of ionic bonds between the neighboring Te atoms. Note that ionic bonds are characterized by large and positive $\rho(\vec{r})$ and positive $\nabla^2\rho(\vec{r})$ at the BCP (see **Table S3**).

With respect to the Sb-Te2 and Sn-Te2 bonds, these possess similar values of $\rho(\vec{r})$ and $\nabla^2\rho(\vec{r})$ at the BCP in the whole pressure range studied. In fact, the value of $\rho(\vec{r})$ of both bonds is an intermediate between that of Sb-Te1 bond and the weak vdW interaction between Te1 atoms along the whole of the studied pressure range. In this context, we have to recall that both parent binary compounds, α-$Sb_2Te_3$ and c-SnTe, have been considered as incipient metals **[35]**, therefore these values of $\rho(\vec{r})$ and $\nabla^2\rho(\vec{r})$ could be characteristic parameters of metavalent bonding with partially delocalized electrons.

As regards the Sb-Te1 bond, it evidences a very high $\rho(\vec{r})$ when compared to the overall interactions and a positive, although close to zero, value of the $\nabla^2\rho(\vec{r})$ at the BCP in the whole pressure range studied. The high $\rho(\vec{r})$ value is typical of covalent or ionic bondings; however, the $\nabla^2\rho(\vec{r})$ value must be negative (positive) for a covalent (ionic) bonding. Therefore, the small positive $\nabla^2\rho(\vec{r})$ value suggests a mixture between the covalent and metavalent bonding, whereas a polar covalent interaction is neglected because typically it should show a closer value of $\nabla^2\rho(\vec{r})$ to zero.

In order to deepen into the analysis of the character of these two bonds, we have plotted the ELF along the different atomic interaction distances of $SnSb_2Te_4$; namely, Te1-Te1, Sb-Te1, Sb-Te2 and Sn-Te2 (see **Figure 10** and **Figure S27**). ELF describes the character of bond formation between the involved atoms. At LPs, the low ELF values (close to 0.2) close to the center of the Te1-Te1 distance, exhibit the typical values of vdW bonds; however, this ELF signal increases rapidly with pressure, which is coherent with the results observed from the BCP electronic topological analysis, and are characteristic of an ionic-type bond. In the case of the Sb-Te1 bond, the ELF value close to the center of its respective interatomic distance is high enough (0.7) to ensure we are dealing with a strong bond, coherent with the polar-covalent results previously obtained from the BCP analysis.

As regards to the ELF signals between the Sn-Te and the Sb-Te2 bonds, these have a medium value (0.5-0.6), which is close to the center of the bond distance. This value has been associated in the literature to be of metavalent bonding character **[34,35,80]**. The intermediate values of $\rho(\vec{r})$ of both bonds, at the BCP and ELF, can be explained by the partial delocalization of electrons of this type of bonding, which stems from the sharing of electrons between several bonds. Therefore, our ELF values support the previous analysis of $\rho(\vec{r})$ and $\nabla^2\rho(\vec{r})$ at the BCP for both interactions. Through such analysis, we are able to establish a new form of identifying metavalent bonds; i.e. these are characterized by an intermediate value of $\rho(\vec{r})$ and a low positive value of $\vec{\nabla}\rho(\vec{r})$. The classification of the different bondings according to $\rho(\vec{r})$, $\nabla^2\rho(\vec{r})$ and ELF are summarized in **Table S3** in SM.

In summary, we think that the changes observed both at the Bader and NCI analyses reflect the IPT occurring from the rhombohedral SnSb$_2$Te$_4$ close to 2 GPa. This IPT is strongly related to the hardening of the Te1-Te1 interlayer interaction and the loss of respective vdW character. Moreover, we have shown that Sb-Te1 bonds are polar covalent bonds, whereas the Sb-Te2 and the Sn-Te2 bonds may fall into what has been recently defined to be the metavalent-type bonding. We have been able to fully characterized these bond types, and a new method by using the concepts of $\rho(\vec{r})$ and $\nabla^2\rho(\vec{r})$ at the BCP has been proposed to identify this types of bonds in complex structures, where a coexistence of several types of interactions occurs.

### III. CONCLUSIONS

This paper describes a joint experimental and theoretical study of the structural, vibrational and electrical properties of the compressed rhombohedral phase of SnSb$_2$Te$_4$ with tetradymite-like layered structure at ambient temperature. Our measurements show that this phase is stable up to 7 GPa and that a decomposition of the sample occurs above this pressure. Such an assumption is supported by the comparison of the evolution of the formation enthalpy of the ternary material and the HP phases of respective parent binary compounds with compression. In this context, the orbital radii map of the tellurium-based ternary chalcogenides have been extended and a possible path at HP has been described for SnSb$_2$Te$_4$ and for the isostructural compounds of the $BA_2$Te$_4$ family.

The analysis of the evolution under pressure of the rhombohedral structure of SnSb$_2$Te$_4$ shows that the compressibility of the interlayer space, governed by vdW interactions between the external Te atoms of the $SLs$, dominates the behavior of the unit cell under compression. With the compression of the structure above 2 GPa, structural variations occur as reflected in the change of sign of the slope of the $c/a$ ratio, which is due to an increase of the strength of vdW interactions between the layers. Such a variation is clearly reflected on the analysis of the pressure dependence of the calculated electron density topology (Bader charges and the electronic density and respective Laplacian at the BCP).

The study of the lattice dynamics of SnSb$_2$Te$_4$ under compression has allowed us to understand the atomic vibrations of the different phonons and assign the mode symmetries of the Raman-active modes. Furthermore, the description of the atomic vibrations has been compared with their parent binary compounds (c-SnTe and $\alpha$-Sb$_2$Te$_3$). A softening of vibrational modes mainly related to the Sn-Te bonds occurs above 3 GPa, and such a feature has been explained within the framework of the Fermi resonance. Our calculations predict that the Fermi resonance must also be observed in the HP dependence of IR-active modes of parent binary c-SnTe around 2 GPa. Our results show strong correlation between the vibrational modes of SnSb$_2$Te$_4$ and those of its parent binary compounds. In fact, the Raman spectrum of SnSb$_2$Te$_4$ shows vibrational modes that are forbidden in SnTe; thus showing a novel way to experimentally observe the forbidden vibrational modes of some compounds.

We have also undertaken a study of the pressure dependence of the electrical properties of SnSb$_2$Te$_4$ with unintentionally p-type semiconducting character. The change in the electrical resistance above 8 GPa has been attributed to the sample decomposition; however, a drastic increase in resistance was observed above 2 GPa and attributed to the generation of defects on the ongoing IPT passing from a p-type degenerate to an non-degenerate semiconductor by the reduction of the hole carrier concentration. This result allows the tuning of the electrical properties to improve the TI capabilities of this compound.

Finally, our electron density topology analysis shows that the IPT around 2 GPa is related to the loss of the vdW character and an increment of the ionic character of the interaction between the Te1 atoms of neighbor *SLs*. The hardening of the Te1-Te1 bond may be the cause of the IPT. In fact, the analysis of the phonon dispersion curves at 0, 2 and 4 GPa reveals the lack of softening of phonon branches with increasing pressure, which confirms that the observed IPT is relate to a phase transition of higher order than 2 (i.e. IPT of electronic origin).

The analysis of the ELF of the Sb-Te1 interaction displays a polar covalent bond character, which remains unalterable under compression. However, the most interesting analysis is obtained when studying the ELF along the Sb-Te2 and Sn-Te interactions, which show the typical intermediate values expected for a metavalent bond. The evaluation of their electronic densities and respective Laplacians at the BCP provides a new criterion to identify these interactions when the material is very complex and different kind of bonds coexist.

In summary, our study highlights the importance of the study of the evolution of the chemical bonds under pressure of topological insulators and the origin of isostructural phase transitions observed on materials belonging to the same family of SnSb$_2$Te$_4$.

## IV. METHODS

### A. Sample preparation

Bulk samples were prepared by melting stoichiometric amounts of the pure elements Sn (99.999%, Smart Elements), Sb (99.999%, Smart Elements) and Te (99.999%, Alfa Aesar) at 950 ºC for 93h in sealed silica glass ampoules under argon atmosphere and subsequent annealing at 450 - 500 °C for two days. **[31]** Representative parts of the samples were crushed to powders and fixed on Mylar foils with silicon grease to collect powder diffraction patterns on a Huber G670 powder diffractometer equipped with an imaging plate detector (Cu-K$_{\alpha 1}$ radiation, Ge monochromator, λ = 1.54051 Å) in Guinier geometry. Rietveld refinement of powder x-ray diffraction data confirmed the high purity of the samples.

### B. Theoretical Calculations

*Ab-initio* calculations have been performed within the density functional theory (DFT) **[81]** using plane-wave basis-sets and the projector-augmented wave (PAW) **[82]** scheme with the Vienna *Ab-initio* Simulation Package (VASP) package **[83]**. Calculations of the electronic-band structures have been considered by employing spin-orbit coupling (SOC). The plane-wave kinetic-energy cutoff was defined with 320 eV, in order to achieve highly converged results. We have used the generalized-gradient approximation (GGA) for the exchange-correlation energy with the Perdew-Burke-Ernzerhof parameterization revised for solids (PBEsol) **[84]**. At each selected volume, the structures were fully relaxed to their equilibrium configuration through the calculation of the forces on atoms and the stress tensor with a dense special k-point sampling Monkhorst-Pack grids. In particular, the electronic band structures along high-symmetry directions and the corresponding electronic density of states (EDOS) were computed with a mesh of 18x18x18 k-points. The application of DFT-based calculations to the study of semiconductor properties under HP has been reviewed in **[85]**.

Lattice-dynamics calculations of phonon modes were performed at the zone center ($\Gamma$ point) of the Brillouin zone. For the calculation of the dynamical matrix at $\Gamma$ we used the direct force-constant approach (or supercell method), **[83,86]** which involves the calculation of all the atomic forces when each non-symmetry related atom in the unit cell is displaced along non-symmetry related directions.

The Bader analysis was performed by partitioning the PBEsol-DFT core and valance charge density grids **[87-91]**. A fine FFT grid was required to accurately reproduce the correct total core charge. The Non-Covalent Interactions (NCI) of the PBEsol-DFT charge densities was computed using the NCIPLOT tool **[92,93]**. Such a tool defines a visualization index based on the electron density and its derivatives, enabling identification of non-covalent interactions, based on the peaks that appear in the reduced density gradient at low densities.

### C. Synchrotron based angle-dispersive X-ray diffraction (ADXRD) under pressure experiments

HP-ADXRD measurements on $SnSb_2Te_4$ at 300 K using a membrane-type diamond-anvil cell (DAC) were carried out in experiment 1 (experiment 2) up to 37 GPa (12 GPa) in beamline I15 (MSPD beamline **[94]**) at Diamond Light Source synchrotron (ALBA synchrotron) using a monochromatic X-ray beam with λ = 0.42408 Å (λ = 0.4246 Å). In experiment 1 (experiment 2) images were collected using a MAR345 image plate (Rayonix MARCCD detector) located at 430 mm (240 mm) from the sample. In experiment 1 (experiment 2), $SnSb_2Te_4$ powder was loaded in a 150-μm diameter hole of a Rhenium (Inconel) gasket in a DAC with diamond-culet sizes of 350 μm using helium (silicone oil) as pressure transmitting medium (PTM). In both experiments, copper was placed inside the pressure cavity and used as the pressure sensor through copper EoS **[95]** and a pinhole placed before the sample position was used as a clean-up aperture for filtering out the tail of the X-ray beam, which was focused down to 20 x 20 μm$^2$ using Kickpatrick-Baez mirrors.

Diffraction patterns obtained in both experiments were integrated as a function of 2θ using FIT2D software in order to give conventional, one-dimensional diffraction profiles **[96].** The refinement of the powder diffraction patterns was performed using GSAS program package **[97,98]**. Due to the resonant excitation energy with Sn K-edge used in both experiments, the relative intensities are not accurate enough to perform Rietveld refinement but a Von Dreele-type Le Bail fit. Therefore, all the experimental structural parameters presented in this work have been obtained by means of a Von Dreele-type Le Bail method. Unfortunately, the lack of Rietveld refinement in our measurements prevents us from validating the degree of cation mixing in our samples.

### D. Raman scattering (RS) measurements under pressure

Unpolarized HP-RS measurements up to 27 GPa using a membrane-type DAC and 16:3:1 methanol/ethanol/water mixture as PTM (quasi-hydrostatic up to 10 GPa) **[99,100]**, were performed with a Horiba Jobin Yvon LabRAM UV HR microspectrometer equipped with a thermoelectrically cooled multichannel charge coupled device detector which allows a spectral resolution better than 2 cm$^{-1}$. The Raman signal was excited with a He-Ne laser (632.8 nm line) with a power of less than 10 mW and collected in backscattering geometry using an edge filter working in perpendicular configuration and cutting at 100 cm$^{-1}$. Raman signals down to 50 cm$^{-1}$ can eventually be detected by adjusting the angle between the edge filter and the light containing the Raman signal (provided that the Rayleigh signal is weak enough and the Raman signal is strong enough). Pressure was determined by the ruby luminescence method **[101,102]**. The frequency of the Raman-active phonons have been experimentally analyzed by fitting Raman peaks with a Voigt profile fixing the Gaussian line width (1.6 cm$^{-1}$) to the experimental setup resolution **[103,104].**

### E. Transport properties under pressure

Electrical resistance of SnSb$_2$Te$_4$ under pressure was measured with the standard four-point probe van der Pauw method using 20μm copper-beryllium wires. Single crystals of SnSb$_2$Te$_4$ of approx. 30 μm thick and 100 x 100 μm$^2$ surface were loaded into a Merrill-Bassett DAC with 400 μm culet diamonds. The electrical average resistance was measured by using four 20 μm copper-beryllium wires. Electrical resistance was measured under two different arrangements. In the first one, the sample was directly in contact with the anvils; i.e. under non-hydrostatic conditions. In the second one, the sample was inside a stainless steel gasket and surrounded by CsI powder as PTM; i.e. under quasi-hydrostatic conditions. Electrical resistance showed similar trends in both arrangements, likely due to the anisotropic (layered) and soft nature of the crystals. Luminescence lines of Ruby powder were used to calibrate the pressure inside the cavity in both methods **[101,102]**.

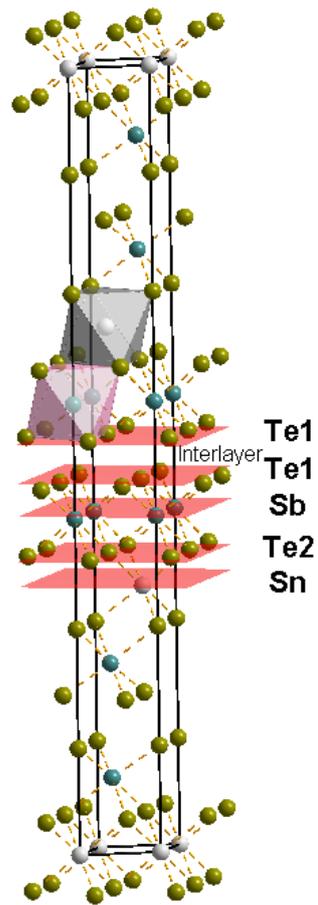

**Figure 1.** 3D structure layout of the SnSb$_2$Te$_4$ compound. Atomic planes are defined.

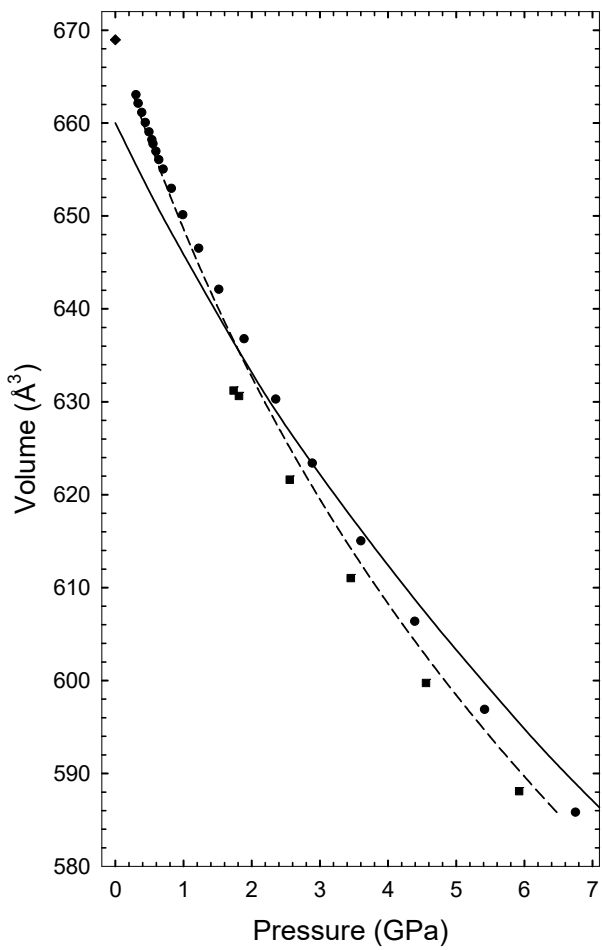 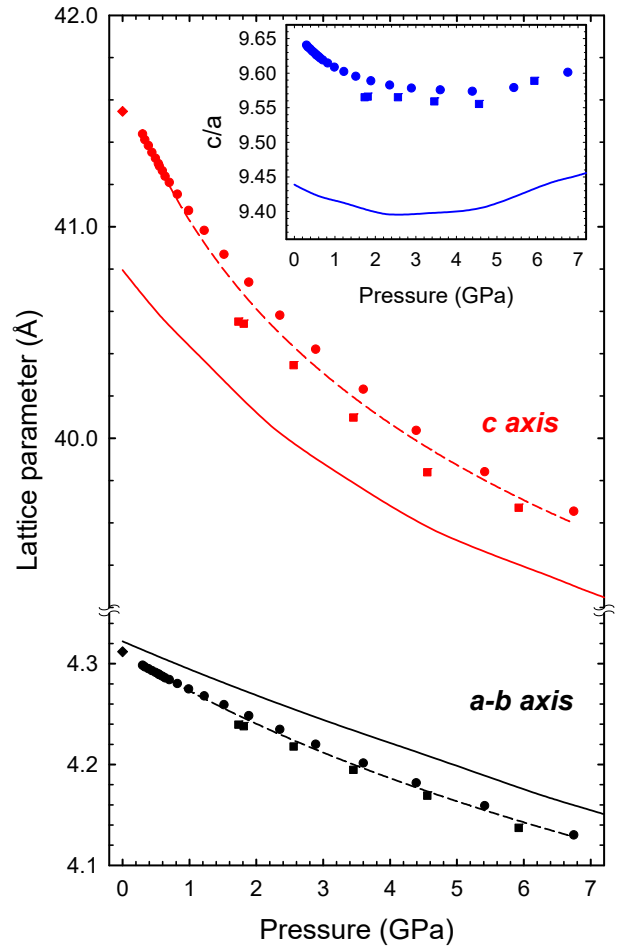

**Figure 2.** Pressure dependence of the unit cell volume (a) and lattice parameters (b). The inset shows the evolution of the c/a ratio with pressure. Solid lines represent the theoretically simulated data, dashed lines represents the fit to EoS equations, solid circles are the experimental data obtained using silicone oil as PTM and solid squares are the experimental data obtained using helium as PTM.

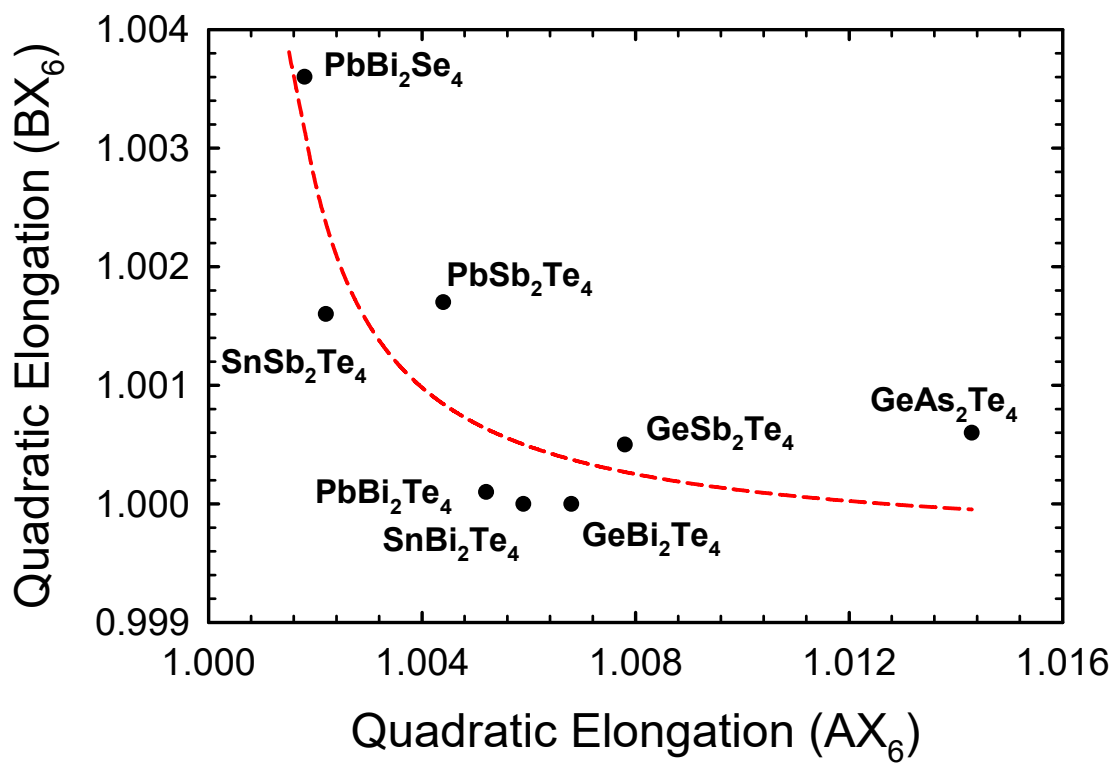

**Figure 3.** Relationship between the quadratic elongation of the $BX_6$ octahedron and the quadratic elongation of the $AX_6$ octahedron in $AB_2X_4$ materials. Structures obtained from Refs. **[105-110]**

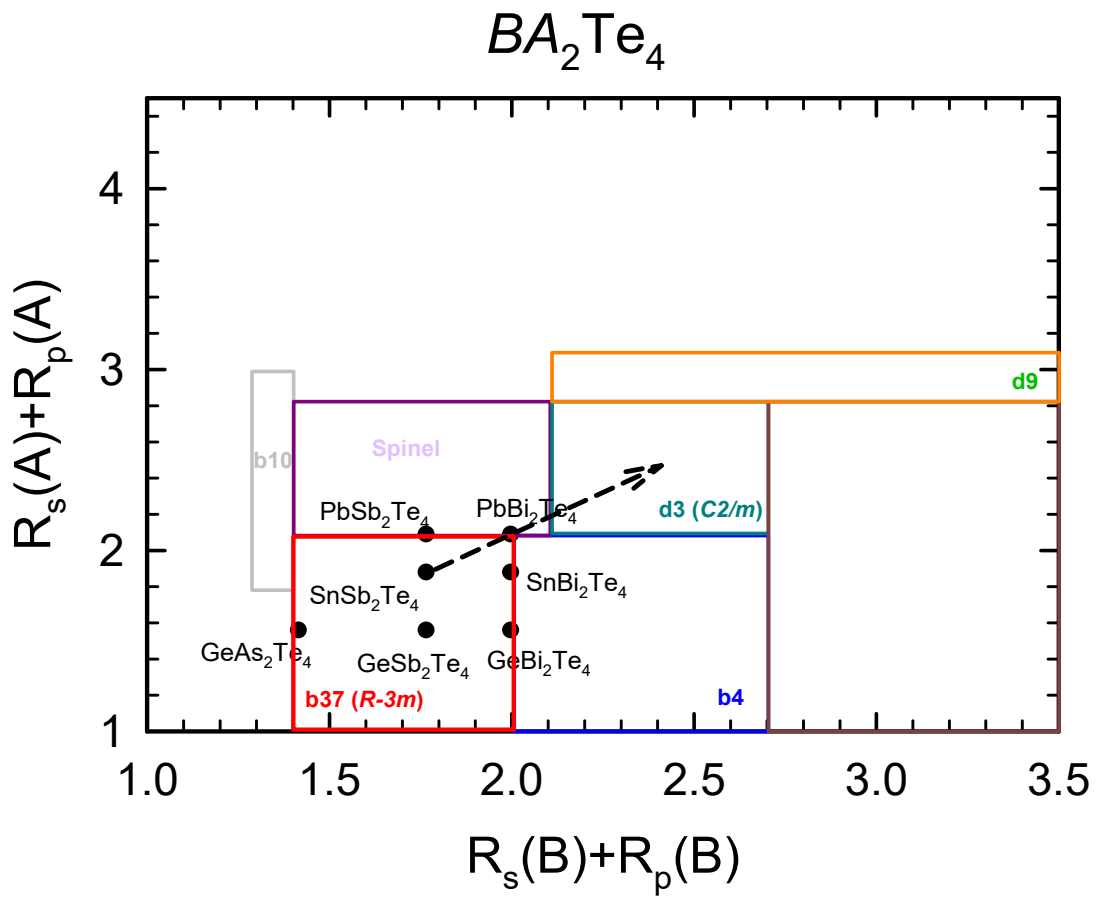

**Figure 4.** Updated orbital radii map of stable *BA₂Te₄* compounds initially proposed by Zhang et al. **[51]**

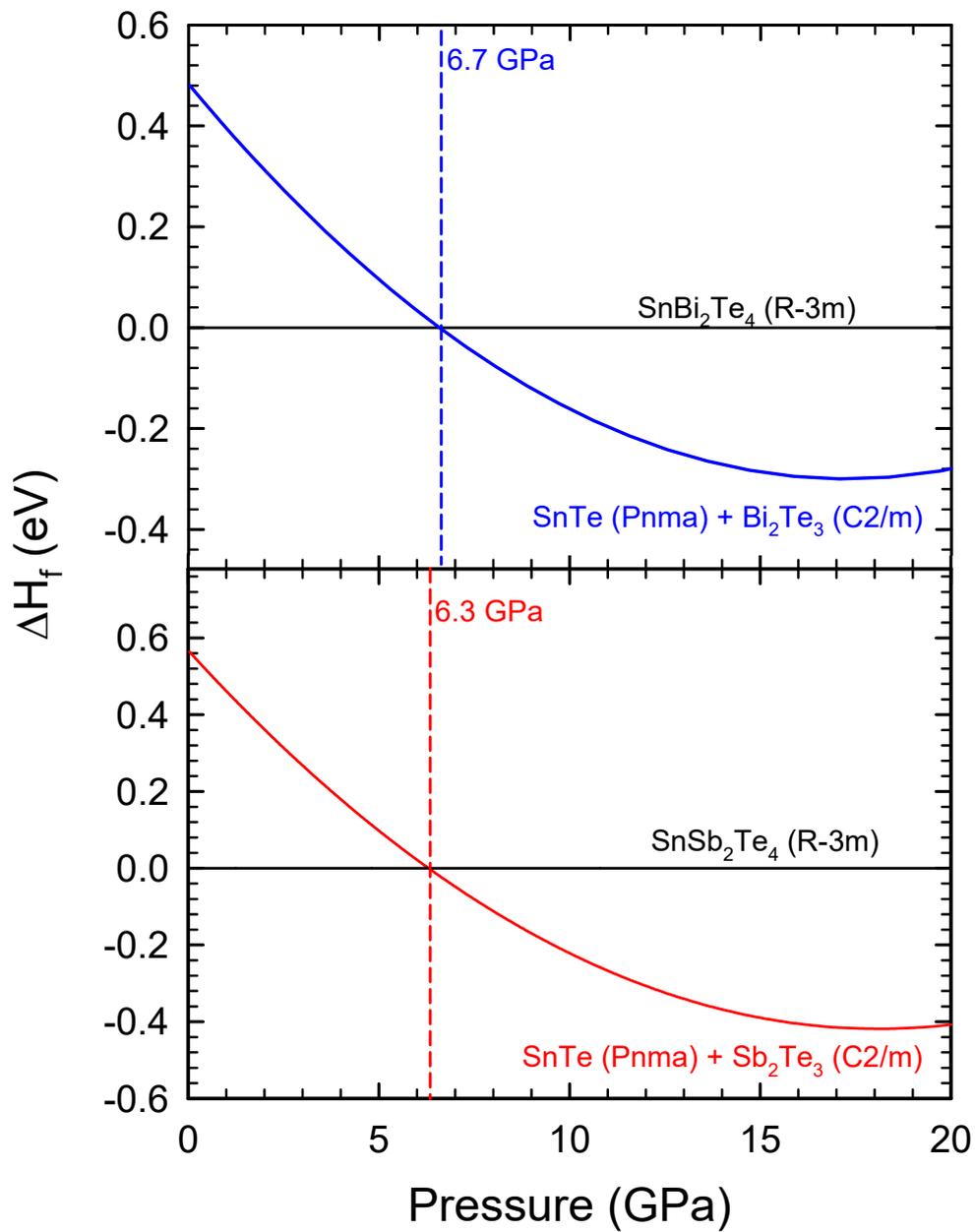

**Figure 5.** Relative formation enthalpy of the HP phases of the parent binary compounds with respect to the $R\bar{3}m$ structure of SnSb$_2$Te$_4$ and SnBi$_2$Te$_4$.

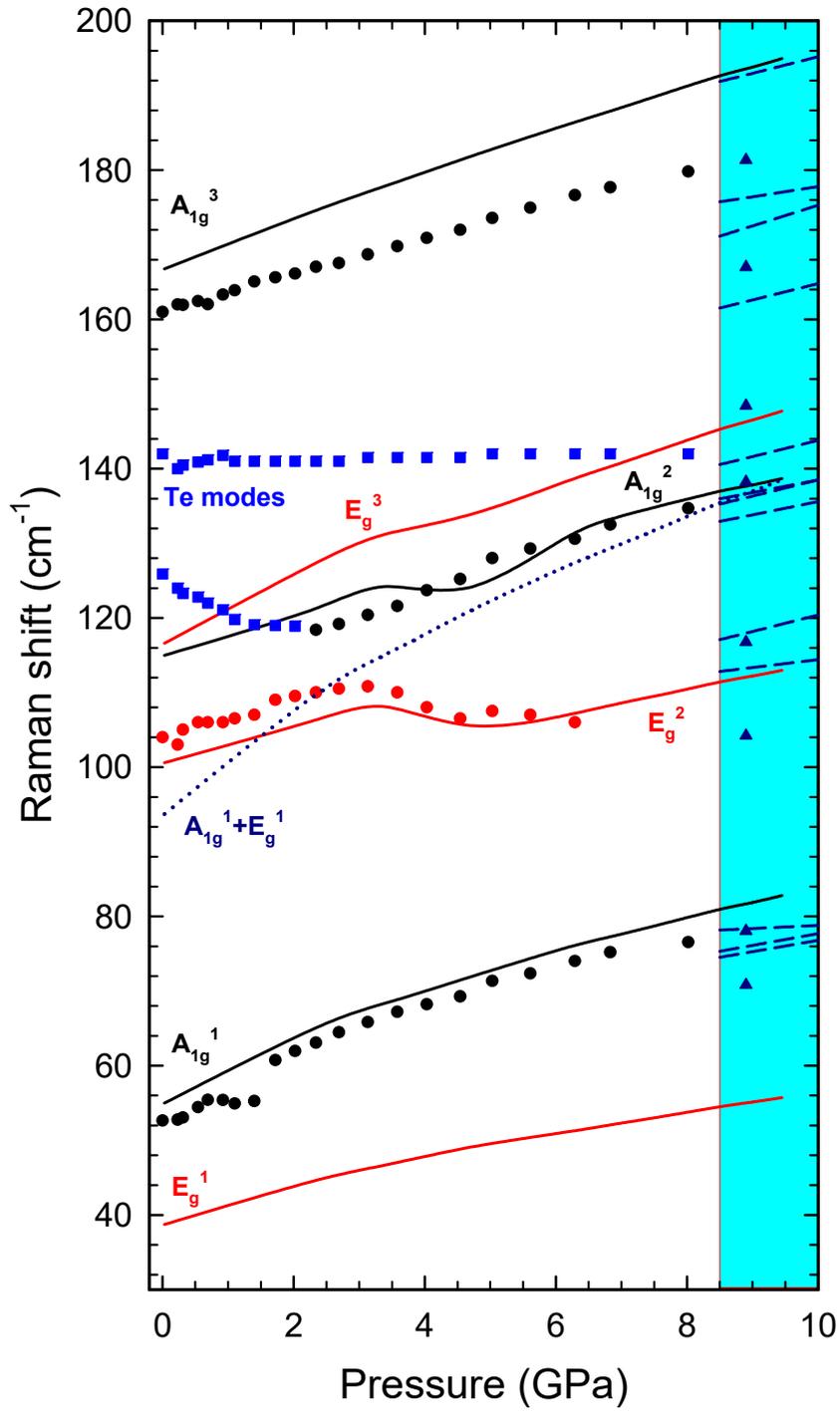

**Figure 6.** Pressure dependence of the experimental (symbols) and theoretical (lines) Raman-active mode frequencies in SnSb$_2$Te$_4$. Dotted line represents the pressure dependence of the A$_{1g}^1$ + E$_g^1$ combination at Γ, while dashed lines represent the Raman-active modes expected for β-Sb$_2$Te$_3$.

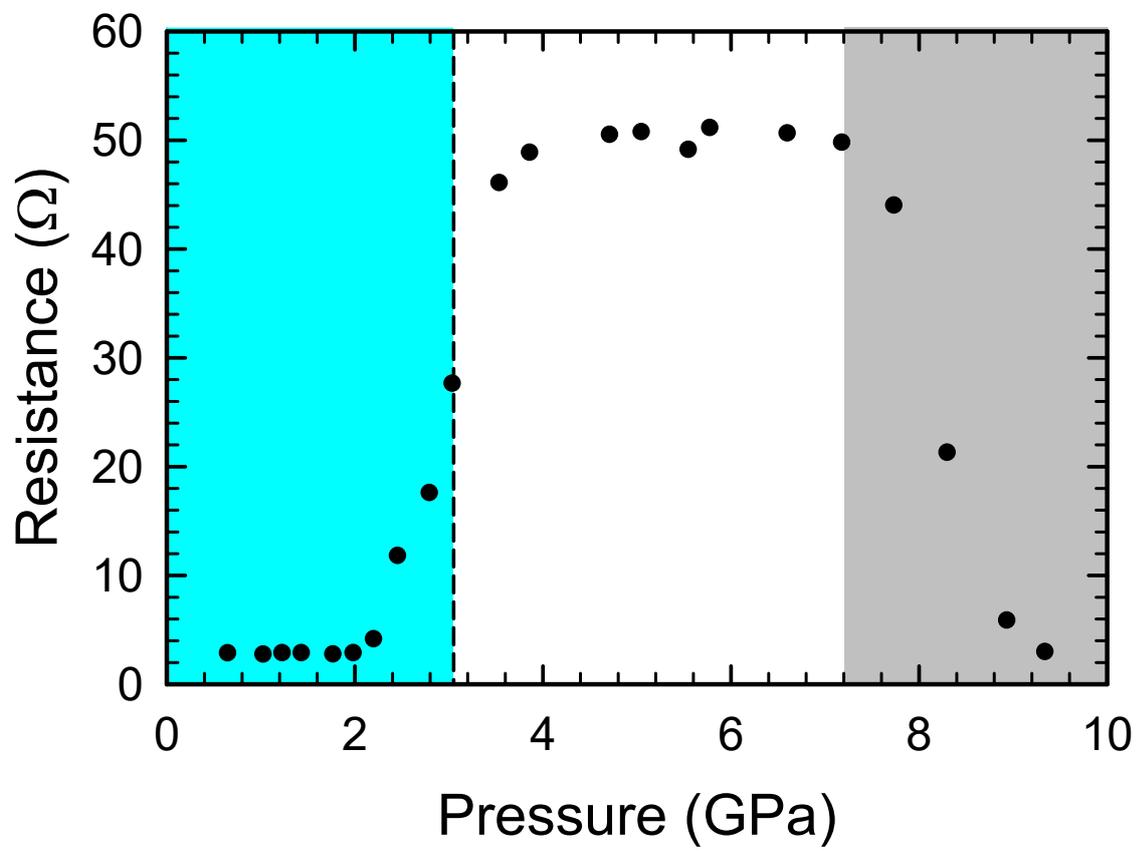

**Figure 7.** Evolution of the resistance of compressed SnSb$_2$Te$_4$ recorded during the upstroke.

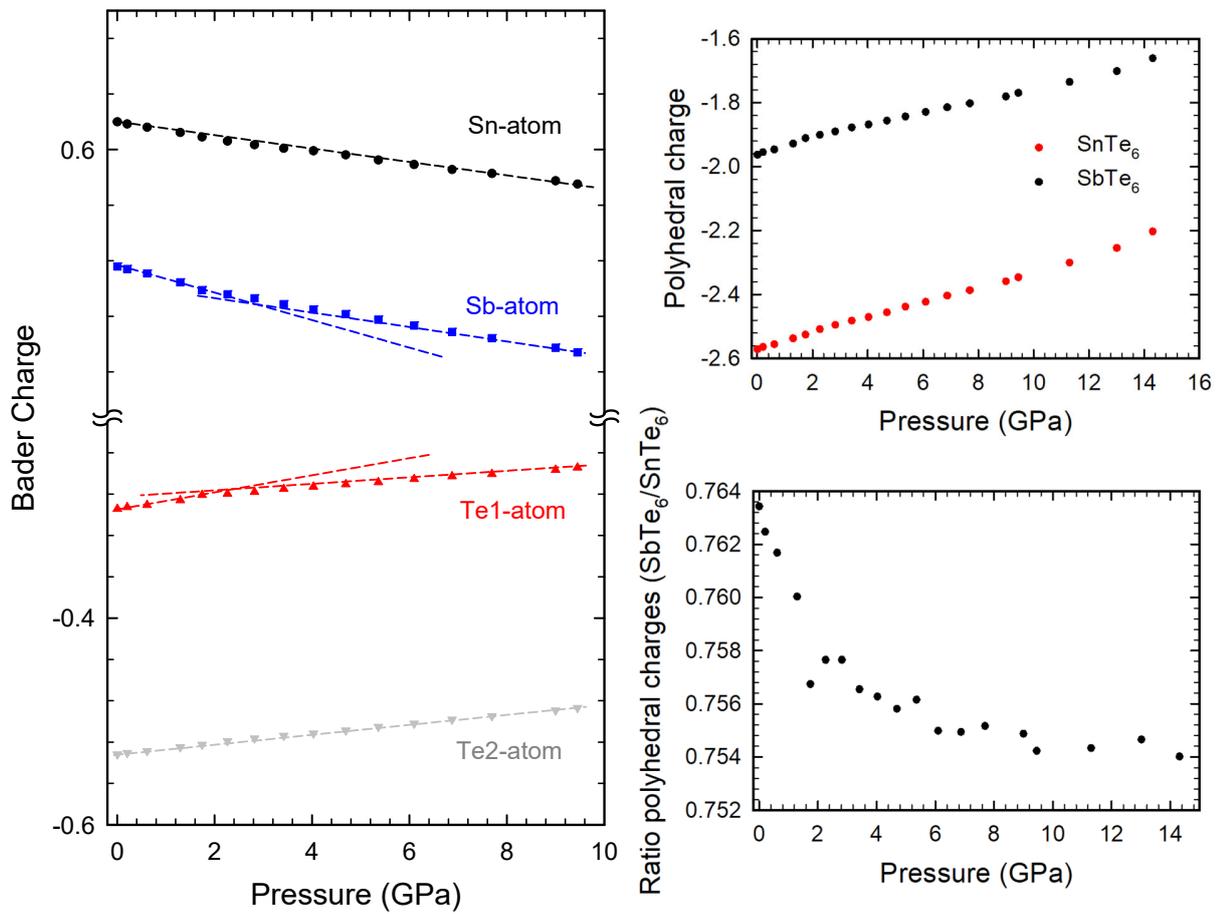

**Figure 8.** Evolution of the (a) Bader charge of the different crystallographic atoms under pressure and (b) the ratio between Te1/Te2 and (c) Sn/Sb Bader charge.

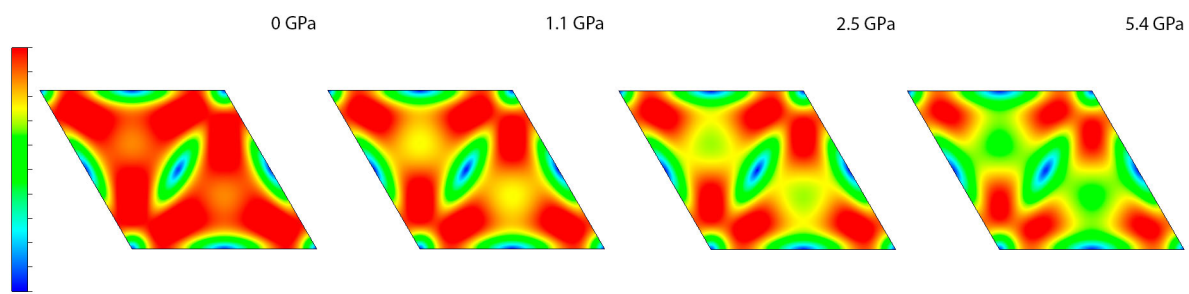

**Figure 9.** Evolution of the electronic distribution in the interlayer plane with increasing pressure.

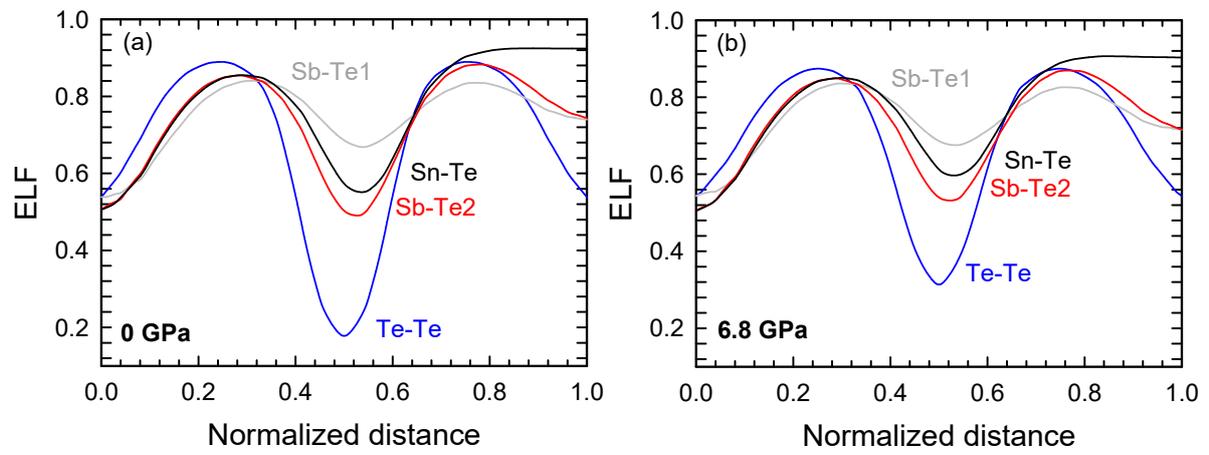

**Figure 10.** Pressure dependence of the ELF along the path of the Sb-Te1, Sb-Te2, Sn-Te and Te-Te bonds at 0 GPa **(a)** and 6.8 GPa **(b)**.

**Table I.** Calculated (th.) and experimental (exp.) Volume ($V_0$), bulk modulus ($B_0$), and its derivative ($B_0$') of SnBi$_2$Te$_4$ and SnSb$_2$Te$_4$ at ambient pressure.

|  | $V_0$ (Å$^3$) | $B_0$ (GPa) | $B_0$' |
|---|---|---|---|
| SnSb$_2$Te$_4$ | | | |
| exp.[a] | 663.1(6) | 31.6(14) | 8(8) |
| th.(GGA-PBESol)[b] | 659.3(6) | 41.0(15) | 6.5(6) |
| α-Sb$_2$Te$_3$ | | | |
| exp.[c] | 481.1(8) | 36.1(9) | 6.2(4) |
| th.(GGA-PBESol)[d] | 473.1(8) | 43(2) | 4.3(5) |

[a] This work; [b] Calculations including SOC in this work; [c] Average experimental value from Ref. 40; [e] Calculations including SOC from Ref. 40.

**Table II.** Calculated (th.) and experimental (exp.) bulk modulus ($B_0$) of the lattice parameters of SnBi$_2$Te$_4$ and SnSb$_2$Te$_4$ and their associated axial compressibilities.

|  | $B_{0a}$ (GPa) | $B_{0c}$ (GPa) | $\kappa_a$ (10$^{-3}$ GPa$^{-1}$) | $\kappa_c$ (10$^{-3}$ GPa$^{-1}$) |
|---|---|---|---|---|
| SnSb$_2$Te$_4$ | | | | |
| exp.[a] | 42.4(2) | 23.8(4) | 7.9(3) | 14(2) |
| th.(GGA-PBESol)[b] | 48(2) | 30(4) | 6.9(7) | 11.1(15) |
| α-Sb$_2$Te$_3$ | | | | |
| exp.[c] | 47.3(12) | 22(3) | 7.0(2) | 15.0(5) |
| th.(GGA-PBESol)[b] | 48(1) | 21(2) | 6.94(14) | 15.9(7) |
| SnBi$_2$Te$_4$ | | | | |
| exp.[d] | 42.3(17) | 25.3(17) | 7.9(3) | 13.1(9) |
| th.(GGA-PBESol)[d] | 48.0(15) | 30(2) | 6.9(7) | 11.1(7) |

[a] This work; [b] Calculations including SOC in this work; [c] Calculated from data of Refs. 40 and 41; [d] Data from Ref. 37, where calculations include SOC.

**Table III.** Theoretical and experimental Raman-active mode frequencies and their pressure coefficients of SnSb$_2$Te$_4$ at room temperature as fitted with equation $\omega(P) = \omega_0 + aP + bP^2$ compared with those of theoretically simulated $\alpha$-Sb$_2$Te$_3$.

| Mode symmetry | SnSb$_2$Te$_4$ | | | | | | $\alpha$-Sb$_2$Te$_3$ | | |
| --- | --- | --- | --- | --- | --- | --- | --- | --- | --- |
| | Experiment | | | Theoretical Calculations[a] | | | Theoretical Calculations[a] | | |
| | $\omega_0$ (cm$^{-1}$) | a (cm$^{-1}$/GPa) | b (cm$^{-1}$/GPa$^2$) | $\omega_0$ (cm$^{-1}$) | a (cm$^{-1}$/GPa) | b (cm$^{-1}$/GPa$^2$) | $\omega_0$ (cm$^{-1}$) | a (cm$^{-1}$/GPa) | b (cm$^{-1}$/GPa$^2$) |
| $E_g^1$ | - | - | - | 38.87(15) | 2.52(8) | -0.081(9) | 50.4 | 2.62 | -0.09 |
| $A_{2g}^1$ | 53.3(8) | 4.2(6) | -0.15(8) | 55.3(2) | 4.25(12) | -0.145(12) | 68.9 | 4.3 | -0.07 |
| $E_g^2$ | 103.3(4) | 3.7(7) | -0.4(2) | 100.51(8) | 2.45(6) | - | 116.6 | 2.11 | |
| $A_{2g}^2$ | 107.8(14) | 4.6(6) | -0.15(6) | 115.1(2) | 2.70(18) | - | 167.6 | 2.57 | |
| $E_g^3$ | - | - | - | 116.65(17) | 4.63(13) | - | | | |
| $A_{2g}^3$ | 160.87(14) | 2.68(11) | -0.035(15) | 167.10(16) | 3.38(8) | -0.041(9) | | | |

[a] This work

## ASSOCIATED CONTENT

This material is available free of charge via the Internet at http://pubs.acs.org.

## AUTHOR INFORMATION

**Corresponding Author**

*(J.A.S.) E-mail: juasant2@upv.es

**Author Contributions**

The manuscript was written through contributions of all authors. All authors have given approval to the final version of the manuscript.

## ACKNOWLEDGMENT

This work has been performed under financial support from Spanish MINECO under projects MALTA Consolider Ingenio 2010 network (MAT2015-71070-REDC) and projects FIS2017-83295-P and MAT2016-75586-C4-1/2/3-P, from Generalitat Valenciana under project PROMETEO/2018/123. This publication is fruit of "Programa de Valoración y Recursos Conjuntos de I+D+i VLC/CAMPUS and has been financed by the Spanish Ministerio de Educación, Cultura y Deporte as part of "Programa Campus de Excelencia Internacional". Supercomputer time has been provided by the Red Española de Supercomputación (RES) and the MALTA cluster. JAS acknowledges "Ramón y Cajal" fellowship (RYC-2015-17482) program for financial support and E.L.d.S acknowledges Marie Sklodowska-Curie grant No. 785789-COMEX from European Union's Horizon 2020 research and innovation programme. We also thank ALBA synchrotron and DIAMOND light source for funded experiments.

REFERENCES

[1] A. R. Mellnik, J. S. Lee, A. Richardella, J. L.Grab, P. J. Mintun, M. H. Fischer, A. Vaezi, A. Manchon, E.-A. Kim, N. Samarth, D. C. Ralph, *Spin-transfer torque generated by a topological insulator*, Nature **511**, 449-451 (2014).

[2] Y. L. Chen, J. G. Analytis, J.-H. Chu, Z. K. Liu, S.-K. Mo, X. L. Qi, H. J. Zhang, D. H. Lu, X. Dai, Z. Fang, S. C. Zhang, I. R. Fisher, Z. Hussain, Z.-X. Shen, *Experimental Realization of a Three-Dimensional Topological Insulator, $Bi_2Te_3$*, Science **325**, 178-181 (2009).

[3] D. Hsieh, Y. Xia, D. Qian, L. Wray, J. H. Dil, F. Meier, J. Osterwalder, L. Patthey, J. G. Checkelsky, N. P. Ong, A. V. Fedorov, H. Lin, A. Bansil, D. Grauer, Y. S. Hor, R. J. Cava, M. Z. Hasan, *A tunable topological insulator in the spin helical Dirac transport regime*, Nature **460**, 1101-1105 (2009).

[4] T. Zhang, Y. Jiang, Z. Song, H. Huang, Y. He, Z. Fang, H. Weng, C. Fang, *Catalogue of topological electronic materials*, Nature **566**, 475-479 (2019).

[5] M. G. Vergniory, L. Elcoro, C. Felser, N. Regnault, B. Andrei Bernevig, Z. Wang, *A complete catalogue of high-quality topological materials*, Nature **566**, 480-485 (2019).

[6] F. Tang, H. C. Po, A. Vishwanath, X. Wan, *Comprehensive search for topological materials using symmetry indicators*, Nature **566**, 486-489 (2019).

[7] A. Zunger, *Beware plausible predictions of fantasy materials*, Nature **566**, 447-449 (2019).

[8] H. Zhang, C.X. Liu, X.L. Qi, X. Dai, Z. Fang, S.-C. Zhang, *Topological insulators in $Bi_2Se_3$, $Bi_2Te_3$ and $Sb_2Te_3$ with a single Dirac cone on the surface*, Nature Physics **5**, 438-442 (2009)

[9] Y. Xia, D. Qian, D. Hsieh, L. Wray, A. Pal, H. Lin, A. Bansil, D. Grauer, Y. S. Hor, R. J. Cava, M. Z. Hasan, *Observation of a large-gap topological-insulator class with a single Dirac cone on the surface*, Nature Physics **5**, 398-402 (2009).

[10] M. Taherinejad, K. F. Garrity, D. Vanderbilt, *Wannier center sheets in topological insulators*, Phys. Rev. B **89**, 115102 (2014).

[11] D. Niesner, S. Otto, V. Hermann, Th. Fauster, T. V. Menshchikova, S. V. Eremeev, Z. S. Aliev, I. R. Amiraslanov, M. B. Babanly, P. M. Echenique, and E. V. Chulkov, *Bulk and surface*

*electron dynamics in a p-type topological insulator SnSb$_2$Te$_4$*, Physical Review B **89**, 081404(R) (2014)

[12] D.M. Rowe, *CRC Handbook of Thermoelectrics*, CRC Press Inc., New York, 1995

[13] R. Venkatasubramanian, E. Siivola, T. Colpitts, B. O'Quinn, *Thin-film thermoelectric devices with high room-temperature figures of merit*, Nature **413**, 597-602 (2001)

[14] S.V. Eremeev, Y.M. Koroteev, E.V. Chulkov, *Effect of the atomic composition of the surface on the electron surface states in topological insulators $A_2^V B_3^{VI}$*, JETP Letters **91**, 387 (2010).

[15] T.V. Menshchikova, S.V. Eremeev, E.V. Chulkov, *On the origin of two-dimensional electron gas states at the surface of topological insulators,* JETP Letters **94**, 106 (2011).

[16] T.V. Menshchikova, S.V. Eremeev, E.V. Chulkov, *Electronic structure of SnSb$_2$Te$_4$ and PbSb$_2$Te$_4$ topological insulators,* Appl. Surf. Sci. **267**, 1-3 (2013).

[17] G. Concas, T.M. de Pascale, L. Garbato, F. Ledda, F. Meloni, A. Rucci, M. Serra, *Electronic and structural properties of the layered SnSb$_2$Te$_4$ semiconductor: Ab initio total-energy and Mössbauer spectroscopy study,* J. Phys, Chem. Solids **53**, 791 (1992).

[18] S. V. Eremev, T. V. Menschikova, I. V. Silkin, M. G. Vergniory, P. M. Echenique, E. V. Chulkov, *Sublattice effect on topological surface states in complex (SnTe)$_{n>1}$(Bi$_2$Te$_3$)$_{m=1}$ compounds,* Phys. Rev. B **91**, 245145 (2015).

[19] K. A. Agaev, A. G. Talybov, *Electron-diffraction analysis of the structure of GeSb$_2$Te$_4$*, Kristallografiya **11**, 454-456 (1966).

[20] R. M. Imamov, S. A. Semiletov, and Z. G. Pinsker, *The crystal chemistry of semiconductors with octahedral and with mixed atomic coordination*, Soviet Phys. Cryst. 15, 239 (1970)

[21] A. Y. Kuznetsov, A. S. Pereira, A. A. Shiryaev, J. Haines, L. Dubrovinsky, V. Dmitriev, P. Pattison, N. Guignot, *Pressure-Induced Chemical Decomposition and Structural Changes of Boric Acid,* J. Phys. Chem. B **110**, 13858-13865 (2006).

[22] L.E. Shelimova, O.G. Karpinskii, P.P. Konstantinov, E.S. Avilov, M.A. Kretova and V.S. Zemskov, *Crystal structures and thermoelectric properties of layered compounds in the ATe-Bi$_2$Te$_3$ (A = Ge, Sn, Pb) systems*, Inorganic Materials **40**, 451 (2004).

[23] B. A. Kuropatwa, A. Assoud, H. Kleinke, *Effects of Cation Site Substitutions on the Thermoelectric Performance of Layered SnBi2Te4 utilizing the Triel Elements Ga, In, and Tl*, Zeitschrift für anorganische und allgemeine Chemie **639**, 2411-2420 (2013).

[24] B. A. Kuropatwa, H. Kleinke, *Thermoelectric Properties of Stoichiometric Compounds in the $(SnTe)_x(Bi_2Te_3)_y$ System*, Z. Anorg. Allg. Chem. **638**, 2640-2647 (2012).

[25] A. Banik, and K. Biswas, *Synthetic of Natural van der Waals Heterostructures,* Angew. Chem. Int. Ed. **56**, 14561-14566 (2017).

[26] L.E. Shelimova, O.G. Karpinskii, T.E. Svechnikova, I.Y. Nikhezina, E.S. Avilov, M.A. Kretova, V.S. Zemskov, *Effect of cadmium, silver, and tellurium doping on the properties of single crystals of the layered compounds PbBi4Te7 and PbSb2Te4*. Inorg. Mater. 44, 371 (2008).

[27] H.W. Shu, S. Jaulmes, J. Flahaut, *Système AsGeTe: III. Étude cristallographique d'une famille de composés a modèles structuraux communs: β-$As_2Te_3$, $As_4GeTe_7$ et $As_2Ge_nTe_{3+n}$ (n = 1 à 5)*, J. Solid State Chem. **74**, 277 (1988).

[28] T.B. Zhukova, A.I. Zaslavskii, *Crystal-Structures of Compounds PbBi4Te, PbBi2Te4, SnBi4Te7, SnBi2Te4, SnSb2Te4, and GeBi4Te7*, Soviet Phys. Crystallogr. **16**, 796 (1972).

[29] K. Adouby, A.A. Toure, G. Kra, J. Olivier-Fourcade, J. C. Jumas, C. P. Vicente, *Phase diagram and local environment of Sn and Te: SnTe-Bi and SnTe-$Bi_2Te_3$ systems*, C. R. Acad. Sci., Chem. **3**, 51 (2000).

[30] T.B. Zhukova, and A.I. Zaslavskii, *Crystal Structures of PbBi4Te7, PbBi2Te4, SnBi4Te7, SnBi2Te4, SnSb2Te4, and GeBi4Te7*, Kristallografiya, **16**, 918 (1971).

[31] O. Oeckler, M. N. Schneider, F. Fahrnbauer, G. Vaughan, *Atom distribution in SnSb2Te4 by resonant X-ray diffraction*. Solid State Sci. **13**, 1157-1161 (2011).

[32] T. Schafer, P. M. Konze, J. D. Huyeng, V. L. Deringer, T. Lesieur, P. Muller, M. Morgenstern, R. Dronkowski and M. Wuttig, *Chemical Tuning of Carrier Type and Concentration in a Homologous Series of Crystalline Chalcogenides*, Chem. Mater. **29**, 6749-6757 (2017).

[33] J. Gallus, *Lattice Dynamics in the SnSb2Te4 Phase Change Material*, Diplomarbeit, Rheinisch-Westfälischen Technischen Hochschule Aachen, 2011.

[34] J.Y. Raty, M. Schumacher, P. Golub, V. L. Deringer, C. Gatti and M. Wuttig, *A Quantum-Mechanical Map for Bonding and Properties in Solids*. Advanced Materials **31**, 1806280 (2019).

[35] M. Wuttig, V. L. Deringer, X. Gonze, C. Bichara, J. Y. Raty, *Incipient Metals: Functional Materials with a Unique Bonding Mechanism*. Advanced Materials **30**, 1803777 (2018).

[36] W.-P. Hsieh, P. Zalden, M. Wuttig, A. M. Lindenberg, W. L. Mao, *HP Raman spectroscopy of phase change materials*. Appl. Phys. Lett. **103**, 191908 (2013).

[37] R. Vilaplana, J. A. Sans, F. J. Manjón, J. Sánchez-Benítez, C. Popescu, O. Gomis, A. L. J Pereira, B. García-Domene, P. Rodríguez-Hernández, A. Muñoz, D. Daisenberger, O. Oeckler, *Structural and electrical study of the topological insulator $SnBi_2Te_4$ at HP*. Journal of Alloys and Compounds **685**, 962-970 (2016).

[38] R. J. Angel, J. Gonzalez-Platas, M. Alvaro, *EosFit-7c and a Fortran module (library) for equation of state calculations*. Zeitschrift für Kristallographie, **229**, 405-419 (2014).

[39] D. Zhou, Q. Li, Y. Ma, Q. Cui, C. Chen, *Unraveling Convoluted Structural Transitions in SnTe at HP*. J. Phys. Chem. C **117**, 5352−5357 (2013).

[40] O. Gomis, R. Vilaplana, F. J. Manjón, P. Rodríguez-Hernández, E. Pérez-González, A. Muñoz. V. Kucek, C. Drasar, *Lattice dynamics of $Sb_2Te_3$ at HPs*. Phys. Rev. B **84**, 174305 (2011).

[41] N. Sakai, T. Kajiwara, K. Takemura, S. Minomura, Y. Fujii, *Pressure-Induced Phase Transition in $Sb_2Te_3$*. Solid State Communications **40**, 1045-1047 (1981).

[42] B.-T. Wang, P. Souvatzis, O. Eriksson, P. Zhang, *Lattice dynamics and chemical bonding in $Sb_2Te_3$ from first-principles calculations*. The Journal of Chemical Physics **142**, 174702 (2015).

[43] A. L. J. Pereira, J. A. Sans, R. Vilaplana, O. Gomis, F. J. Manjon, P. Rodríguez-Hernandez, A. Muñoz, C. Popescu, A. Beltran, *Isostructural Second-Order Phase Transition of β-$Bi_2O_3$ at HPs: An Experimental and Theoretical Study*. J. Phys. Chem. C **118**, 23189−23201 (2014).

[44] V.P. Cuenca-Gotor, J.A. Sans, J. Ibáñez, C. Popescu, O. Gomis, R. Vilaplana, F.J. Manjón, A. Leonardo, E. Sagasta, A. Suárez-Alcubilla, I.G. Gurtubay, M. Mollar, and A. Bergara. *Structural, vibrational, and electronic study of α-$As_2Te_3$ under compression*. J. Phys. Chem. C **120**, 19340 (2016).

[45] K. Robinson, G. V. Gibbs, P. H. Ribbe, *Quadratic Elongation: A Quantitative Measure of Distortion in Coordination Polyhedra*. Science **172**, 567-570 (1971).

[46] W. H. Baur, *The geometry of polyhedral distortions. Predictive relationships for the phosphate group*, Acta Crystallogr., Sect. B: Struct. Sci. **30**, 1195 (1974).

[47] A. Walsh, and W. Watson, *Influence of the Anion on Lone Pair Formation in Sn(II) Monochalcogenides: A DFT Study*, J. Phys. Chem. B **109**, 18868-18875 (2005).

[48] A. Skowron, F. W. Boswell, J. M. Corbett, N. J. Taylor, *Structure Determination of $PbSb_2Se_4$*, J. Sol. Stat. Chem. **112**, 251-254 (1994).

[49] P. K. Smith, J. B. Parise, *Structure Determination of $SnSb_2S_4$ and $SnSb_2Se_4$ by High-Resolution Electron Microscopy*, Acta Cryst. **B41**, 84-87 (1985).

[50] Y. Iitaka, W. Nowacki, *A Redetermination of the Crystal Structure of Galenobismutite, $PbBi_2S_4$*, Acta Cryst. **15**, 691 (1962).

[51] X. Zhang, V. Stevanović, M. d'Avezac, S. Lany, A. Zunger, *Prediction of $A_2BX_4$ metal-chalcogenide compounds via first-principles thermodynamics*, Physical Review B 86, 014109 (2012).

[52] A. Zunger, *Systematization of the stable crystal structure of all AB-type binary compounds: A pseudopotential orbital-radii approach*, Physical Review B 22, 5839-5872 (1980).

[53] F. J. Manjón, R. Vilaplana, O. Gomis, E. Pérez-González, D. Santamaría-Pérez, V. Marín-Borrás, A. Segura, J. González, P. Rodríguez-Hernández, A.Muñoz, C. Drasar, V. Kucek, and V. Muñoz-Sanjosé. *HP studies of topological insulators $Bi_2Se_3$, $Bi_2Te_3$, and $Sb_2Te_3$*. Phys. Stat. Sol. (b) **250**, 669 (2013).

[54] A. V. Kolobov, J. Haines, A. Pradel, M. Ribes, P. Fons, J. Tominaga, Y. Katayama, T. Hammouda, and T. Uruga, *Pressure-Induced Site-Selective Disordering of $Ge_2Sb_2Te_5$: A New Insight into Phase-Change Optical Recording*, Phys. Rev. Lett. **97**, 035701 (2006).

[55] A. K. Arora, *Pressure-induced amorphization versus decomposition*, Solid State Commun. **115**, 665 (2000).

[56] W. A. Basset, L.-C. Ming, *Disproportionation of $Fe_2SiO_4$ to $2FeO+SiO_2$ at pressures up to 250 kbar and temperatures up to 3000 °C*. Phys. Earth Planet. Interiors **6**, 154-160 (1972).

[57] Y. Fei, H.-K. Mao, *Static Compression of Mg(OH)$_2$ to 78 GPa at High Temperature and Constraints on the Equation of State of Fluid H$_2$O*. Journal of Geophysical Research **98**, 11875-11884 (1993).

[58] A. Y. Kuznetsov, A. S. Pereira, A. A. Shiryaev, J. Haines, L. Dubrovinsky, V. Dmitriev, P. Pattison, N. Guignot, *Pressure-Induced Chemical Decomposition and Structural Changes of Boric Acid,* J. Phys. Chem. B **110**, 13858-13865 (2006).

[59] J. Catafesta, P. R. Rovani, C. A. Perottoni, A. S. Pereira, *Pressure-enhanced decomposition of Ag$_3$[Co(CN)$_6$]*, Journal of Physics and Chemistry of Solids **77**, 151–156 (2015).

[60] D. Duan, X. Huang, F. Tian, D. Li, H. Yu, Y. Liu, Y. Ma, B. Liu, and T. Cui, *Pressure-induced decomposition of solid hydrogen sulfide*, Phys. Rev. B **91**, 180502(R) (2015).

[61] E. Kroumova, M. I. Aroyo, J. M. Perez Mato, A. Kirov, C. Capillas, S. Ivantchev, H. Wondratschek, *Bilbao Crystallographic Server: Useful Databases and Tools for Phase Transitions Studies*, Phase Transitions **76**, 155-170 (2003).

[62] P. Canepa, R. M. Hanson, P. Ugliengo, M. Alfredsson, *J-ICE: A New Jmol Interface for Handling and Visualizing Crystallographic and Electronic Properties*, J. Appl. Cryst. 44, 225-229 (2011).

[63] C. Marini, D. Chermisi, M. Lavagnini, D. Di Castro, C. Petrillo, L. Degiorgi, S. Scandolo, P. Postorino, *HP phases of crystalline tellurium: A combined Raman and ab initio study*. Physical Review B **86**, 064103 (2012).

[64] R. Vilaplana, O. Gomis, E. Pérez-González, H.M. Ortiz, F.J. Manjón, P. Rodríguez-Hernández, A. Muñoz, P. Alonso-Gutiérrez, M.L. Sanjuán, V.V. Ursaki and I.M. Tiginyanu, *Lattice dynamics study of HgGa$_2$Se$_4$ at HPs*, J. Phys. Chem. C **117**, 15773 (2013).

[65] G.A.S. Ribeiro, L. Paulatto, R. Bianco, I. Errea, F. Mauri, and M. Calandra, *Strong anharmonicity in the phonon spectra of PbTe and SnTe from first principles*, Phys. Rev. B 97, 014306 (2018).

[66] J. Pellicer-Porres, A. Segura, Ch. Ferrer-Roca, J. A. Sans, P. Dumas, *Investigation of lattice dynamical and dielectric properties of MgO under high pressure by means of mid- and far-infrared spectroscopy*, J. Phys.: Condens. Matter. **25**, 505902 (2013).

[67] C.-H. Wang, X.-P. Jing, W. Feng, J. Lu, *Assignment of Raman-active vibrational modes of MgTiO$_3$*, Journal of Applied Physics **104**, 034112 (2008)

[68] K. Zhao, Y. Wang, Y. Sui, C. Xin, X. Wang, Y. Wang, Z. Liu, and B. Li, *First principles study of isostructural phase transition in Sb$_2$Te$_3$ under HP,* Phys. Status Solidi RRL **9**, 379–383 (2015).

[69] B.-T. Wang, P. Souvatzis, O. Eriksson, and P. Zhang, *Lattice dynamics and chemical bonding in Sb$_2$Te$_3$ from first-principles calculations,* The Journal of Chemical Physics **142**, 174702 (2015)

[70] M. Cardona, M. L. Thewalt, *Isotope effects on the optical spectra of semiconductors*, Reviews of Modern Physics **77**, 1173-1224 (2005).

[71] F. J. Manjón, J. Serrano, I. Loa, K. Syassen, C. T. Lin, M. Cardona, *Effect of Pressure on the Anomalous Raman Spectrum of CuBr*, Physica Status Solidi B. **223**, 331-336 (2001).

[72] M. Krauzman, R. M. Pick, H. Poulet, G. Hamel, B. Prevot, *Raman Detection of one-Phonon—Two-Phonon Interactions in CuCl*, Physical Review Letters **33**, 528-530 (1974).

[73] G. Kanellis, W. Kress, H. Bilz, *Fermi Resonance in the Phonon Spectra of Copper Halides*, Physical Review Letters **56**, 938-940 (1986).

[74] V. M. Agranovich, *Spectroscopy and Excitation Dynamics of Condensed Moiecular Systems*. Edited by North-Holland, Amsterdam, 83 (1983).

[75] M. A. Hakeem, D. E. Jackson, J. J. Hamlin, D. Errandonea, J. E. Proctor, and M. Bettinelli, *High Pressure Raman, Optical Absorption, and Resistivity Study of SrCrO$_4$.* Inorg. Chem. **57**, 7550-7557 (2018).

[76] N. Sakai, T. Kajiwara, K. Takemura, S. Minomura, and Y. Fujii, *Pressure-induced Phase Transition in Sb$_2$Te$_3$*, Solid State Commun. **40**, 1045 (1981).

[77] D. Errandonea, A. Segura, D. Martínez-García and V. Muñoz-San Jose, *Hall-effect and resistivity measurements in CdTe and ZnTe at high pressure: Electronic structure of impurities in the zinc-blende phase and the semimetallic or metallic character of the high-pressure phases*, Physical Review B **79**, 125203 (2009).

[78] A. Segura, V. Panchal, J.F. Sánchez-Royo, V. Marín-Borrás, V. Muñoz-Sanjosé, P. Rodríguez-Hernández, A. Muñoz, E. Pérez-González, F.J. Manjón, and J. González, *Trapping of three-dimensional electrons and transition to two-dimensional transport in the three-dimensional topological insulator Bi2Se3 under high pressure*, Physical Review B **85**, 195139 (2012).

[79] C. Gatti, *Chemical bonding in crystals: new directions*, Z. Kristallogr. **220**, 390-457 (2005).

[80] M. Xu, S. Jakobs, R. Mazzarello, J.-Y. Cho, Z. Yang, H. Hollermann, D. Shang, X.S. Miao, Z.H. Yu, L. Wang, and M. Wuttig, *Impact of Pressure on the Resonant Bonding in Chalcogenides*, J. Phys. Chem. C **121**, 25447 (2017).

[81] P. Hohenberg W. Kohn, *Inhomogeneous Electron Gas*, Phys. Rev. B **136**, 864-871 (1964).

[82] P. E. Blöchl, *Projector Augmented-Wave Method*, Phys. Rev. B **50**, 17953-17979 (1994).

[83] G. Kresse, J. Hafner, *Ab Initio Molecular Dynamics for Liquid Metals*, Phys. Rev. B **47**, 558-561 (1993).

[84] J. P. Perdew, A. Ruzsinszky, G. I. Csonka, O. A. Vydrow, G. E. Suseria, L. A. Constantin, X. Zhou, K. Burke, *Restoring the Density-Gradient Expansion for Exchange in Solids and Surfaces*, Phys. Rev. Lett. **100**, 136406 (2008).

[85] A. Mujica, A. Rubio, A. Muñoz, R. J. Needs, *HP phases of group-IV, III–V, and II–VI compounds*. Rev. Mod. Phys. **79**, 863-912 (2003).

[86] K. Parlinski, see: *http://www.computingformaterials.com/index.html*

[87] W. Tang, E. Sanville, and G. Henkelman, *A grid-based Bader analysis algorithm without lattice bias*, J. Phys.: Compute Mater. **21** 084204 (2009).

[88] E. Sanville, S. D. Kenny, R. Smith, and G. Henkelman, *An improved grid-based algorithm for Bader charge allocation*, J. Comp. Chem. **28**, 899-908 (2007).

[89] G. Henkelman, A. Arnaldsson, and H. Jónsson, *A fast and robust algorithm for Bader decomposition of charge density*, Comput. Mater. Sci. **36**, 254-360 (2006).

[90] M. Yu and D. R. Trinkle, *Accurate and efficient algorithm for Bader charge integration*, J. Chem. Phys. **134**, 064111 (2011).

[91] http://theory.cm.utexas.edu/henkelman/code/bader/

[92] E. R. Johnson, S. Keinan, P. Mori-Sanchez, J. Contreras-Garcia, A. J. Cohen, and W. Yang, *Revealing Noncovalent Interactions*, J. Am. Chem. Soc. **132**, 6498-6506 (2010).

[93] J. Contreras-Garcia, E. R. Johnson, S. Keinan, R. Chaudret, J-P. Piquemal, D. N. Beratan, and W. Yang, *NCIPLOT: A Program for Plotting Noncovalent Interaction Regions*, J. Chem. Theory Comput. **7**, 625- 632 (2011).

[94] F. Fauth, I. Peral, C. Popescu, M. Knapp, *The New Material Science Powder Diffraction Beamline at ALBA Synchrotron*, Powder Diffr., **28**, 360 (2013).

[95] A. Dewaele, P. Loubeyre, M. Mezouar, *Equations of State of Six Metals above 94 GPa*, Phys. Rev. B **70**, 094112 (2004).

[96] A. P. Hammersley, S. O. Svensson, M. Hanfland, A. N. Fitch, D. Häusermann, *Two-dimmensional detector software: From real detector to idealized image or two-theta scan*, HP Research **14**, 235-248 (1996).

[97] B. H. Toby, *EXPGUI, A Graphical User Interface for GSAS*, J. Appl. Crystallogr. **34**, 210−213 (2001).

[98] A. C. Larson, R. B. Von Dreele, *General Structure Analysis System (GSAS)*, Los Alamos National Laboratory Report LAUR **86**, 748 (1994).

[99] S. Klotz, J.-C. Chervin, P. Munsch and G. Le Marchand, *Hydrostatic limits of 11 pressure transmitting media*, J. Phys. D: Appl. Phys. **42**, 075413 (2009)

[100] D. Errandonea, A. Muñoz, J. González-Platas, *Comment on "High-pressure x-ray diffraction study of $YBO_3/Eu^{3+}$, $GdBO_3$, and $EuBO_3$: Pressure-induced amorphization in $GdBO_3$"*, Journal of Applied Physics **115**, 216101 (2014)

[101] H. K. Mao, J. Xu, P. M. Bell, *Calibration of the Ruby Pressure Gauge to 800 kbar under Quasi-Hydrostatic Conditions*, J. Geophys. Res. **91**, 4673−4676 (1986).

[102] K. Syassen, *Ruby Under Pressure*, HP Res. **28**, 75−126 (2008).

[103] A. Debernardi, C. Urlich, M. Cardona, K. Syassen, *Pressure Dependence of Raman Linewidth in Semiconductors*, Phys. Status Solidi B **223**, 213−223 (2001).

[104] B. García-Domene, H. M. Ortiz, O. Gomis, J. A. Sans, F. J. Manjón, A. Munoz, P. Rodríguez-Hernández, S. N. Achary, D. Errandonea, D. Martínez-García, A. H. Romero, A.

Singhal, A. K. Tyagi, *HP Lattice Dynamical Study of Bulk and Nanocrystalline In$_2$O$_3$*, J. Appl. Phys. **112**, 123511 (2012).

[105] H. W. Shu, S. Jaulmes, J. Flahaut, *Système AsGeTe: III.Ètude cristallographique d'une famille de composés a modèles structuraux communs: β-As$_2$Te$_3$, As$_4$GeTe$_7$ et As$_2$Ge$_n$Te$_{3+n}$ (n = 1 à 5)*, Journal of Solid State Chemistry **74**, 277-286 (1988).

[106] T. Matsunaga and N. Yamada, *Structural investigation of GeSb$_2$Te$_4$: A high-speed phase-change material*, Physical Review B **69**, 104111 (2004).

[107] O. G. Karpinskii, L. E. Shelimova, M. A. Kretova, *Crystal structure and point defects of Ge$_{1+delta}$Bi$_2$Te$_4$*, Inorganic Materials **33**, 793-797 (1997).

[108] T. B. Zhukova, and A. I. Zaslavskii, *Crystal Structures of PbBi$_4$Te$_7$, PbBi$_2$Te$_4$, SnBi$_4$Te$_7$, SnBi$_2$Te$_4$, SnSb$_2$Te$_4$, and GeBi$_4$Te$_7$*, Kristallografiya **16**, 918–922 (1971).

[109] L. E. Shelimova, O. G. Karpinskii, T. E. Svechnikova, I. Y. Nikezhina, E. S. Avilov, M. A. Kretova, V. S. Zemskov, *Effect of cadmium, silver and tellurium doping on the properties of single crystals of the layered compounds PbBi$_4$Te$_7$ and PbSb$_2$Te$_4$*, Neorganicheskie Materialy **44**, 438-442 (2008).

[110] K. A. Agaev, S. A. Semiletov, *Electron-diffraction study of the structure of PbBi$_2$Se$_4$*, Kristallografiya **13**, 258-260 (1968).

# Supplementary Information of "Characterization and decomposition of the natural van der Waals heterostructure SnSb$_2$Te$_4$ under compression"

Juan A. Sans,$^{a,*}$ Rosario Vilaplana,$^b$ E. Lora Da Silva,$^a$ Catalin Popescu,$^c$ Vanesa P. Cuenca-Gotor,$^a$ Adrián Andrada-Chacón,$^d$ Javier Sánchez-Benitez,$^d$ Oscar Gomis,$^b$ André L. J. Pereira,$^{a,e}$ Plácida Rodríguez-Hernández,$^f$ Alfonso Muñoz,$^f$ Dominik Daisenberger,$^g$ Braulio García-Domene,$^h$ Alfredo Segura,$^h$ Daniel Errandonea,$^h$ Ravhi S. Kumar,$^i$ Oliver Oeckler,$^j$ Julia Contreras-García,$^k$ and Francisco J. Manjón$^a$

$^a$ Instituto de Diseño para la Fabricación y Producción Automatizada, MALTA-Consolider Team, Universitat Politècnica de València, Valencia, Spain
$^b$ Centro de Tecnologías Físicas, MALTA-Consolider Team, Universitat Politècnica de València, Valencia, Spain
$^c$ ALBA-CELLS, Barcelona, Spain
$^d$ Departamento de Química-Física, MALTA-Consolider Team, Universidad Complutense de Madrid, Madrid, Spain
$^e$ Grupo de Pesquisa de Materiais Fotonicos e Energia Renovavel - MaFER, Universidade Federal da Grande Dourados, Dourados, MS 79825-070, Brazil
$^f$ Departamento de Física, MALTA-Consolider Team, Instituto de Materiales y Nanotecnología, Universidad de La Laguna, Tenerife, Spain
$^g$ Diamond Light Source Ltd, Oxon, England
$^h$ Departamento de Física Aplicada-ICMUV, MALTA-Consolider Team, Universidad de Valencia, Valencia, Spain
$^i$ Department of Physics, University of Illinois at Chicago, Chicago IL 60607-7059, USA
$^j$ Institut für Mineralogie, Kristallographie und Materialwissenschaft, Universität Leipzig, Germany
$^k$ CNRS, UMR 7616, Laboratoire de Chimie Théorique, F-75005, Paris, France

# Structural features of SnSb$_2$Te$_4$

In SnSb$_2$Te$_4$, roughly 50% of Sb cations are mixed with Sn cations in the 3a Wyckoff site and 25% of Sn are mixed with Sb cations in 6c atomic position. This result will not affect to the interlayer character featured by van der Waals interactions between Te sublayers. The similar covalent radii of Sn and Sb (1.39 Å in both) **[1]** and ionic radii in an octahedral distribution, with a value of 83 Å for Sn and 90 Å for Sb **[2]** suggests that the perturbation in the Sb-Te and Sn-Te octahedral units will be mostly influence by the electronic interactions, instead of geometrical effects. On the other hand, the isostructural SnBi$_2$Te$_4$ shows a similar mixed cationic occupancy as its counterpart SnSb$_2$Te$_4$. According to Kuropatawa and Kleinke,**[3]** Sn remains mostly on the 3a atomic position with an occupancy of 74% like Bi that is in 6c atomic position with an occupancy of 68%.

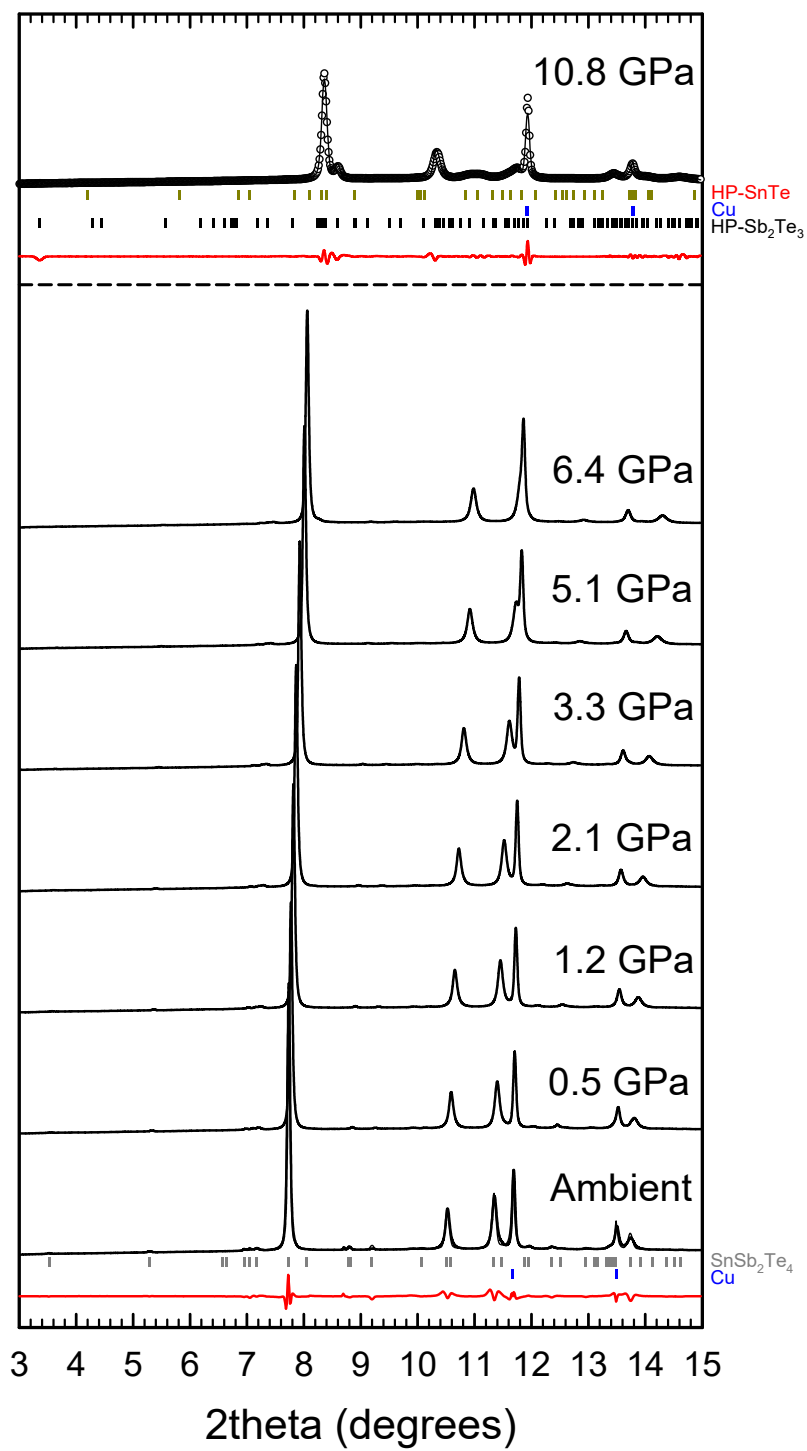

**Figure S1.** HP-ADXRD patterns of SnSb$_2$Te$_4$ at room temperature up to 11 GPa.

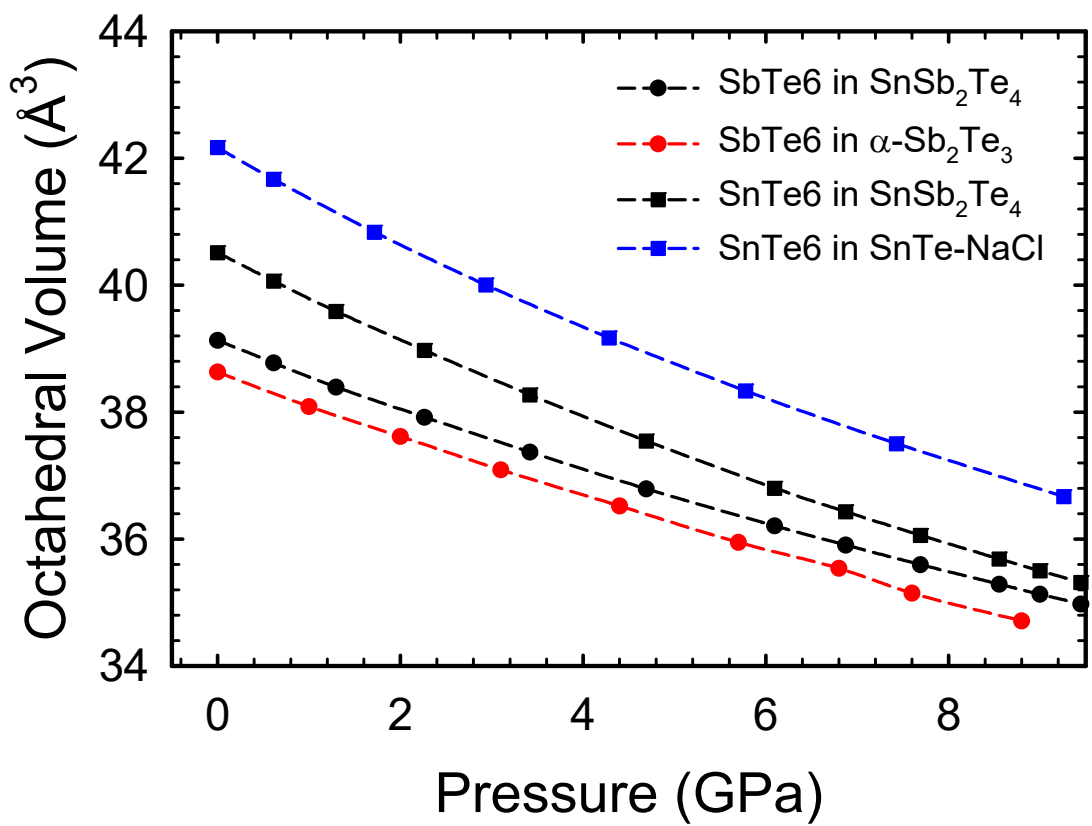

**Figure S2.** Pressure dependence of the theoretical volumes of SbTe$_6$ and SnTe$_6$ octahedra in SnSb$_2$Te$_4$ and in α-Sb$_2$Te$_3$ [Gomis11] and cubic SnTe [Zhou7].

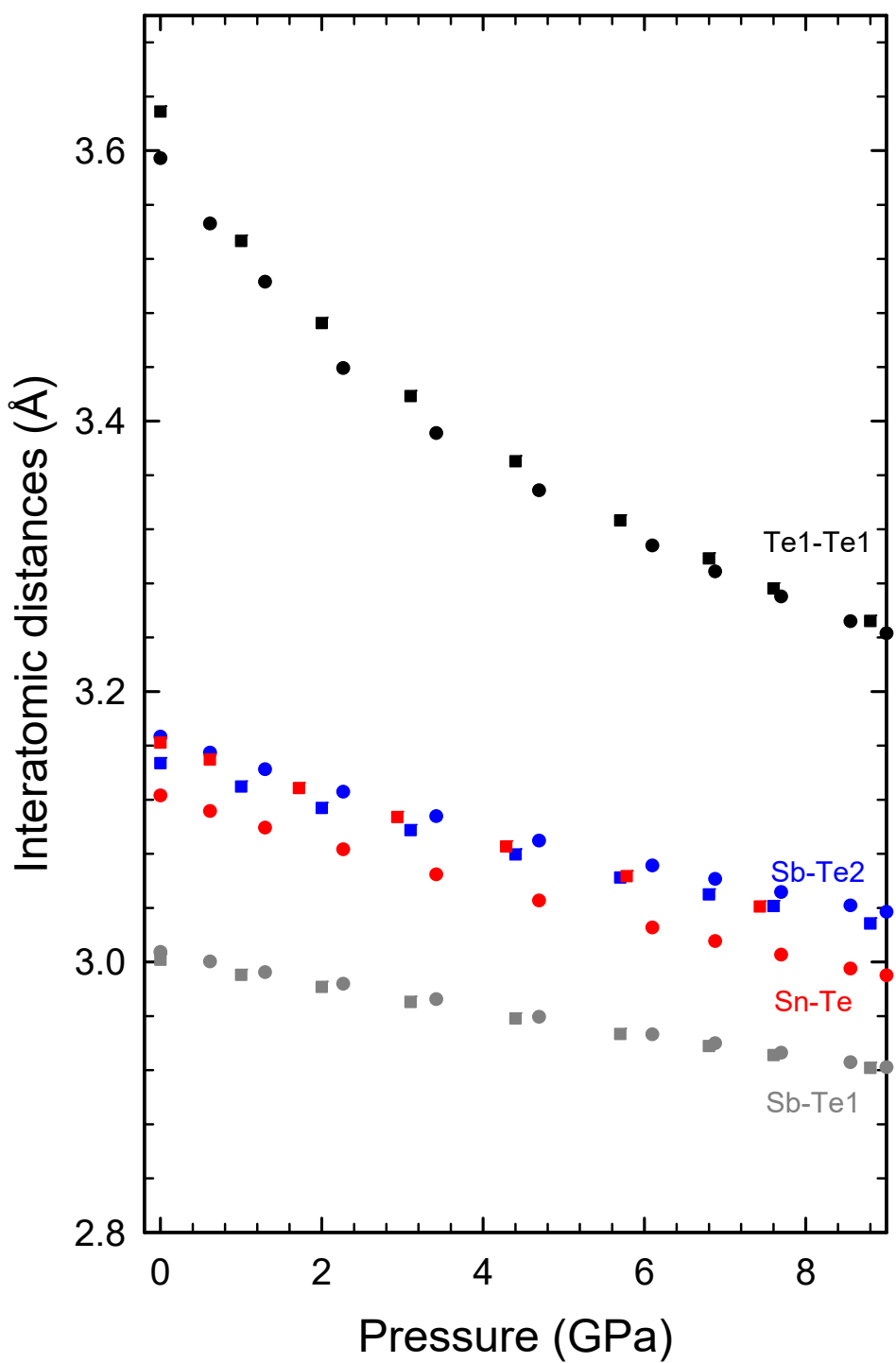

**Figure S3.** Pressure dependence of the theoretical interatomic distances in SnSb$_2$Te$_4$ (circles) and its binary constituents (squares), α-Sb$_2$Te$_3$ (blue, grey and black) and c-SnTe (red).

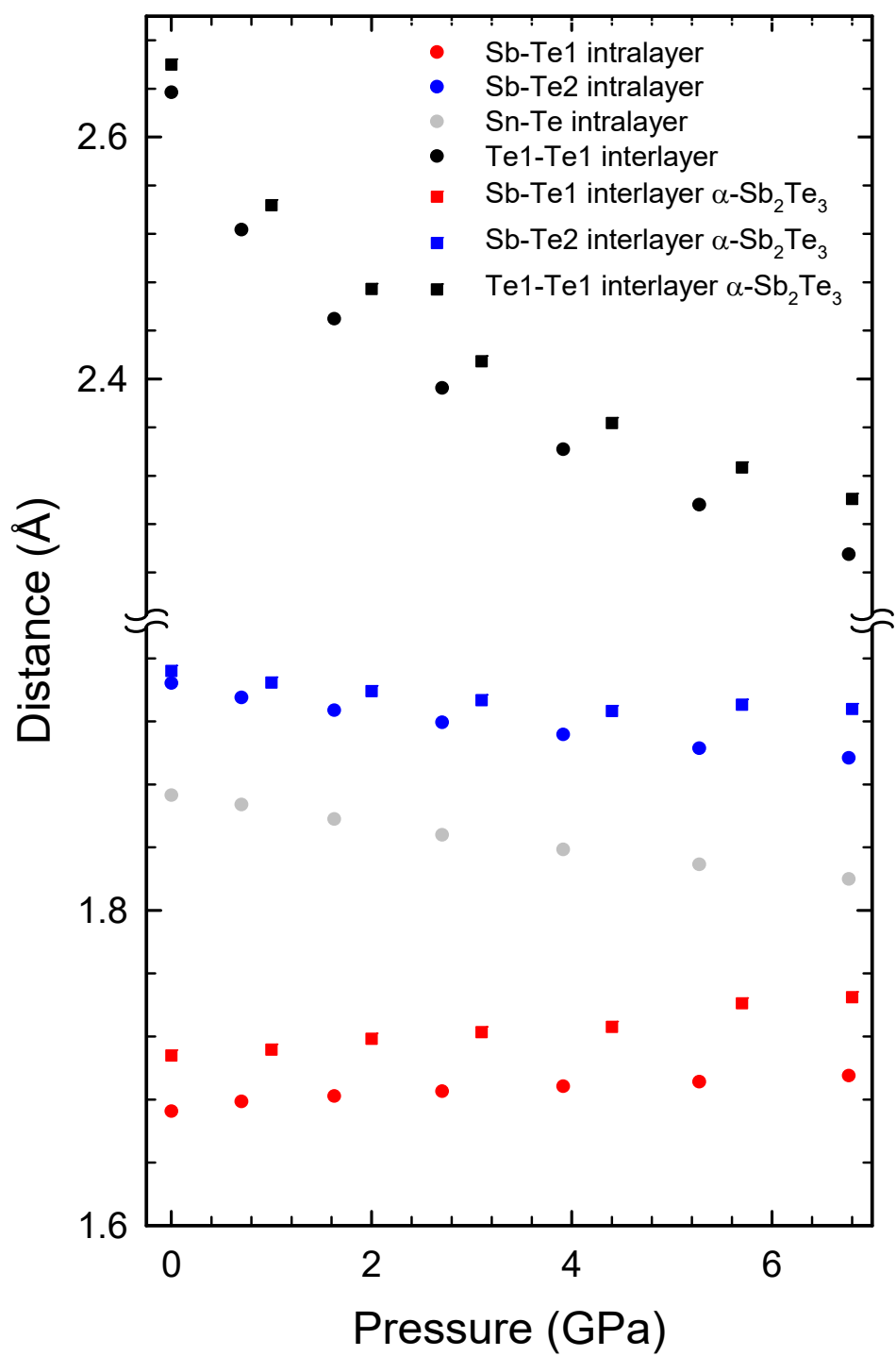

**Figure S4.** Pressure dependence of the theoretical interplanar distances: Te1-Te1 interlayer distance and the different intralayer distances in SnSb$_2$Te$_4$ (circles), and in α-Sb$_2$Te$_3$ and c-SnTe (squares).

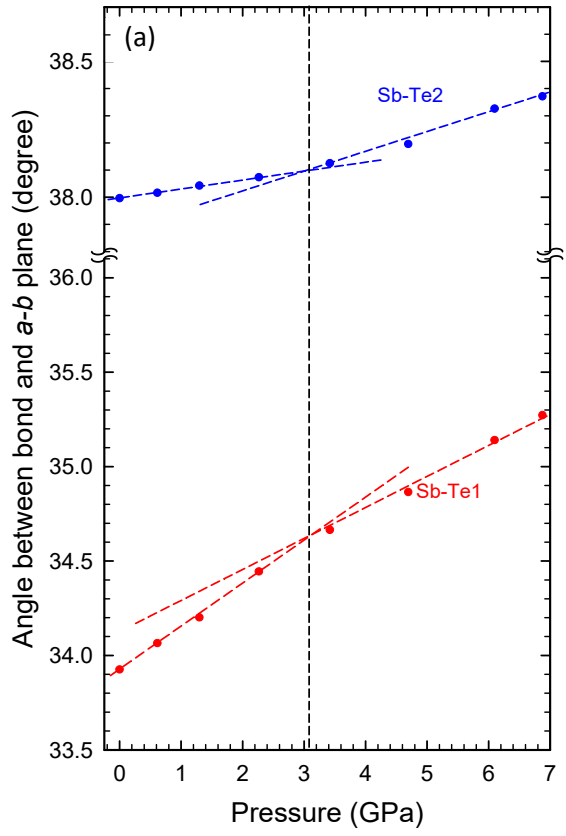 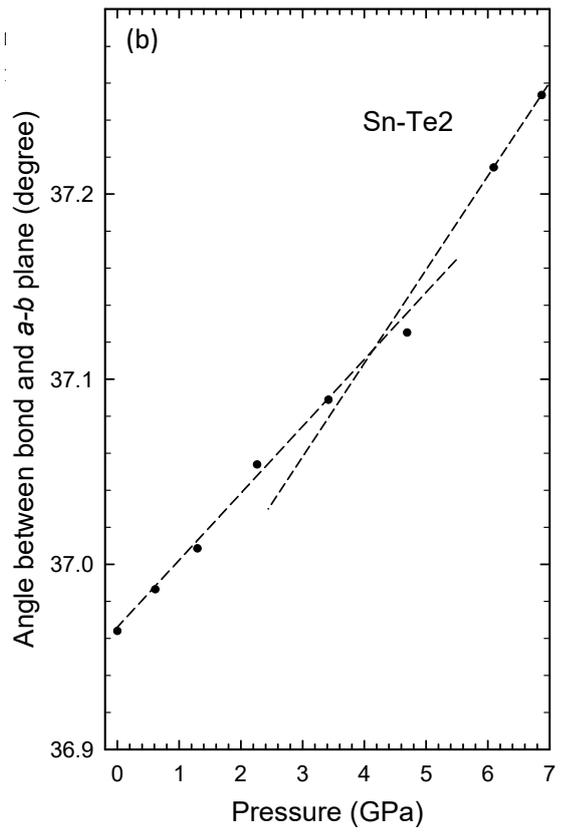

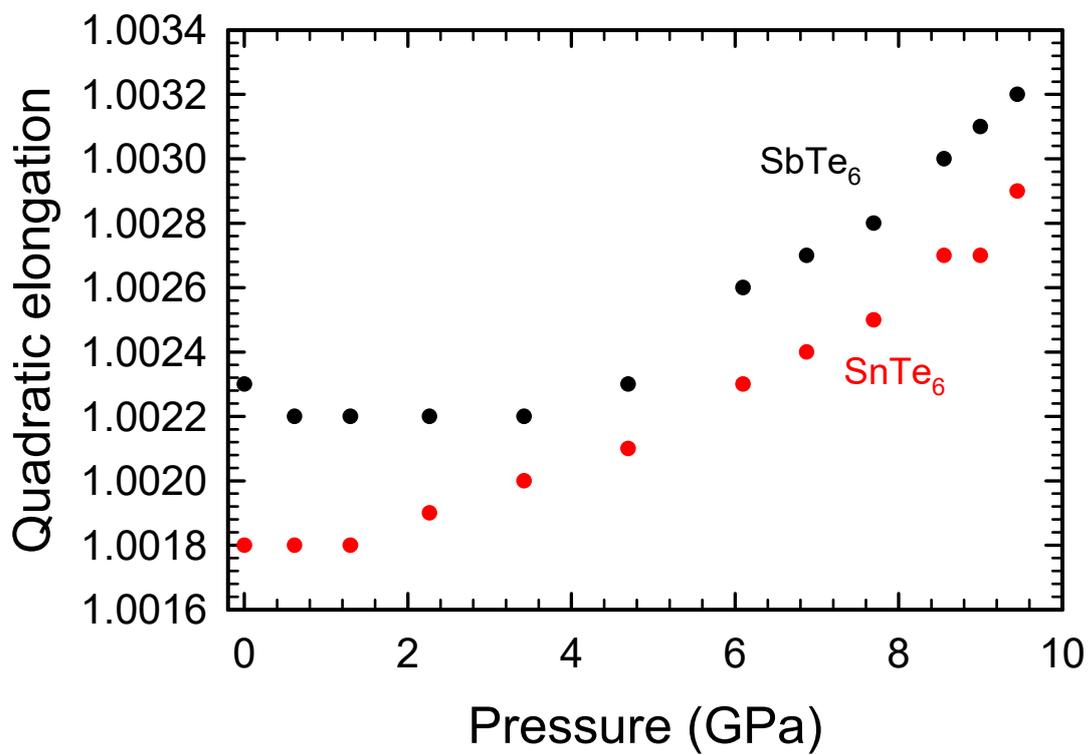

**Figure S6.** Pressure dependence of the theoretical quadratic elongation in the $SbTe_6$ and $SnTe_6$ octahedral units of $SnSb_2Te_4$.

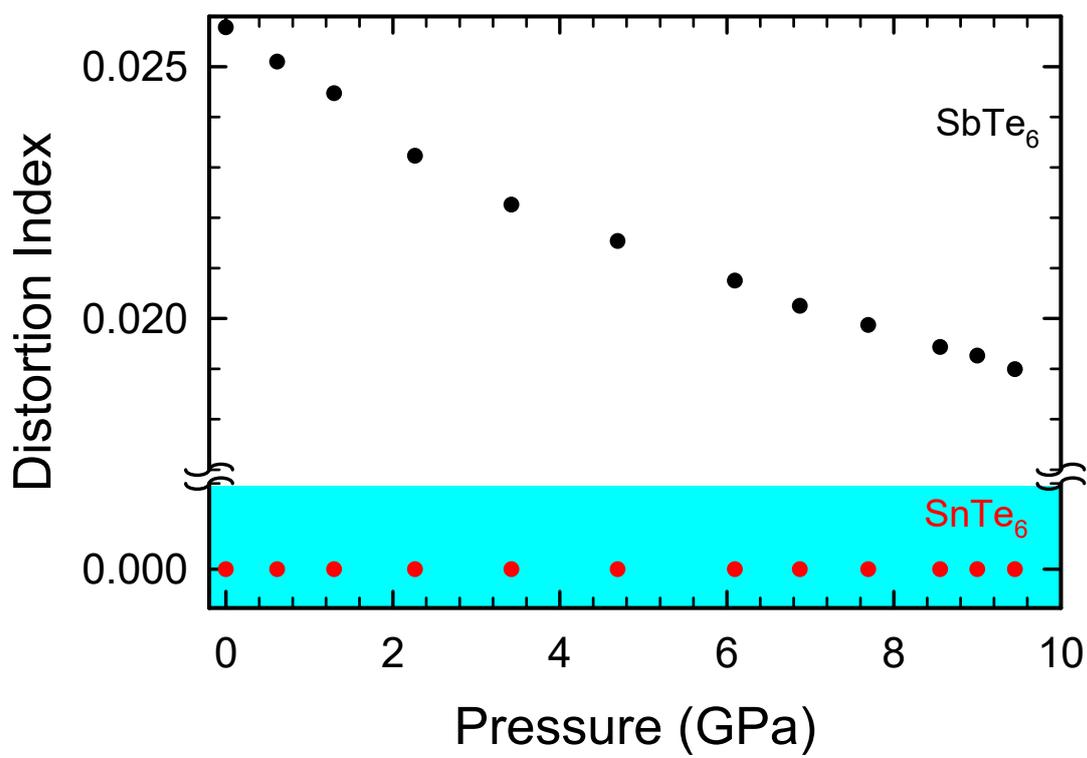

**Figure S7.** Pressure dependence of the theoretical distortion index of the SbTe$_6$ and SnTe$_6$ octahedral units of SnSb$_2$Te$_4$.

# Vibrational modes in SnSb$_2$Te$_4$ at the Γ point

It is well-known that in layered materials, which usually crystallize either in rhombohedral, hexagonal or tetragonal space groups, the lowest-frequency E (doubly degenerated) and A (or B) modes at the Γ point can be classified as interlayer modes (low-frequency phonons mainly characterized by out-of-phase vibrations of atoms corresponding to adjacent layers) or intralayer modes (medium- and high-frequency phonons mainly characterized by out-of-phase vibrations of atoms inside the layers). Interlayer E and A (or B) modes are grouped by pairs and are usually related to shear or transversal vibrations between adjacent layers along the layer plane (*a-b*) and to longitudinal vibrations of one layer against the adjacent ones (along the *c* axis), respectively. These are also known as rigid layer modes and both E and A (or B) interlayer modes arises from transversal acoustic (TA) and longitudinal acoustic (LA) modes, respectively, due to the folding of the Brillouin-zone (BZ) border into the Γ-point due to the decreasing symmetry from cubic to hexagonal or tetragonal. Similarly, E and A (or B) intralayer modes come from transversal optic (TO) and longitudinal optic (LO) modes at Γ and from additional modes due to the folding of the BZ border into the Γ–point.

The number of interlayer and intralayer modes in layered materials depends on the complexity of the unit cell. In the simplest case, there should be two interlayer modes (one of E symmetry and one A or B symmetry) and two intralayer modes, such as what occurs SnS$_2$ **[4]**. In the case of SnSb$_2$Te$_4$, there are two almost pure interlayer modes (E$_g^1$ and A$_{1g}^1$), which have the lowest frequencies and are Raman-active and correspond to out-of-phase movements of the neighbor layers both along the *a-b* plane (E$_g^1$ mode) and along *c* axis (A$_{1g}^1$ mode). Similar to other Raman-active modes, these modes are characterized by the immobility of the central Sn atom located in a highly symmetric Wyckoff site, the in-phase movements of all atoms of each sublayer above and below the central Sn plane and the out-of-phase movement of the atoms in the two sublayers (see **Fig. S8**). Furthermore, it can be observed that both the frequency and pressure coefficients of the interlayer A mode is larger than that of the interlayer E mode as what typically occurs in van der Waals-type layered compounds (see **Tables S1 and S2 and Figs. 6 and S19**).

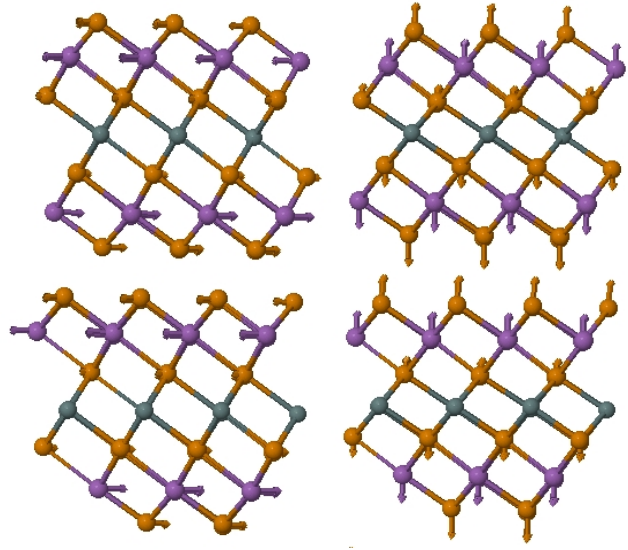

**Figure S8.** Atomic movements (see arrows) of low-frequency interlayer Raman-active modes $E_g^1$ (left) and $A_{1g}^1$ (right) located near 38 and 55 cm$^{-1}$ at room pressure, respectively. Sn, Sb and Te atoms are depicted in green, purple and orange colors, respectively.

The next couple of E and A modes, discussed in order of increasing frequency, is formed by the low-frequency $E_u^1$ mode and the $A_{2u}^1$ mode (see **Fig. S9**). These two intralayer modes are characterized by an out-of-phase vibration of the central Sn atoms and the external Sb atoms. The $E_u^1$ mode is characterized by a vibration of the central SnTe$_6$ unit against the external SbTe$_3$ units in the a-b plane. The $A_u^1$ mode is the complementary mode to the $E_u^1$ mode and it is characterized by an out-of-phase vibration of the central SnTe$_6$ unit against the external SbTe$_3$ units along the *c*-axis.

We may observe that while all $A_{2u}$ and $E_u$ modes show an in-phase vibration of the neighbor Te atoms on adjacent layers, all $A_{1g}$ and $E_g$ modes evidence out-of-phase vibrations of neighbor Te atoms on adjacent layers similar to those modes of the pure interlayer modes. Note, however, that the intralayer modes are clearly dominated by the strong vibration amplitudes of intralayer structures. Similarly, it can be observed that the vibration of the central Sn atom is observed in all ungerade (IR-active) modes, whereas the Sn atom is mainly static in characterized by the gerade (Raman-active) modes.

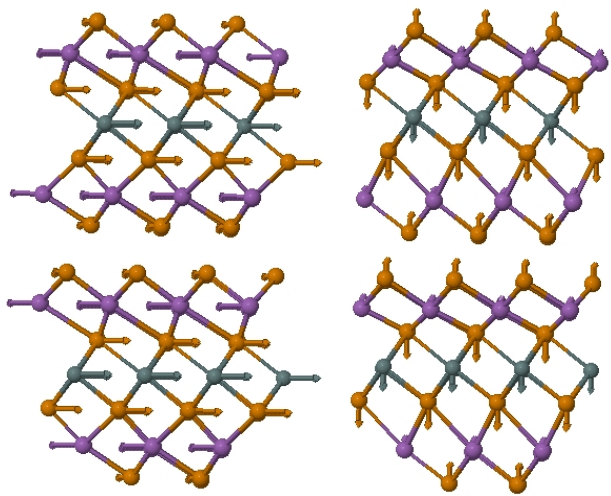

**Figure S9.** Atomic movements of IR-active modes $E_u^1$ (left) and $A_{2u}^1$ (right) located near 62 and 81 cm$^{-1}$ at room pressure, respectively. Sn, Sb and Te atoms are depicted in green, purple and orange colors, respectively.

The following couple of E and A modes, in order of increasing frequency, is formed by the low-frequency $E_u^2$ mode and the middle-frequency $A_{2u}^2$ mode (see **Fig. S10**). These two intralayer modes are characterized by an in-phase vibration of the central Sn atoms and the external Sb atoms against the Te atoms. The $E_u^2$ mode is characterized by the vibration of the network of Sn and Sb atoms against the network of Te atoms along the a-b plane; i.e., it is the main asymmetric bending mode of the Sn-Te bond in the central SnTe$_6$ unit. On the other hand, the $A_{2u}^2$ mode is the complementary mode to the $E_u^2$ mode and it is characterized by an in-phase vibration of the central Sn atom and the Sb atoms against the network of Te atoms along the *c*-axis; i.e., it is the main asymmetric stretching mode of the Sn-Te bond in the central SnTe$_6$ unit. Therefore, these two modes are characteristic of the SnTe$_6$ octahedron and do not occur in Sb$_2$Te$_3$ as we will comment later.

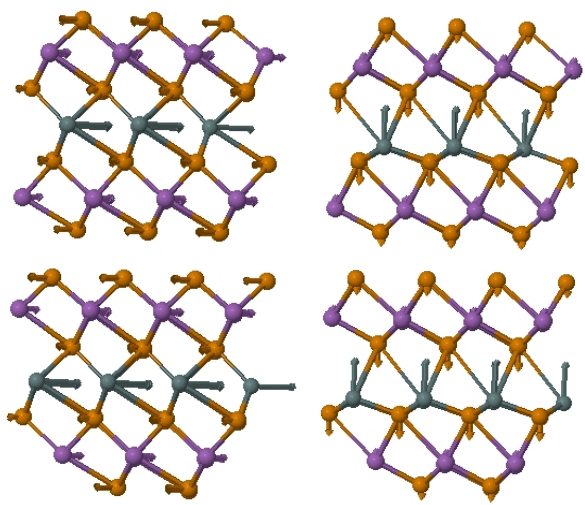

**Figure S10.** Atomic movements of IR-active mode $E_u^2$ (left) and $A_{2u}^2$ (right) located near 65 and 104 cm$^{-1}$ at room pressure, respectively. Sn, Sb and Te atoms are depicted in green, purple and orange colors, respectively.

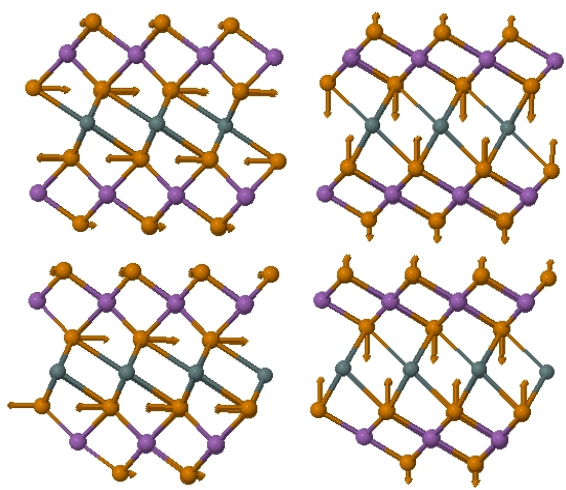

**Figure S11.** Atomic movements of middle-frequency Raman-active modes $E_g^2$ (left) and $A_{1g}^2$ (right) located near 100 and 115 cm$^{-1}$ at room pressure, respectively. Sn, Sb and Te atoms are depicted in green, purple and orange colors, respectively.

The next two modes, discussed in order of increasing frequency, is formed by the medium-frequency $E_g^2$ and $A_{1g}^2$ modes (see **Fig. S11**). The $E_g^2$ mode is characterized by the strong out-of-phase vibration of the Te atoms close to the central Sn atoms along the a-b plane; i.e., it is the main symmetric bending mode of the Sn-Te of the central SnTe$_6$ unit. The $A_{1g}^2$ mode is the complementary mode to the $E_g^2$ and it is characterized by the strong vibration of Te atoms against Sn and Sb atoms alternately along the c axis; i.e., it is the main symmetric stretching mode of the Sn-Te bond in the central SnTe$_6$ unit. Again, these two modes are characteristic of the SnTe$_6$ octahedron and do not occur in Sb$_2$Te$_3$ as we will comment further on.

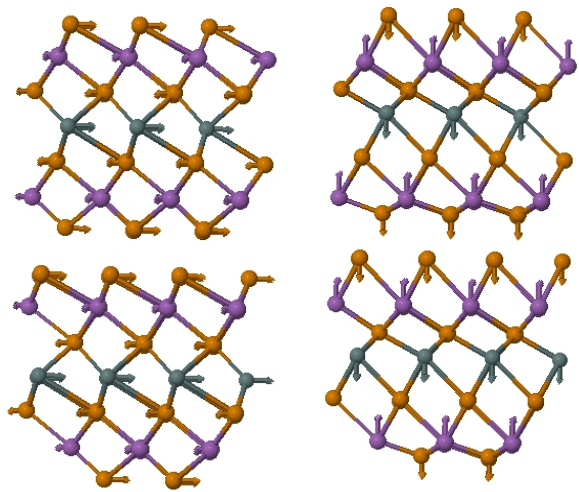

**Figure S12.** Atomic movements of IR-active modes $E_u^3$ (left) and $A_u^3$ (right) located near 112 and 157 cm$^{-1}$ at room pressure, respectively. Sn, Sb and Te atoms are depicted in green, purple and orange colors, respectively.

The next pair of frequencies are formed by the medium-frequency $E_u^3$ mode and the high-frequency $A_u^3$ mode (see **Fig. S12**). These two intralayer modes are characterized by an out-of-phase vibration of the central Sn atoms and the external Sb atoms as in the $E_u^1$ and $A_u^1$ modes. Regarding the $E_u^1$ mode, the central Te atoms show an in-phase vibration with the central Sn atom leading to a vibration of the central SnTe$_6$ unit against the external SbTe$_3$ units in the a-b plane; however, for the $E_u^3$ mode the central Sn atom shows an out-of-phase vibration with respect to the adjacent Te atoms as what occurs for a Sn-Te bending mode of the SnTe$_6$ unit. Additionally, since the external Te atoms vibrate out-of-phase with respect to the Sb atoms, such a vibrational mode results also in a symmetric Sb-Te bending mode of the SbTe$_6$ unit. Similarly, the $A_u^1$ mode is an out-of-phase vibration of the central SnTe$_6$ unit against the external SbTe$_3$ units along the c axis; however, the $A_u^3$ mode corresponds solely to the central Sn atom that vibrates against the external Sb atoms in an asymmetric way (central Te atoms are static), thus leading to a coupled asymmetric Sn-Te and Sb-Te stretching mode of both SnTe$_6$ and SbTe$_6$ units.

Finally, the last two intralayer modes, referenced in order of increasing frequency is formed by the medium-frequency $E_g^3$ mode and the high-frequency $A_g^3$ mode (see **Fig. S13**). Both the $E_g^3$ and $A_g^3$ modes are characterized by the small vibration of the central $SnTe_6$ unit, such as what occurs for the $E_g^1$ and $A_g^1$ modes; however, for the $E_g^1$ and $A_g^1$ modes, the Sb atoms vibrate in phase with adjacent Te atoms, whereas in the $E_g^3$ and $A_g^3$ modes, external Te and Sb atoms move out-of-phase. Additionally, for both four modes there is an out-of-phase movement of all atoms in the two sublayers. In this way, atomic movements of Te and Sb atoms along the a-b plane evidence the $E_g^3$ mode as being the asymmetric bending mode of Sb-Te in the $SbTe_6$ units. Alternatively, the complementary $A_g^3$ mode shows Te and Sb atoms moving out-of-phase along the c axis, therefore this mode can be viewed as the asymmetric stretching mode of Sb-Te of the $SbTe_6$ units.

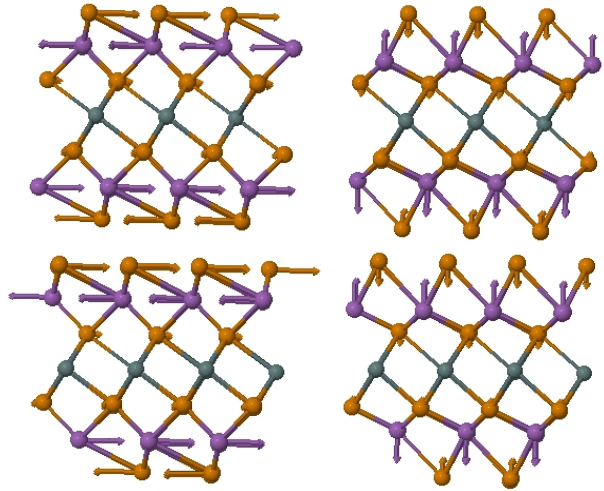

**Figure S13.** Atomic movements of Raman-active modes $E_g^3$ and $A_g^3$ located near 117 and 167 cm$^{-1}$ at room pressure, respectively. Sn, Sb and Te atoms are depicted in green, purple and orange colors, respectively.

A first comparison can be stablished between the vibrational modes at Γ in $SnSb_2Te_4$ and the parent compound $Sb_2Te_3$. In $Sb_2Te_3$ there are four Raman-active modes ($E_g^1$, $A_{1g}^1$, $E_g^2$ and $A_{1g}^2$) and four IR-active modes ($E_u^1$, $E_u^2$, $A_{2u}^1$ and $A_{2u}^2$) and referenced in order of increasing frequency **[5,6]**. As regards to the Raman-active modes, the interlayer modes of $Sb_2Te_3$ are the $E_g^1$ and $A_{1g}^1$ modes (**Fig. S14**) and these are similar to the $E_g^1$ and $A_g^1$ modes of $SnSb_2Te_4$ (**Fig. S8**). For all these modes the central part of the layer remains almost static and the external adjacent Sb and Te atoms in a sublayer vibrate in-phase and also out-of-phase with respect to the atoms of the other adjacent sublayer. Similarly, the intralayer $E_g^2$ and $A_{1g}^2$ modes of $Sb_2Te_3$ (**Fig. S15**) are similar to the intralayer $E_g^3$ and $A_g^3$ modes of SnSb$_2$Te$_4$ (**Fig. S13**) since for all these modes the central part of the layer remains almost static and the external adjacent Sb and Te atoms in a sublayer vibrate, among them and with respect to atoms of the other sublayer, out-of-phase. The similarity of $E_g^2$ and $A_{1g}^2$ modes in Sb$_2$Te$_3$ and $E_g^3$ and $A_g^3$ modes in SnSb$_2$Te$_4$ is so remarkable that these modes possess practically the same theoretically predicted frequency values (see **Table S1**). With respect to the Raman-active $E_g^2$ and $A_g^2$ modes of SnSb$_2$Te$_4$ (**Fig. S11**), these have no analog on Sb$_2$Te$_3$ since both modes involve out-of-phase vibrations of the internal Te atoms, which cannot occur in Sb$_2$Te$_3$ with only one internal Te atom.

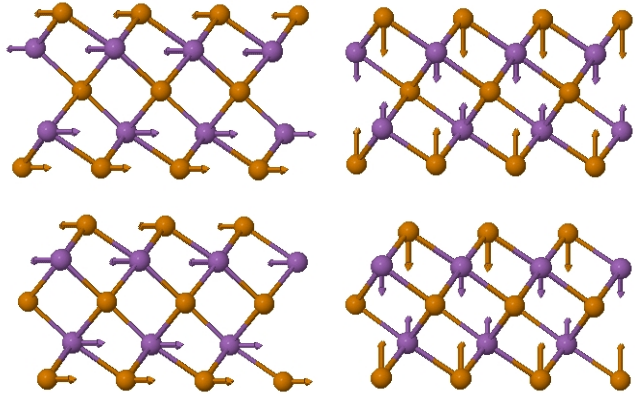

**Figure S14.** Atomic movements (see arrows) of low-frequency interlayer Raman-active modes $E_g^1$ and $A_{1g}^1$ in Sb$_2$Te$_3$ located near 50.4 and 68.9 cm$^{-1}$ at room pressure, respectively. Sb and Te atoms are depicted in purple and orange colors, respectively.

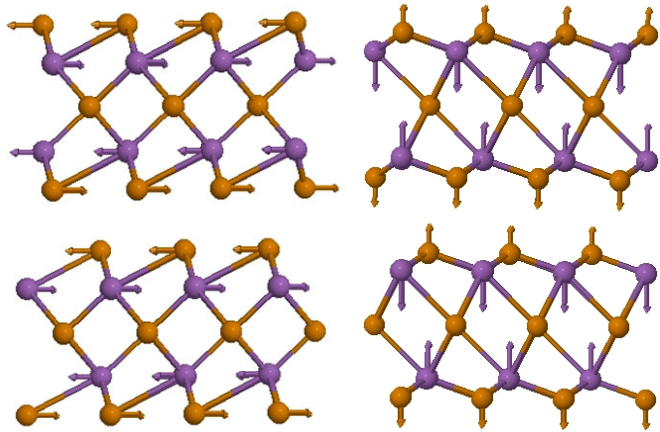

**Figure S15.** Atomic movements (see arrows) of high-frequency intralayer Raman-active modes $E_g^2$ and $A_{1g}^2$ in Sb$_2$Te$_3$ located near 116.6 and 167.6 cm$^{-1}$ at room pressure, respectively. Sb and Te atoms are depicted in purple and orange colors, respectively.

With respect to the IR-active modes, the $E_u^1$ ($E_u(2)$ in **[5]**) and $A_{2u}^2$ ($A_{2u}(3)$ in **[5]**) modes of Sb$_2$Te$_3$ are similar to the $E_u^2$ and $A_u^2$ modes of SnSb$_2$Te$_4$. For both modes of

Sb$_2$Te$_3$, Sb atoms vibrate in-phase in the two sublayers and vibrate out-of-phase with respect to all Te atoms as what occurs for the E$_u^2$ and A$_u^2$ modes of SnSb$_2$Te$_4$. Note that the movement of the central Te atoms of the A$_{2u}^2$ mode is very low (not shown in **Fig. S16**) but in phase with the other Te atoms, similar to the A$_u^2$ mode of SnSb$_2$Te$_4$. On the other hand, the E$_u^2$ (E$_u$(3) in **[5]**) and A$_{2u}^1$ (A$_{2u}$(2) in **[5]**) modes of Sb$_2$Te$_3$ are similar to the E$_u^3$ and A$_u^3$ modes of SnSb$_2$Te$_4$. For both E$_u^2$ and A$_{2u}^1$ modes of Sb$_2$Te$_3$ central Te atoms vibrate out-of-phase with respect to the external Te atoms as to what is observed for the E$_u^3$ and A$_u^3$ modes of SnSb$_2$Te$_4$.

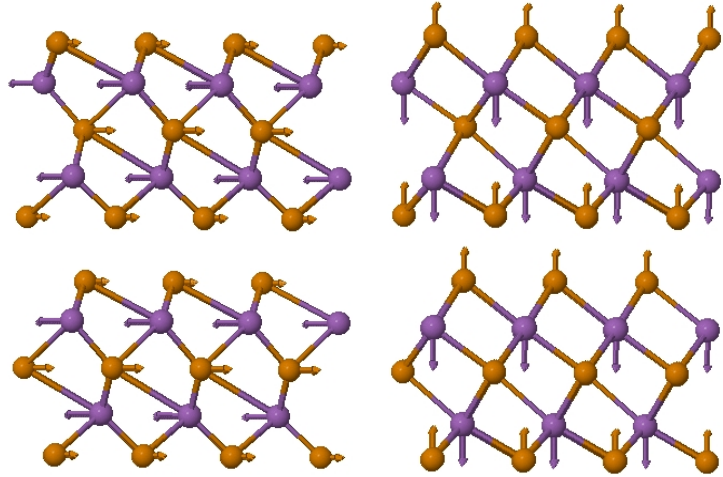

**Figure S16.** Atomic movements (see arrows) of intralayer IR-active modes E$_u^1$ and A$_{2u}^2$ in Sb$_2$Te$_3$ located near 78.0 and 138.7 cm$^{-1}$ at room pressure, respectively. Sb and Te atoms are depicted in purple and orange colors, respectively.

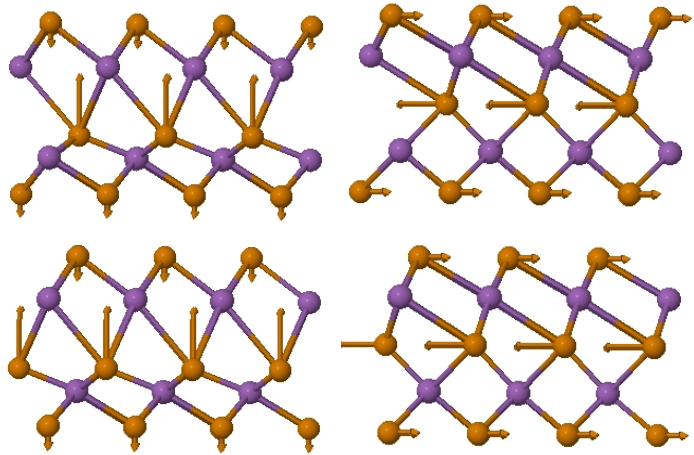

**Figure S17.** Atomic movements (see arrows) of intralayer IR-active modes E$_u^2$ and A$_{2u}^1$ in Sb$_2$Te$_3$ located near 100.4 and 109.9 cm$^{-1}$ at room pressure, respectively. Sb and Te atoms are depicted in purple and orange colors, respectively.

Finally, it is noteworthy of mentioning that the IR-active E$_u^1$ and A$_u^1$ modes of SnSb$_2$Te$_4$ have no resemblance with IR-active modes in Sb$_2$Te$_3$. Note that these two modes refer to the Sb atoms vibrating in-phase with their adjacent external Te atoms, a feature that does not occur in any of the IR-active modes of $Sb_2Te_3$. Finally, it must be stressed that for all IR-active modes of $Sb_2Te_3$, Sb atoms of the two sublayers vibrate in-phase, while for all Raman-active modes of $Sb_2Te_3$ vibrate out-of-phase. The same behavior is observed in $SnSb_2Te_4$. This is the main characteristic to discern between Raman-active and IR-active modes of both compounds.

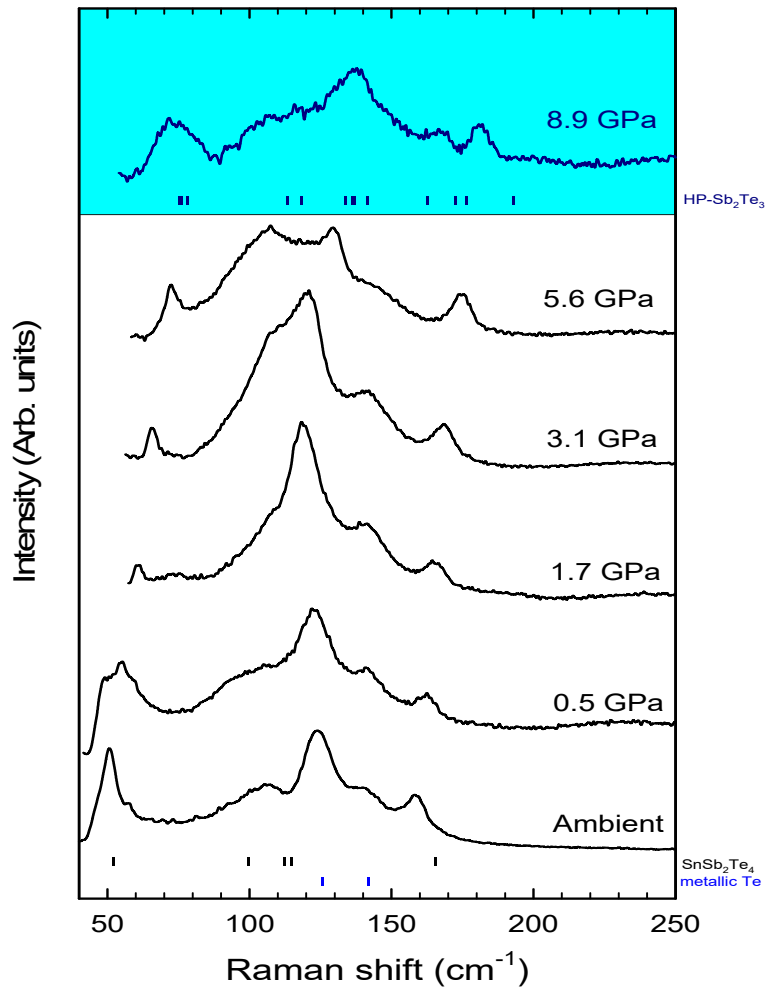

**Figure S18.** Raman scattering spectra of rhombohedral $SnSb_2Te_4$ at different pressures up to 8.9 GPa. Black (blue) vertical ticks correspond to theoretically predicted Raman-active mode frequency of $SnSb_2Te_4$ (Metallic Te).

Referring to the pressure coefficients of the different Raman-active and IR-active modes, it can be observed that generally the A modes have larger pressure coefficients than their associated E modes, as it is expected in non-polar layered compounds with van der Waals forces between their layers (see **Tables S1 and S2**). This has been already commented for interlayer Raman-active $E_g^1$ and $A_{2g}^1$ modes and it applies to both SnSb$_2$Te$_4$ and Sb$_2$Te$_3$. Usually, the small pressure coefficient of the low-frequency E mode in layered materials is ascribed to the weak bending force constant due to weak van der Waals forces between the neighboring layers. On the other hand, the large pressure coefficient of the low-frequency A mode is due to the extraordinary increase of the stretching force constant between neighboring layers due to the strong decrease of the interlayer distance [4,7]. This behavior is also found for the low-frequency interlayer modes in layered Sb$_2$Te$_3$ and SnSb$_2$Te$_4$, and it is also valid for the other pairs of intralayer E and A modes, previously commented. This can be understood if intralayer E modes are mainly associated to bending Sb-Te (Sb-Te and Sn-Te) modes in Sb$_2$Te$_3$ (SnSb$_2$Te$_4$), while intralayer A modes are mainly associated to stretching Sb-Te (Sb-Te and Sn-Te) modes in Sb$_2$Te$_3$ (SnSb$_2$Te$_4$). This reasoning allows also to explain the reason for which the A modes always possess larger frequencies than their associated E modes.

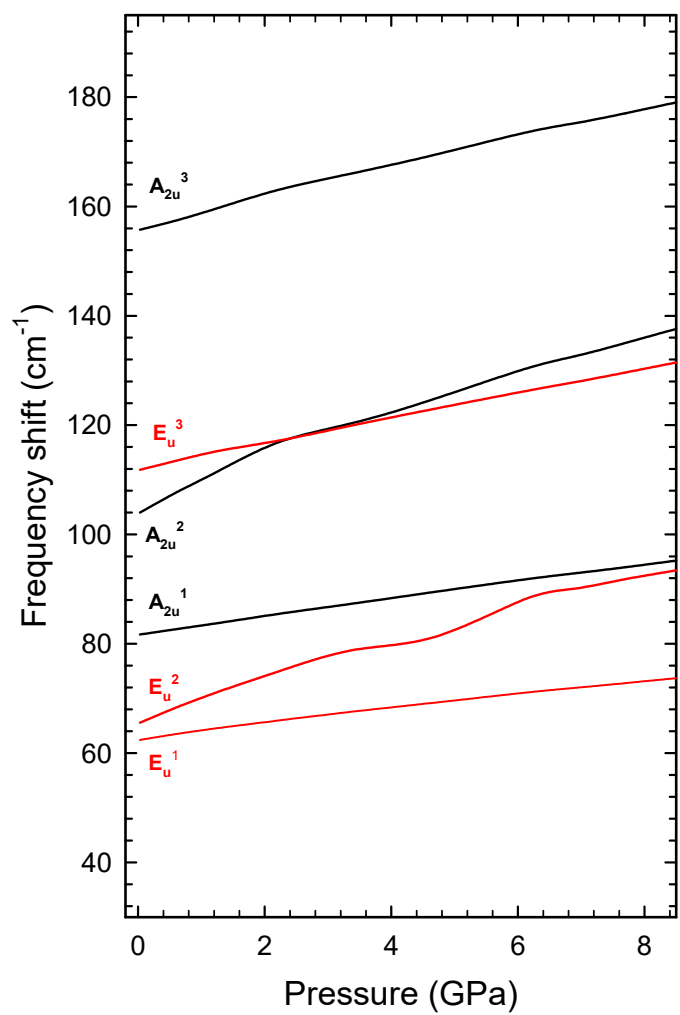

**Figure S19.** Pressure dependence of the theoretical IR-active modes of SnSb$_2$Te$_4$. A modes and doubly-degenerate E modes are depicted in black and red, respectively.

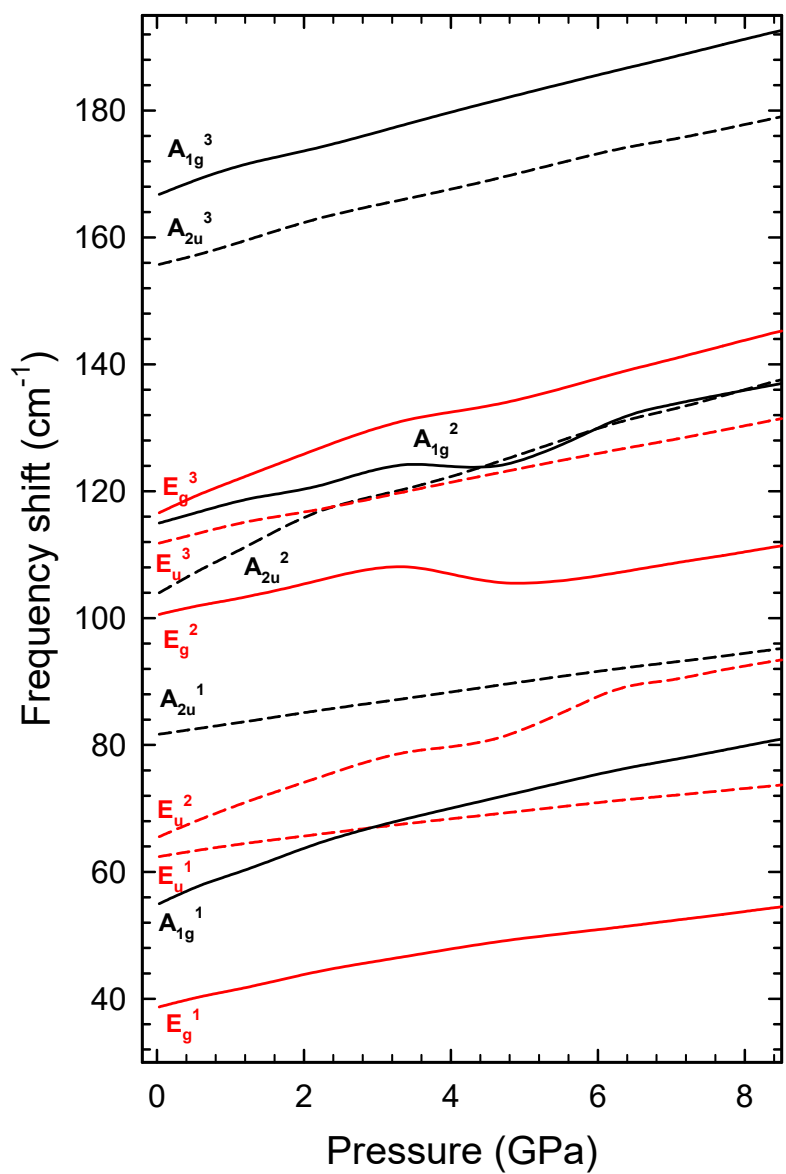

**Figure S20.** Pressure dependence of the theoretical (solid lines) Raman-active and (dashed lines) infrared-active mode frequencies of SnSb$_2$Te$_4$.

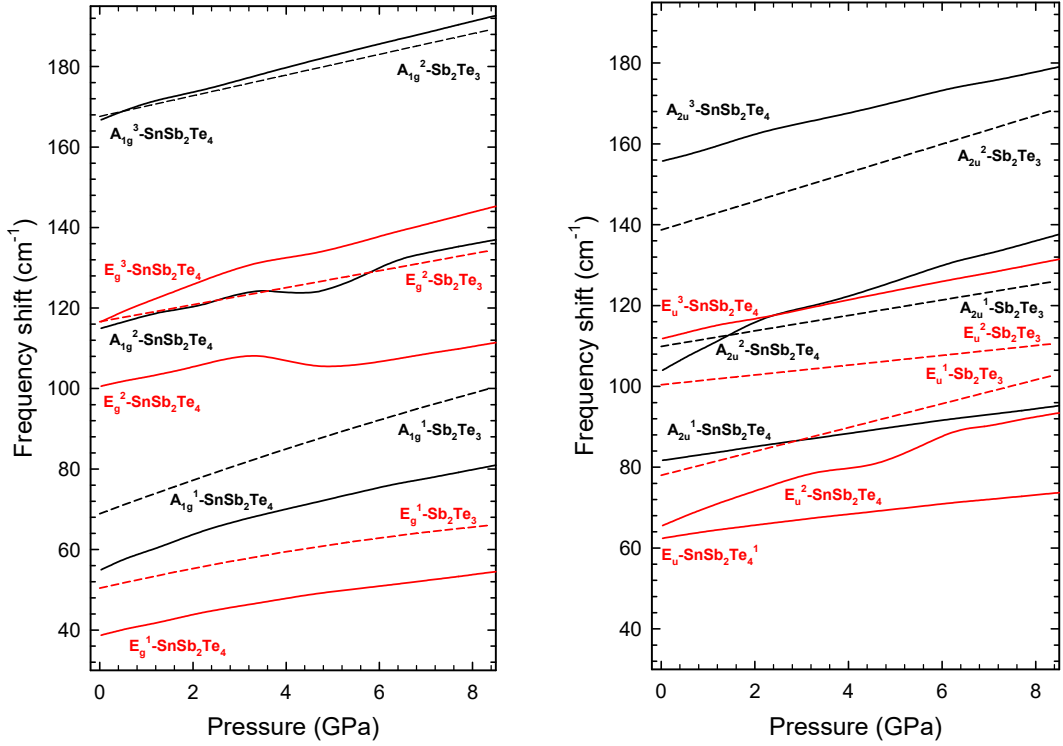

**Figure S21.** Pressure dependence of the theoretical (left) Raman-active and (right) infrared-active mode frequencies of SnSb$_2$Te$_4$ and Sb$_2$Te$_3$.

The most notable deviation of the rule mentioned above of the Raman-active modes, is that of the $E_g^3$ mode in SnSb$_2$Te$_4$. The theoretical pressure coefficient of this mode is larger than its associated $A_{2g}^3$ mode. This feature contrasts with Sb$_2$Te$_3$ where the equivalent modes $E_g^2$ and $A_{2g}^2$ show a normal behavior. Additionally, it must be noted that the pressure coefficient at zero pressure obtained for the $A_{2g}^2$ mode in SnSb$_2$Te$_4$ is quite high because the fit has been performed with high-pressure data due to the lack of values near room pressure.

The larger pressure coefficient of the A modes when compared to their corresponding E modes also applies for IR-active modes. Note that in Sb$_2$Te$_3$ the pressure coefficient of $E_u^1$ mode is smaller than its associated $A_{2u}^2$ mode and that of the $E_u^2$ mode is smaller than its associated $A_{2u}^1$ mode. This reasoning also applies to their similar IR-active modes in SnSb$_2$Te$_4$; i.e., the $E_u^2$ and $A_{2u}^2$ and the $E_u^3$ and $A_{2u}^3$ modes, respectively. Moreover, the same rule applies to $E_u^1$ and $A_{2u}^1$ modes in SnSb$_2$Te$_4$ that have no correspondence in Sb$_2$Te$_3$. This similarity between Raman and IR modes both in Sb$_2$Te$_3$ and SnSb$_2$Te$_4$ remarks the strangely large pressure coefficient of the theoretical $E_g^3$ mode

(which is almost double from its expected value) in SnSb$_2$Te$_4$, since the Raman-active A$_{2g}^3$ mode has a similar value of the pressure coefficient than its IR-active counterpart (the A$_{2u}^3$ mode).

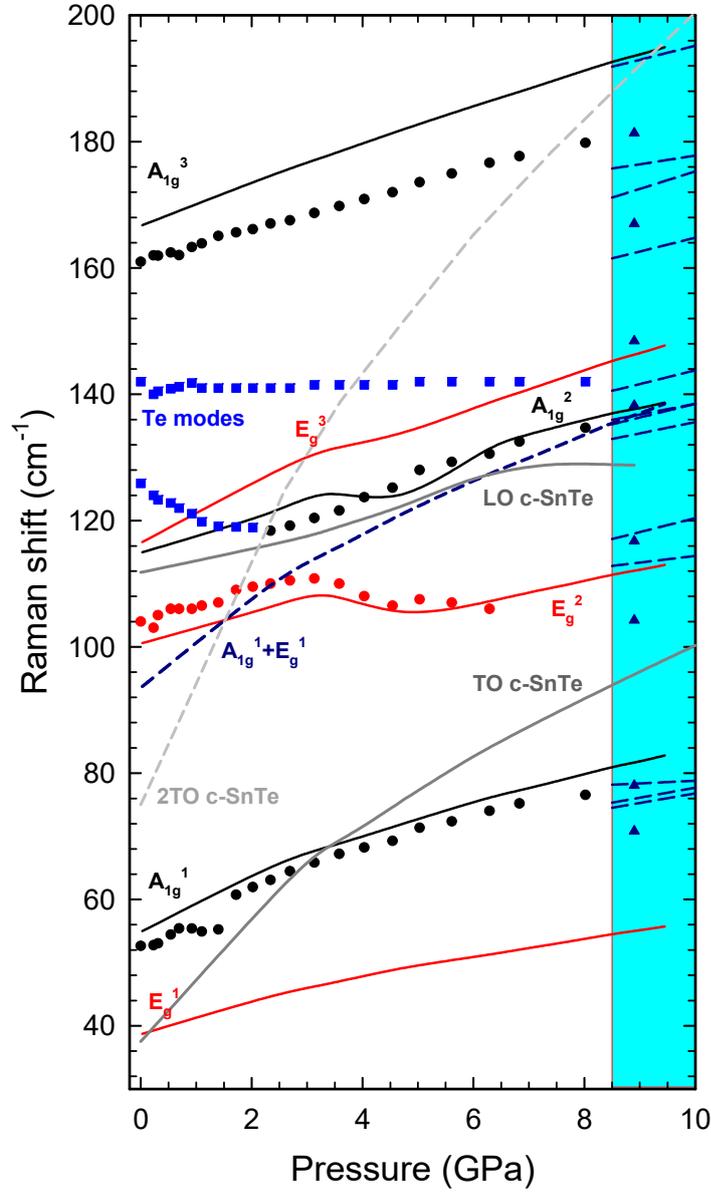

**Figure S22.** Pressure dependence of the experimental (symbols) and theoretical (lines) Raman-active mode frequencies in SnSb$_2$Te$_4$ together with the representation of theoretical LO and TO IR-active modes of c-SnTe. Dashed lines represent the pressure dependence of the A$_{1g}^1$ + E$_g^1$ combination at Γ in SnSb$_2$Te$_4$ and the 2TO mode at Γ in c-SnTe.

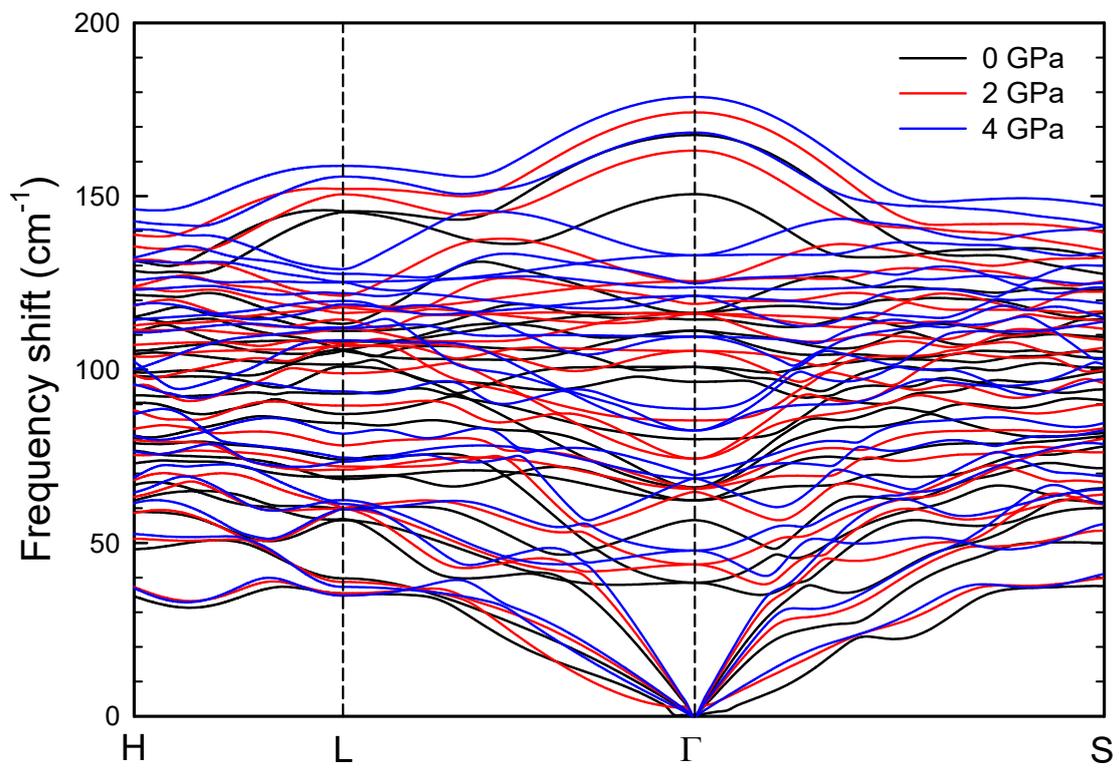

**Figure S23.** Phonon dispersion curves of SnSb$_2$Te$_4$ at 0, 2 and 4 GPa.

# Evolution of the electronic topology under pressure

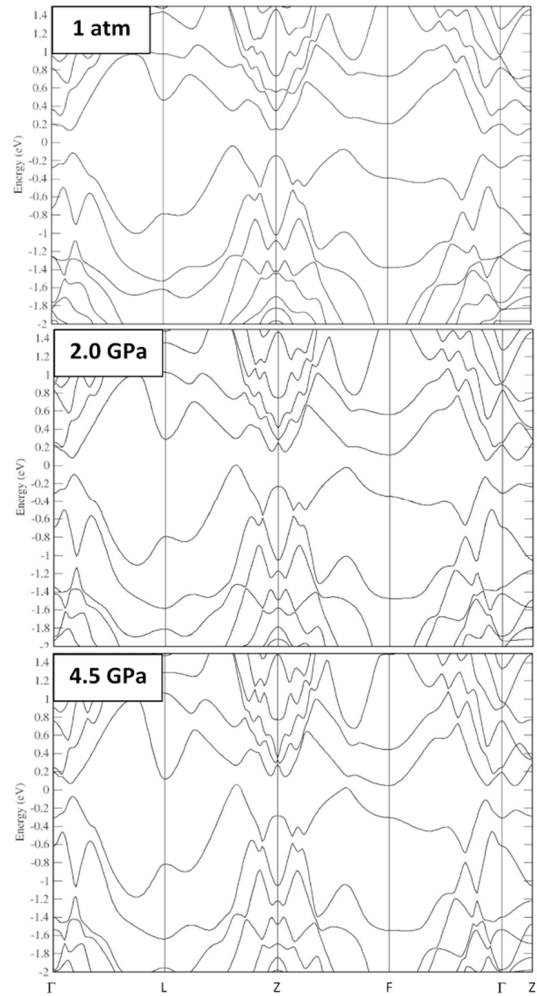

**Figure S24.** *Ab-initio* calculated band electronic structure of SnSb$_2$Te$_4$ theoretically predicted at 1 atm (top), 2 GPa (middle), and 4.5 GPa (bottom).

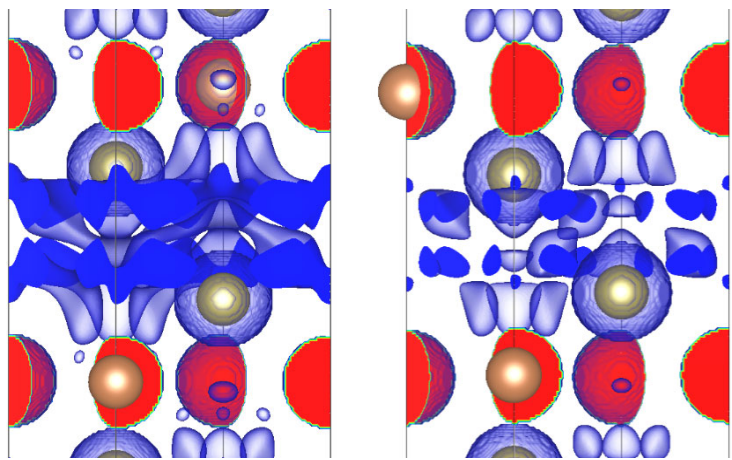

**Figure S25.** Reduced density gradient map of rhombohedral SnSb$_2$Te$_4$ around the interlayer space at 1 atm (left) and 2.5 GPa (right).

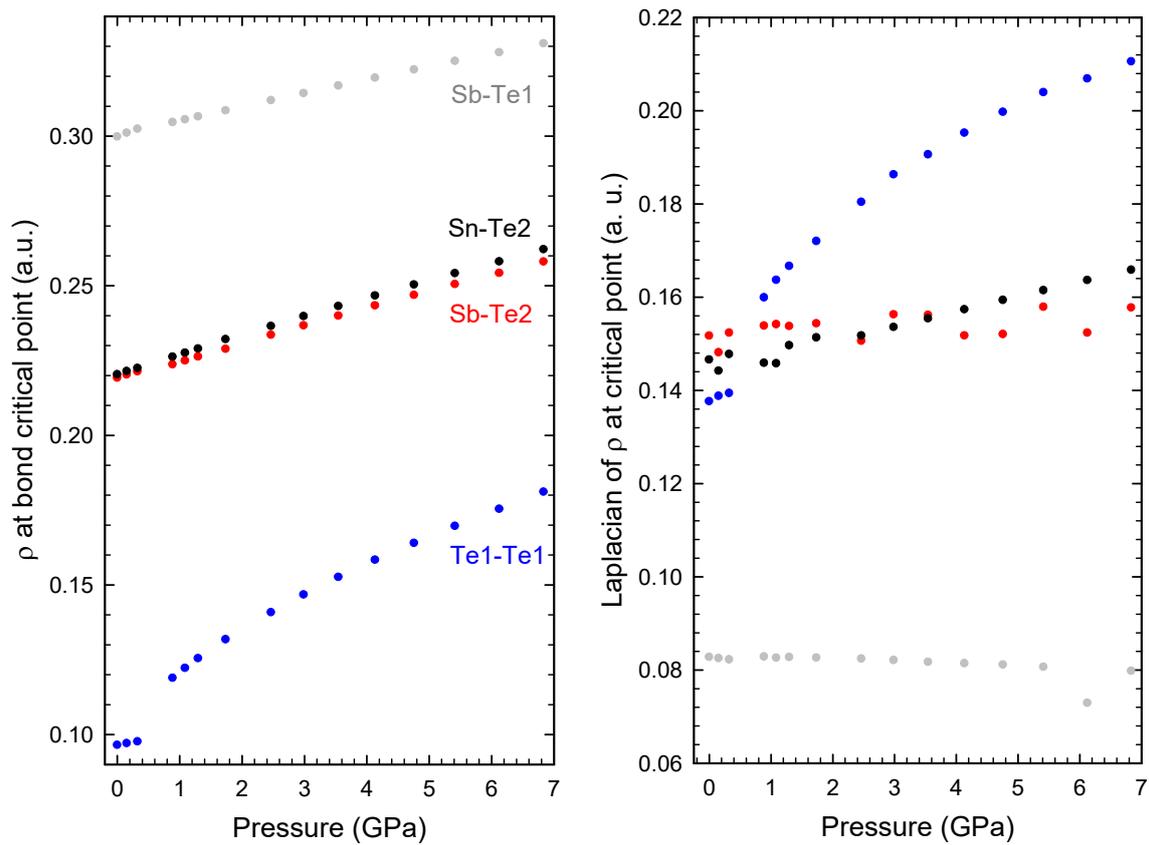

**Figure S26.** Pressure dependence of the electron density (left) and Laplacian of the electron density (right) at the bond critical point of the interactions of SnSb$_2$Te$_4$.

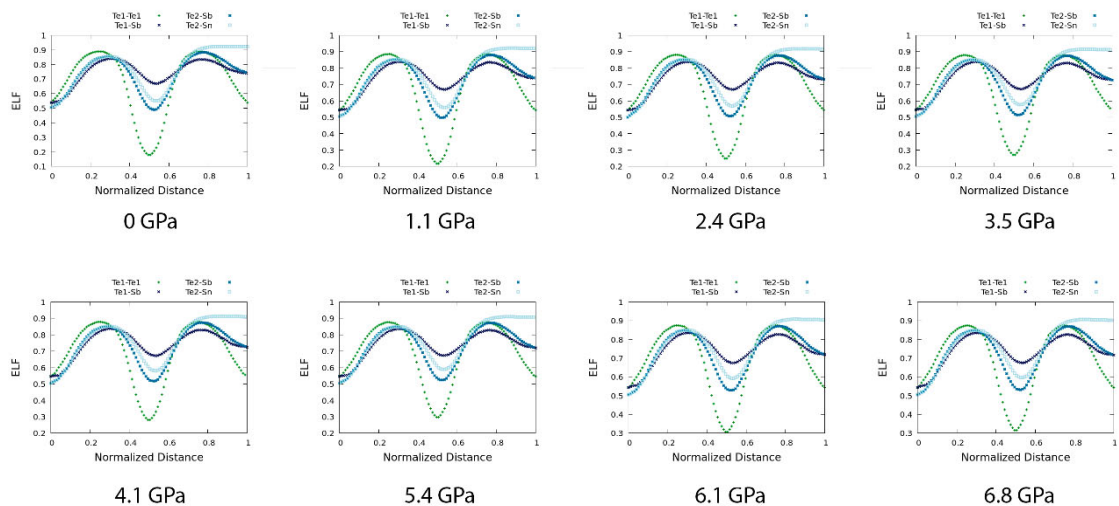

**Figure S27.** Pressure dependence of the ELF along the different bonds of SnSb$_2$Te$_4$.

**Table S1.** Frequencies and pressure coefficients at zero pressure of the experimental and theoretical Raman-active modes in SnSb$_2$Te$_4$. Theoretical values for Sb$_2$Te$_3$ are also given for comparison. Spin-orbit coupling has been included in all theoretical calculations.

| Mode symmetry | SnSb$_2$Te$_4$ Experiment | | | SnSb$_2$Te$_4$ Theory | | | α-Sb$_2$Te$_3$ Theory | | | Mode symmetry |
|---|---|---|---|---|---|---|---|---|---|---|
| | $\omega_0$ (cm$^{-1}$) | a (cm$^{-1}$/GPa) | b (cm$^{-1}$/GPa$^2$) | $\omega_0$ (cm$^{-1}$) | a (cm$^{-1}$/GPa) | b (cm$^{-1}$/GPa$^2$) | $\omega_0$ (cm$^{-1}$) | a (cm$^{-1}$/GPa) | b (cm$^{-1}$/GPa$^2$) | |
| $E_g^1$ | - | - | - | 38.9(2) | 2.5(3) | -0.08(1) | 50.4(2) | 2.6(2) | -0.09(1) | $E_g^1$ |
| $A_{1g}^1$ | 53.3(8) | 4.2(4) | -0.15(8) | 55.3(2) | 4.3(4) | -0.15(2) | 68.9(3) | 4.3(5) | -0.07(1) | $A_{1g}^1$ |
| $E_g^2$ | 103.3(4) | 3.7(3) | -0.4(2) | 100.5(3) | 2.5(2) | - | - | - | - | |
| $A_{1g}^2$ | 107.8(14) | 4.6(4) | -0.15(6) | 115.1(2) | 2.7(2) | - | - | - | - | |
| $E_g^3$ | - | - | - | 116.7(2) | 4.6(4) | - | 116.6(4) | 2.1(3) | - | $E_g^2$ |
| $A_{1g}^3$ | 160.9(4) | 2.7(2) | -0.04(1) | 167.1(3) | 3.4(3) | -0.04(1) | 167.6(6) | 2.6(4) | - | $A_{1g}^2$ |

**Table S2.** Frequencies and pressure coefficients at zero pressure of the theoretical IR-active modes in SnSb$_2$Te$_4$, Sb$_2$Te$_3$ and SnTe. Spin-orbit coupling has been included in all theoretical calculations.

| | SnSb$_2$Te$_4$ | | | α-Sb$_2$Te$_3$ and c-SnTe | | | |
|---|---|---|---|---|---|---|---|
| Mode symmetry | $\omega_0$ (cm$^{-1}$) | a (cm$^{-1}$/GPa) | b (cm$^{-1}$/GPa$^2$) | $\omega_0$ (cm$^{-1}$) | a (cm$^{-1}$/GPa) | b (cm$^{-1}$/GPa$^2$) | Mode symmetry |
| $E_u^1$ | 62.5(5) | 1.6(3) | -0.037(2) | 78.0 | 2.9 | - | $E_u^1$ |
| $E_u^2$ | 65.6(2) | 4.3(4) | - | 100.4 | 1.2 | - | $E_u^2$ |
| $A_{2u}^1$ | 81.6(3) | 1.8(2) | -0.024(2) | 39.9 | 9.1(5) | -0.31(4) | $T_{1u}$ (TO) |
| $A_{2u}^2$ | 104.0(2) | 5.8(5) | - | 112.9(8) | 1.98(17) | - | $T_{1u}$ (LO) |
| $E_u^3$ | 111.9(4) | 2.5(2) | -0.021(5) | 109.9 | 1.9 | - | $A_{2u}^1$ |
| $A_{2u}^3$ | 155.7(5) | 3.3(7) | -0.063(7) | 138.7 | 3.5 | - | $A_{2u}^2$ |

**Table S3.** Values of the electronic charge density and its gradient at the BCP as well as the ELF along different types of bondings.

| | Ionic | Covalent | Metavalent | Metallic | van der Waals |
|---|---|---|---|---|---|
| $\rho(\vec{r})$ | Large | Large | Medium | Small | Small |
| $\nabla^2 \rho(\vec{r})$ | Positive | Negative | Small (positive or negative) | Positive | Positive (small) |
| ELF | Large | Large | Medium | Small | Small |

**References**

[1] B. Cordero, V. Gómez, A. E. Platero-Prats, M. Revés, J. Echevarría, E. Cremades, F. Barragán, S. Alvarez. Covalent radii revisited. *Dalton Transactions* **2008**, *21*, 2832–2838

[2] J. E. Huheey, E. A. Keiter, R. L. Keiter. Inorganic Chemistry: Principles of Structure and Reactivity, 4th edition, HarperCollins, New York, USA, 1993.

[3] B. A. Kuropatwa, H. Kleinke. Thermoelectric Properties of Stoichiometric Compounds in the $(SnTe)_x(Bi_2Te_3)_y$ System. Z. Anorg. Allg. Chem. **2012**, *638*, 2640-2647.

[4] J.M. Skelton, L.A. Burton, A.J. Jackson, F. Oba, S.C. Parker and A. Walsh, Lattice dynamics of the tin sulphides SnS2, SnS and Sn2S3: vibrational spectra and thermal transport. Phys. Chem. Chem. Phys. **2017**, *19*, 12452.

[5] G.C. Sosso, S. Caravati and M. Bernasconi, Vibrational properties of crystalline $Sb_2Te_3$ from first principles. J. Phys.: Condens. Matter **2009**, *21*, 095410.

[6] O. Gomis, R. Vilaplana, F.J. Manjón, P. Rodríguez-Hernández, E. Pérez-González, A. Muñoz, C. Drasar, and V. Kucek, Study of the lattice dynamics of $Sb_2Te_3$ at high pressures, Physical Review B **2011**, *84*, 174305.

[7] A. M. Kulibekov, H. P. Olijnyk, A. Jephcoat, Z. Y. Salaeva, S. Onari, K. R. Allakverdiev. Raman Scattering under Pressure and the Phase Transition in ε-GaSe Phys. Stat. Sol (b) **2003**, *235*, 517-520.